\documentclass{article}
\usepackage{graphicx}
\usepackage{amsmath}
\usepackage{amsfonts}
\usepackage{amssymb}
\usepackage{ifthen}
\usepackage{bm,bbm}

\newtheorem{theorem}{Theorem}[section]
\newtheorem{lemma}[theorem]{Lemma}

\newtheorem{remark}[theorem]{Remark}
\newtheorem{definition}[theorem]{Definition}

\newcommand{\ENDofSUBSECTION}{\bigskip \bigskip}

\newcommand{\SSS}{\scriptscriptstyle}

\newcommand{\DEFINED}[1]{{\bf #1}}

\newcommand{\AUFspace}[1]{\mathcal{#1}}

\newcommand{\Ii}{{\bf i}}

\newcommand{\ZEROvect}{\mathbf{0}}

\newcommand{\PHASE}[1]{e^{\Ii #1}}

\newcommand{\STbasis}[1]{e_{#1}}

\newcommand{\bO}{\bullet}

\newcommand{\REALS}{\mathbbm{R}}

\newcommand{\CONJ}[1]{\overline{#1}}

\newcommand{\diag}{\mathbf{diag}}

\newcommand{\RELofEQUI}{\simeq}

\renewcommand{\Re}{\mathbf{Re}}
\renewcommand{\Im}{\mathbf{Im}}

\newcommand{\DIFFERENTIAL}[2]{{D \! #1}_{#2}}

\newcommand{\HADprod}{\circ}

\newcommand{\EXPentrywise}{\mathbf{EXP}}

\newcommand{\ELEMENTof}[3]{{\left[ #1 \right]}_{#2,#3}}

\newcommand{\DEPHASED}{dephased}

\newcommand{\TRANSPOSE}[1]{{\left( #1 \right)}^T}

\newcommand{\HADAMARDS}[1]{{\cal H}_{#1}}

\newcommand{\HADreprs}[1]{{\cal G}_{#1}}

\newcommand{\AUF}{affine Hadamard family}
\newcommand{\AUFs}{affine Hadamard families}
\newcommand{\maxAUF}{maximal affine Hadamard family}
\newcommand{\maxAUFs}{maximal affine Hadamard families}

\newcommand{\AUFfamily}[4][]{{#2}^{(#4)}_{#3#1}}
\newcommand{\AUFphases}[4][]{R_{{#2}^{(#4)}_{#3#1}}}

\newcommand{\PERMUTATIONequivalent}{permutation equivalent}

\newcommand{\COGNATE}{cognate}
\newcommand{\selfCOGNATE}{self--cognate}

\newcommand{\ISOLATED}{isolated}

\newcommand{\DITAtwoBYtwo}[3]
{
  \left[ \begin{array}{c|c}
            \ELEMENTof{#1}{1}{1} \cdot #2 & \ELEMENTof{#1}{1}{2} \cdot #3 \\
            \hline
            \ELEMENTof{#1}{2}{1} \cdot #2 & \ELEMENTof{#1}{2}{2} \cdot #3
         \end{array} \right]
}

\newcommand{\CSP}[2]{{\mathbf{#1}}^{(#2)}}

\begin{document}

\title{A concise guide to complex {H}adamard matrices}
 \author{Wojciech Tadej$^{1}$ and Karol \.Zyczkowski$^{2,3}$\\
\smallskip
$^1${\small Faculty of Mathematics and Natural Sciences, College of Sciences,}\\
    {\small Cardinal Stefan Wyszy{\'n}ski University, Warsaw,
      Poland}\\
\\
$^2${\small Institute of Physics, Jagiellonian University, Krak{\'o}w,
     Poland}\\
$^3${\small Center for Theoretical Physics, Polish Academy of Sciences,
     Warsaw, Poland}\\
\smallskip
        {\small e-mail: wtadej@wp.pl \quad \quad 
               karol@tatry.if.uj.edu.pl}\\
         }
\date{January 26, 2006}

\maketitle

\begin{abstract}
Complex Hadamard matrices, consisting of unimodular entries
with arbitrary phases, play an important role
in the theory of quantum information.
We review basic properties of complex Hadamard matrices and
present a catalogue of inequivalent cases known for dimension $N=2, \dots, 16$.
In particular, we explicitly write down some families of complex Hadamard
 matrices for $N=12,14$ and $16$,
which we could not find in the existing literature.
\end{abstract}

%
%

\section{Introduction}

In a 1867 paper on
'simultaneous sign-successions, tessellated pavements in two or more colors,
 and ornamental tile-work'
Sylvester used {\sl self--reciprocial} matrices,
defined as a square array of elements of
which each is proportional to its first minor \cite{Sy67}.
This wide class of matrices includes in particular
these with orthogonal rows and columns.
In 1893 Hadamard proved that such matrices
attain the largest value of the determinant
among all matrices with entries bounded by unity \cite{Ha93}.
After this paper the matrices with entries equal to $\pm 1$
and mutually orthogonal rows and columns 
were called \DEFINED{Hadamard}.

 Originally there was an interest in 
\DEFINED{real} Hadamard matrices, $H_{ij}\in {\mathbb R}$, which found
diverse mathematical applications, in particular
in error correction and coding theory \cite{Ag85}
and in the design of statistical experiments \cite{HSS99}.
Hadamard proved that such matrices
may exist only for $N=2$ or for size $N$ being a multiple of $4$
and conjectured that they exist for all such $N$.
A huge collection of real Hadamard matrices
for small values of $N$ is known (see e.g. \cite{Slo05,Seb05}),
but the original Hadamard conjecture remains unproven.
After a recent discovery of $N=428$ real Hadamard matrix \cite{KT04},
the case $N=668$ is the smallest order, for which
the existence problem remains open.

Real Hadamard matrices may be generalized in various ways.
Butson introduced a set $H(q,N)$ of Hadamard matrices
of order $N$, the entries of which are $q$-th roots
of unity \cite{Bu62,Bu63}. Thus $H(2,N)$ represents real Hadamard matrices,
while $H(4,N)$ denotes Hadamard matrices\footnote{These matrices
were also called \DEFINED{complex} Hadamard matrices
\cite{Tu70,Wa73,MS96,CHK97}.}
with entries $\pm 1$ or $\pm i$. If $p$ is prime, then $H(p,N)$
can exist if $N=mp$ with an  integer $m$ \cite{Bu62}, and it is conjectured
that they exist for all such  cases \cite{Be98}.

In the simplest case $m=1$  Hadamard matrices $H(N,N)$
exist for any dimension.
An explicit construction based on \DEFINED{Fourier matrices}
\begin{equation}
  \label{Fourier}
  \ELEMENTof{F_N'}{j}{k}:= 
       \frac{1}{\sqrt{N}} 
       \PHASE{(j-1)(k-1) \frac{2\pi}{N}}
       {\rm \quad with \quad} j,k \in \{1,\ldots,N\}
\end{equation}
works for an arbitrary $N$, not necessarily prime. A Fourier
matrix is unitary, but a rescaled matrix $F_N=\sqrt{N}F_N'$
belongs to $H(N,N)$. In the following we are going to use Hadamard
matrices with unimodular entries, but for
convenience we shall also refer to $F_N$ as Fourier matrix.

For $m=2$ it is not difficult \cite{Bu62}
to construct matrices $H(p,2p)$.
In general, the problem of finding all pairs $\{q,N \}$
for which Butson--type matrices $H(q,N)$
do exist, remains unsolved \cite{HC98},  even though
some results on non-existence are available \cite{Wi00}.
The set of  $p$--th roots of unity
forms  a finite group and it is possible to generalize the
notion of Butson--type matrices for any finite
group \cite{Dr79,Ha97,PH98}.
\bigskip

In this work we will be interested in a more general case of
\DEFINED{complex Hadamard matrices},
for which there are no restrictions on phases
of each entry of $H$. Such a
matrix, also called \DEFINED{biunitary} \cite{Ni04},
corresponds to taking
the limit $q\to \infty$ in the definition of Butson.
In this case there is no analogue of Hadamard conjecture,
since complex Hadamard matrices exists for any dimension.
However, in spite of many years of research,
the problem of finding all complex
Hadamard matrices of a given size $N$
is still open \cite{Ha96,Di04}.

Note that this problem may be considered as a particular
case of a more general issue 
of specifying all unitary matrices
such that their squared moduli give a fixed 
doubly stochastic matrix \cite{AMM91,BEKTS05}.
This very problem was intensively studied 
by high energy physicists
investigating the parity violation
and analyzing the Cabibbo--Kobayashi--Maskawa matrices
\cite{Ja85,BD87,NP87,Di05}.

On one hand, the search for complex Hadamard matrices is
closely related to various mathematical problems, including
construction of some  $*$-subalgebras in finite von Neumann
algebras \cite{Po83,HJ90,MW92,Ha96}, analyzing bi-unimodular
sequences or finding cyclic $n$--roots \cite{BF94,BS95} and
equiangular lines \cite{GR05}. Complex Hadamard matrices were
used to construct error correcting codes \cite{HC98} and to
investigate the spectral sets and Fuglede's conjecture
\cite{Ta04,Ma04,KM04a,KM04}.

On the other hand,
Hadamard matrices find numerous applications
in several problems of theoretical physics.
For instance, Hadamard matrices
(rescaled by $1/\sqrt{N}$ to achieve unitarity),
are known in quantum optics as
symmetric multiports \cite{RZBB94,JSZ95}
(and are sometimes called Zeilinger matrices)
and may be used 
to construct spin models \cite{GR05},
or schemes of selective coupling
of a multi--qubit system \cite{Le02}.

Complex Hadamard matrices play a crucial role in the 
{\sl theory of quantum information}
as shown in a seminal paper of  Werner \cite{We01}.
They are used 
in solving the Mean King Problem \cite{VAA87,EA01,KR05}, 
and in finding 'quantum designs' \cite{Za99}.
Furthermore, they allow one to construct

{\bf a) Bases of unitary operators}, i.e. the set 
of mutually orthogonal unitary operators,
$\{ U_k \}_{k=1}^{N^2}$ such that 
 $U_k \in {\cal U}(N)$ and
Tr$U_k^{\dagger}U_l=N\delta_{kl}$ for $k,l=1,\dots, N^2$, 

{\bf b) Bases of maximally entangled states}, 
i.e. the set 
$\{ |\Psi_k\rangle \}_{k=1}^{N^2}$ such that 
each $| \Psi_k\rangle $ belongs to a composed Hilbert space
with the partial trace  
Tr$_N(| \Psi_k \rangle \langle \Psi_k|)={\mathbbm 1}/N$, 
and they are mutually orthogonal, 
$\langle \Psi_k|\Psi_l \rangle =\delta_{kl}$ \cite{WGC03},

{\bf c) Unitary depolarisers}, i.e. the set  
$\{ U_k \}_{k=1}^{N^2}$ such that for any 
bounded linear operator $A$ 
the property 
 $\sum_{k=1}^{N^2} U_k^{\dagger} A U_k =N $Tr$A\  {\mathbbm 1}$
holds. 

\smallskip

The problems {\bf a)}-{\bf c)} are equivalent in the sense that
given a solution to one problem one can find a 
solution to the other one, as well as a 
corresponding scheme of 
teleportation or dense coding \cite{We01}.
In particular Hadamard matrices are usefull 
to construct a special class of unitary bases 
of a group type, 
also called 'nice error basis' \cite{Kn96,KR02}

Another application of Hadamard matrices
is related to quantum tomography:
To determine all $N^2-1$ parameters 
characterizing a density matrix of size $N$
one needs to perform $k\ge N+1$ orthogonal measurements.
Each measurement can be specified by an orthogonal basis
$\Phi_u=\{|\phi_i^{(u)}\rangle \}_{i=1}^N$ set for $u=1,\dots,k$. 
Precision of  such a measurement scheme is optimal
if the bases are \DEFINED{mutually unbiased}, i.e. they are such that
\begin{equation}
  \label{MUB}
|\langle \phi_i^{(u)} | \phi_j^{(s)}\rangle|^2 =  \frac{1}{N}(1-\delta_{us})
 + \delta_{us}\delta_{ij} \ .
\end{equation}
If the dimension $N$ is prime or a power of prime 
the number of maximally unbiased bases (MUB) is equal to $N+1$
\cite{Iv81,WF91},
but for other dimensions the answer to this question 
is still unknown \cite{KR03,Gr04,BSTW05b}. The task  
of finding $(k+1)$ MUBs is equivalent to finding
a collection of $k$ 
\DEFINED{mutually unbiased Hadamards} (MUH), 
\begin{equation}
\label{MUH}
\{H_i\in {\cal H}_N\}_{i=1}^k : \quad 
\frac{1}{\sqrt{N}}  H_i^{\dagger} H_j 
\in {\cal H}_N,
\quad i>j=1,\dots , k-1,
\end{equation}
since the set $\{{\mathbbm 1}, H_1/\sqrt{N}, \dots, H_k/\sqrt{N} \}$
forms a set of MUBs. Here ${\cal H}_N$
denotes the set of
complex\footnote{Similarly, knowing unbiased real Hadamard matrices
one constructs real MUBs \cite{BSTW05}.}
 Hadamard matrices of size $N$.

\bigskip

The aim of this work is to review properties
 of complex Hadamard matrices and
to provide a handy collection
of these matrices of size ranging from $2$ to $16$.
Not only we list concrete Hadamard matrices,
the existence of which follows from recent
papers by Haagerup \cite{Ha96} and Di{\c t}{\v a} \cite{Di04},
but also we present several other Hadamard matrices
which have not appeared in the literature so far.

%
%
%

\section{Equivalent Hadamard matrices and the \DEPHASED\ form}

We shall start this section providing some formal definitions.


\begin{definition}
  \label{def_Hadam1}

  A square matrix $H$ of size $N$ consisting of unimodular entries,
  $|H_{ij}|=1$,
  is called a \DEFINED{Hadamard matrix} if
  \begin{equation}
    \label{Hadam11}
    HH^{\dagger}=N \: {\mathbbm 1} \ ,
  \end{equation}
  where $^{\dagger}$ denotes 
the Hermitian transpose. One distinguishes
  
  \begin{description}
     \item[a)] \DEFINED{real Hadamard matrices}, $H_{ij}\in {\mathbb  R}$, %
for $i,j=1,\dots,N$,
     \item[b)] \DEFINED{Hadamard matrices of \DEFINED{Butson type}} $H(q,N)$,
for which $(H_{ij})^q=1$,
     \item[c)] \DEFINED{complex Hadamard matrices}, $H_{ij}\in {\mathbb
          C}$. 
  \end{description}

\end{definition}

The set of all Hadamard matrices of size $N$ will be denoted by
$\HADAMARDS{N}$.
\bigskip


\begin{definition}
  \label{def_equiv1}

  Two Hadamard matrices $H_1$ and $H_2$
  are called \DEFINED{equivalent},
  written $H_1 \RELofEQUI H_2$,
  if there exist diagonal unitary matrices $D_1$ and $D_2$
  and permutations matrices $P_1$ and $P_2$ such that \cite{Ha96}
  \begin{equation}
    \label{equival}
    H_1=D_1 P_1\; H_2\; P_2 D_2 \ .
  \end{equation}
\end{definition}

This equivalence relation may be considered
as a generalization of the Hadamard equivalence
in the set of real Hadamard matrices, in which
permutations and negations of rows and columns are 
allowed.\footnote{Such an equivalence relation may be extended
to include also transposition and complex conjugation \cite{Di05}.
Since the transposition of a matrix is not realizable in 
physical systems we prefer to stick to the original definition
of equivalence.}
\medskip


\begin{definition}
  \label{def_dephased}

  A complex Hadamard matrix is called \DEFINED{\DEPHASED}\ when the
  entries of its first row and column are all equal to 
unity\footnote{in case of real Hadamard matrices such a form is called
\DEFINED{normalised}.},
  \begin{equation}
    \label{dephas1}
    H_{1,i} = H_{i,1} = 1 \quad  {\rm \quad for \quad} i=1,\dots,N \ .
\end{equation}
\end{definition}


\begin{remark}
  For any complex $N \times N$ Hadamard matrix $H$ there exist
  uniquely determined diagonal unitary matrices,
  $D_r=\diag({\bar H}_{11}, {\bar H}_{21},\dots, {\bar H}_{N1})$,
  and
  $D_c=\diag(1, H_{11}{\bar H}_{12},\dots, H_{11}{\bar H}_{1N})$,
  such that
  $\ELEMENTof{D_c}{1}{1} = 1$ and
  \begin{equation}
    \label{eq_dephasing}
    D_r \cdot H \cdot D_c
  \end{equation}
  is \DEPHASED.
\end{remark}

Two Hadamard matrices with the same \DEPHASED\ form are equivalent.
Thus the relevant information on a Hadamard matrix
is carried by the lower right submatrix of size $N-1$,
called the \DEFINED{core} \cite{Bu63}.

It is often useful to define a \DEFINED{log--Hadamard} matrix $\Phi$,
such that
\begin{equation}
  \label{logHad}
  H_{kl}=\PHASE{\Phi_{kl}} \ ,
\end{equation}
is Hadamard. The {\bf phases} $\Phi_{kl}$
 entering a log-Hadamard matrix
may be chosen to belong to $[0,2\pi)$.
This choice of phases implies
that the matrix $q\Phi /2\pi$ corresponding to a Hadamard
matrix of the Butson type $H(q,N)$ consists
of zeros and integers smaller than $q$.
All the entries of the first row and column of a log Hadamard
matrix $\Phi$, corresponding to a \DEPHASED\ Hadamard matrix, are equal to
zero.

To illustrate the procedure of dephasing consider the Fourier-like
matrix of size four, $\ELEMENTof{{\tilde F}_4}{j}{k}:=  \PHASE{jk 2\pi/4}$
where $j,k \in \{1,2,3,4\}$. Due to this choice of entries the
matrix ${\tilde F}_4$ is not \DEPHASED, but after operation
(\ref{eq_dephasing}) it takes the \DEPHASED\ form $F_4$,
\begin{equation}
  \label{G4}
  {\tilde F}_4=
  \left[
    \begin{array}{rrrr}
      i  & -1 & -i & 1 \\
      -1 & 1 & -1 & 1 \\
      -i & -1 & i & 1 \\
      1  & 1  & 1 & 1
    \end{array}
  \right]\ ,
  \quad\quad \quad
  F_4= 
  \left[
    \begin{array}{rrrr}
      1 & 1  & 1 &   1\\
      1 & i  & -1 & -i\\
      1 & -1 &  1 & -1\\
      1 & -i & -1 & i
    \end{array}
  \right] \ .
\end{equation}
The corresponding log--Hadamard matrices read
\begin{equation}
  \label{R4}
  {\tilde \Phi}_4=
  \frac{2 \pi}{4} 
  \left[
    \begin{array}{rrrr}
      1  & 2 & 3 & 0 \\
      2 & 0 & 2 & 0 \\
      3 & 2 & 1 & 0 \\
      0  & 0  & 0 & 0
    \end{array}
  \right] \ ,
  \quad\quad \quad
  \Phi_4=
  \frac{2 \pi}{4}
  \left[
    \begin{array}{rrrr}
      0 & 0  & 0 & 0\\
      0 & 1  & 2 & 3\\
      0 & 2 &  0 & 2\\
      0 & 3 &  2 & 1
    \end{array}
  \right] \ .
\end{equation}
Note that in this case dephasing is equivalent
to certain permutation of rows and columns, but in general
the role of both operations is different.
It is straightforward to perform  (\ref{eq_dephasing})
which brings any Hadamard matrix into the \DEPHASED\ form.
However, having two matrices in such a form it might not be easy to
verify, whether there exist permutations $P_1$ and $P_2$
necessary to establish
the equivalence relation (\ref{equival}).
This problem becomes difficult for larger
matrix sizes, since the number of possible permutations grows as
$N!$. What is more, combined multiplication by unitary diagonal and permutation
matrices  may still be necessary.

To find necessary conditions for equivalence one
may introduce various invariants of operations allowed in
(\ref{equival}), 
and compare them for two matrices investigated.
In the case of real Hadamard matrices and permutations only,
the number of negative elements in the \DEPHASED\ form
may serve for this purpose.
Some more advanced methods for detecting inequivalence
were recently proposed by Fang and Ge \cite{FG04}.
It has been known for many years that
for $N=4,8$ and $12$ all real Hadamard matrices are equivalent,
while the number of equivalence classes for
$N=16,20,24$ and $28$ is equal to $5,3,60$ and $487$, respectively.
For higher dimensions this number grows dramatically:
For $N=32$ and $36$ the number of inequivalent
matrices is not smaller than $3,578,006$ and $4,745,357$,
but the problem of enumerating all of them remains open
-- see e.g. \cite{Or05}.

To characterize a complex Hadamard matrix $H$
let us define a set of coefficients,
\begin{equation}
  \label{invHag1}
  \Lambda =
  \{
    \Lambda_{\mu}:= H_{ij}\CONJ{H_{kj}} H_{kl}\CONJ{H_{il}}: 
    \ \ \ {\mu} = (i,j,k,l) \in {\{1,\dots,N\}}^{\times 4}
  \} \ ,
\end{equation}
where no summation over repeating indeces is assumed and
$\mu$ stands for a composite index $(i,j,k,l)$. Due to
complex conjugation of the even factors in the above definition
any concrete value of $\Lambda_{\mu}$ is invariant with respect to
multiplication by diagonal unitary 
matrices.\footnote{Such invariants
were used by physicists investigating unitary Kobayashi--Maskawa matrices of
size $3$ and $4$ \cite{Ja85,BD87,NP87,Di05}.}
Although this value
may be altered by a permutation, the entire set $\Lambda$ is
invariant with respect to operations allowed in (\ref{equival}).
This fact allows us to state the Haagerup condition for
inequivalence  \cite{Ha96},


\begin{lemma}
  \label{haglemm}
   If two Hadamard matrices have different sets  $\Lambda$
 of invariants  (\ref{invHag1}), they are not equivalent.
\end{lemma}

The above criterion  works in one direction only.
For example, any Hadamard matrix $H$ and its transpose,  $H^T$,
possess the same sets $\Lambda$, but they need not to be
permutation equivalent.

Some other equivalence criteria are
dedicated to certain special cases.
The tensor product of two Hadamard
matrices is also a Hadamard matrix:
\begin{equation}
  H_1 \in \HADAMARDS{M}\ \ \ \mbox{and}\ \ \  H_2 \in \HADAMARDS{N} 
  \ \ \ \Longrightarrow\ \ \  
  H_1 \otimes H_2 \in \HADAMARDS{MN}
\end{equation}
and this fact will be used  to construct
Hadamard matrices of a larger size.
Of particular importance are tensor products
of Fourier matrices,
$F_M \otimes F_N \in \HADAMARDS{MN}$.
For arbitrary dimensions both
tensor products are equivalent,
$F_N \otimes F_M \RELofEQUI  F_M \otimes F_N$.
However, their equivalence with $F_{MN}$
depends on the number theoretic property of the product $M \cdot N$ :
the equivalence holds if $M$ and $N$ are relatively prime \cite{Ta05}.

To classify and compare various Hadamard matrices
we are going to use the \DEPHASED\ form  (\ref{dephas1})
thus fixing the first row and column in each compared matrix.
However, the freedom of permutation
inside the remaining submatrix of size $(N-1)$
does not allow us to specify a unique 'canonical form'
for a given complex Hadamard matrix.
In other words we are going to list only
certain representatives of each equivalence class known,
but the reader has to be warned
that other choices of equivalent representatives are
equally legitimate.
For instance, the one parameter family
of $N=4$ complex Hadamard matrices
presented in \cite{Ha96,Di04}
contains all parameter-dependent elements of $H$
in its lower-right corner,
while our choice (\ref{F4_maxAUF_explicit})
with variable phases in second and fourth row
is due to the fact
that such an orbit stems  from the Fourier  matrix $F_4$.
\medskip

Identifying matrices equivalent with
respect to (\ref{equival})
we denote by
\begin{equation}
  \HADreprs{N} = \HADAMARDS{N} / \RELofEQUI
\end{equation}
every set of representatives of different equivalence classes.
\medskip

Interestingly $\HADreprs{N}$ is known only for $N=2,3,4,5$, while
the problem of finding \DEFINED{all} complex Hadamard matrices for $N
\ge 6$ remains unsolved. In particular, compiling our list of
Hadamard matrices we took into account all continuous families known to us,
but in several cases it is not clear, whether there exist any
equivalence relations between them.

%
%

\section{Isolated Hadamard matrices and continuous orbits of inequivalent
matrices}

In this and the following sections we shall use the symbol
$\HADprod$ to denote the Hadamard product of two matrices,
\begin{equation}
 \ELEMENTof{H_1 \HADprod H_2}{i}{j} =
                      \ELEMENTof{H_1}{i}{j} \cdot
                      \ELEMENTof{H_2}{i}{j}\ ,
\label{Hadprod}
\end{equation}
and the $\EXPentrywise$ symbol to denote
the entrywise $\exp$ operation on a
matrix,
\begin{equation}
\ELEMENTof{\EXPentrywise(R)}{i}{j} =
 \exp\left(\ELEMENTof{R}{i}{j}\right) \ .
\label{EXPfunct}
\end{equation}

%
%

\subsection{Isolated Hadamard matrices}


\begin{definition}
  \label{def_isolated}

  A \DEPHASED\ $N \times N$ complex Hadamard matrix $H$ is called
  \DEFINED{\ISOLATED}\ if there is a neighbourhood $\mathcal{W}$
  around $H$ such that there are no other \DEPHASED\ complex Hadamard
  matrices in $\mathcal{W}$.

\end{definition}

To have a tool useful in determining
whether a given \DEPHASED\ complex Hadamard matrix of size
$N$ is isolated, we introduce the notion of \DEFINED{defect}:

%
\begin{definition}
  \label{def_defect}
  \newcommand{\SPACES}{\ \ \ \ \ \ \ \ }

  The \DEFINED{defect} $d(H)$ of an $N \times N$ complex Hadamard 
  matrix $H$ is the
  dimension of the solution space of the real linear system with
  respect to a matrix variable $R \in {\mathbbm R}^{N^2}$\ \ :
  \begin{eqnarray}
    R_{1,j} & = \ 0 & \SPACES j \in \{2,\ldots,N\}   
\label{eq_first_row_deph_system} \\
    R_{i,1} & = \ 0 & \SPACES i \in \{1,\ldots,N\}   
\label{eq_first_col_deph_system} \\
    \sum_{k=1}^{N} H_{i,k} \CONJ{H}_{j,k} \left( R_{i,k} - R_{j,k}
    \right)  & = \ 0 &  \SPACES 1 \leq i < j \leq N  
\label{eq_rows_orthogonality_system}
  \end{eqnarray}
   
\end{definition}

The defect allows us to formulate a 
{\sl one way} criterion:


\begin{lemma}
  \label{lem_zero_defect_isolation}
   A \DEPHASED\ complex Hadamard matrix $H$ is isolated if the
   defect of $H$ is equal to zero.
\end{lemma}

If N is prime then the defect of $F_N$ is zero,
so the Fourier matrix is isolated,
as earlier shown in \cite{Ni04,Pe97}.
For any composed $N$ the defect 
of the Fourier matrix is positive.
For instance, if the dimension $N$
is a product of two distinct primes, then
$d(F_{pq})=2(p-1)(q-1)$,
while for powers of a prime, $N=p^k$ with $k\ge 2$, 
the defect reads
\begin{equation}
\label{powprime}
d(F_{p^k})\ = \ p^{k-1}[k(p-1)-p]+1 \ .
\end{equation}
An explicit formula for $d(F_N)$
for an arbitrary composed $N$ 
is derived elsewhere \cite{Ta06b}.
In that case the Fourier matrix belongs to a
continuous family, as examples show in section \ref{sec_catalogue}.
\bigskip

The reasoning behind the criterion of 
Lemma \ref{lem_zero_defect_isolation} runs as follows:
\medskip

Any \DEPHASED\ complex Hadamard matrices, in particular those in a
neighbourhood of a \DEPHASED\ complex Hadamard matrix $H$, must be of the form:
\begin{equation}
  \label{eq_H_generated_Hadamard}
  H \HADprod \EXPentrywise (\Ii \cdot R)
\end{equation}
where an $n \times n$ real matrix $R$ satisfies 'dephased property'
and unitarity conditions for (\ref{eq_H_generated_Hadamard}):
{
\newcommand{\SPACES}{\ \ \ \ \ \ \ \ }
\begin{eqnarray}
  R_{1,j} & = 0 & \SPACES j \in \{2,\ldots,N\} \label{eq_first_row_dephased} \\
  R_{i,1} & = 0 & \SPACES i \in \{1,\ldots,N\} \label{eq_first_col_dephased} \\
  -\Ii \cdot \sum_{k=1}^{N} H_{i,k} \CONJ{H}_{j,k} \PHASE{\left( R_{i,k} -
      R_{j,k} \right)}  
  & = 0 &  \SPACES 1 \leq i < j \leq N     \label{eq_rows_orthogonality}
\end{eqnarray}
}

We will rewrite these conditions using a real vector function $f$, whose
coordinate functions will be indexed by the values (symbol sequences)
from the set $\mathcal{I}$, related to the standard index set by a
fixed bijection (one to one map) 
$\beta:\ \mathcal{I} \longrightarrow \{1,2,\ldots,(2N-1)+(N^2-N)\}$. The set
$\mathcal{I}$ reads:
\begin{eqnarray}
  \mathcal{I} &  =  & \{\ (1,2),\ (1,3),\ \ldots\ ,\ (1,N)\ \}\ \cup\ 
                      \{\ (1,1),\ (2,1),\ \ldots\ ,\ (N,1)\  \} \nonumber\\
              &     & \cup\ \ \ 
                      \{\ \ \ (i,j,t):\ \ \  1 \leq i < j \leq N\ \ \
                      \mbox{and}\ \ \  
                      t \in \{\Re,\ \Im\}\ \ \ \}
\end{eqnarray}
For simplicity of notation we further write $f_i,\ \ i\in
\mathcal{I}$\ \ to denote $f_{\beta(i)}$.

Similarly, a fixed bijection 
$\alpha:\ \{1,\ldots,N\}\times\{1,\ldots,N\} \longrightarrow
\{1,\ldots,N^2\}$ allows matrix indexing the components of a real $N^2$
element vector variable $R$, an argument to $f$, and we write
$R_{k,l}$ to denote $R_{\alpha(k,l)}$
\medskip

The $(2N-1) + (N^2-N)$ element function vector $f$ is defined by the
formulas:
\begin{eqnarray}
  f_{(1,j)} = R_{1,j}  &  &  j \in \{2,\ldots,N\}
  \label{eq_f_1st_row_deph} \\
  f_{(i,1)} = R_{i,1}  &  &  i \in \{1,\ldots,N\}
  \label{eq_f_1st_col_deph} \\
  f_{(i,j,\Re)} = 
  \Re\left(
    -\Ii \cdot \sum_{k=1}^{N} H_{i,k} \CONJ{H}_{j,k} \PHASE{\left( R_{i,k} -
        R_{j,k} \right)}
    \right)            &  &  1 \leq i < j \leq N
  \label{eq_f_Re_unitarity} \\
  f_{(i,j,\Im)} = 
  \Im\left(
    -\Ii \cdot \sum_{k=1}^{N} H_{i,k} \CONJ{H}_{j,k} \PHASE{\left( R_{i,k} -
        R_{j,k} \right)}
    \right)            &  &  1 \leq i < j \leq N
  \label{eq_f_Im_unitarity} 
\end{eqnarray}

Conditions
(\ref{eq_first_row_dephased},\ref{eq_first_col_dephased},\ref{eq_rows_orthogonality})
can now be rewritten as:
\begin{equation}
  \label{eq_f_system}
  f(R) = \ZEROvect
\end{equation}
where the $\alpha(i,j)$-th coordinate of a real variable vector $R$
represents the $i,j$-th entry of the corresponding matrix $R$ sitting in
(\ref{eq_H_generated_Hadamard}).
\medskip

The value of the linear map $\DIFFERENTIAL{f}{\ZEROvect}:
\REALS^{N^2} \longrightarrow \REALS^{(2N-1)+(N^2-1)}$, being the
differential of $f$ at $\ZEROvect$, at $R$ is the vector:
\begin{eqnarray}
  {[\DIFFERENTIAL{f}{\ZEROvect}(R)]}_{(1,j)} = R_{1,j}  
  &  &  j \in \{2,\ldots,N\}\ \ \ \ \ \ \ \ 
  \label{eq_Dfat0_1st_row_deph} \\
  {[\DIFFERENTIAL{f}{\ZEROvect}(R)]}_{(i,1)} = R_{i,1}  
  &  &  i \in \{1,\ldots,N\}\ \ \ \ \ \ \ \
  \label{eq_Dfat0_1st_col_deph} \\
  {[\DIFFERENTIAL{f}{\ZEROvect}(R)]}_{(i,j,\Re)} = 
  \Re\left(
    \sum_{k=1}^{N} H_{i,k} \CONJ{H}_{j,k} \left( R_{i,k} - R_{j,k} \right)
  \right)            
  &  &  1 \leq i < j \leq N\ \ \ \ \ \ \ \ 
  \label{eq_Dfat0_Re_unitarity} \\
  {[\DIFFERENTIAL{f}{\ZEROvect}(R)]}_{(i,j,\Im)} = 
  \Im\left(
    \sum_{k=1}^{N} H_{i,k} \CONJ{H}_{j,k} \left( R_{i,k} - R_{j,k} \right)
  \right)            
  &  &  1 \leq i < j \leq N\ \ \ \ \ \ \ \ 
  \label{eq_Dfat0_Im_unitarity}
\end{eqnarray}
where again indexing for $f$ defined by $\beta$ is used.
\medskip

It is clear now that the kernel of the differential, 
$\{R \in \REALS^{N^2}:\ \DIFFERENTIAL{f}{\ZEROvect}(R) = \ZEROvect\}$, corresponds to the
solution space of system 
(\ref{eq_first_row_deph_system},\ref{eq_first_col_deph_system},\ref{eq_rows_orthogonality_system}),
in which $R$ now takes the meaning of an input variable vector to
$f$, with indexing determined by $\alpha$

Note that the $N^2 - N$ equation subsystem
(\ref{eq_rows_orthogonality_system}):
\begin{equation}
  \sum_{k=1}^{N} H_{i,k} \CONJ{H}_{j,k} \left( R_{i,k} - R_{j,k}
    \right)\ \  =\ \  0\ \ \ \ \ \  \ \ \ \ \   1 \leq i < j \leq N
\end{equation}
is solved at least by the
$(2N-1)$-dimensional real space spanned by $2N-1$ vectors, defined by:
\begin{equation}
  R_{k,l} = 
  \left\{
      \begin{array}{cc}
        1 & \mbox{  for  } (k,l) \in \{1,\ldots,N\} \times \{j\} \\
        0 & \mbox{  otherwise  }
      \end{array}
  \right.
  ,\ \ \ \ j \in \{2,\ldots,N\}
\end{equation}
and
\begin{equation}
  R_{k,l} = 
  \left\{
      \begin{array}{cc}
        1 & \mbox{  for  } (k,l) \in \{i\} \times \{1,\ldots,N\} \\
        0 & \mbox{  otherwise  }
      \end{array}
  \right.
  ,\ \ \ \ i \in \{1,\ldots,N\}
\end{equation}
that is, by vectors, if treated as matrices (with the $i,j$-th
entry being equal to the $\alpha(i,j)$-th coordinate of the corresponding variable
vector), forming matrices with either a row or a column filled all
with $1$'s and the other entries being $0$'s.
\medskip

If the defect of $H$ equals $0$, then the overall system
(\ref{eq_first_row_deph_system},\ref{eq_first_col_deph_system},\ref{eq_rows_orthogonality_system})
is solved only by $\ZEROvect$, the differential
$\DIFFERENTIAL{f}{\ZEROvect}$ has full rank $N^2$, i.e. 
$\dim\left(\DIFFERENTIAL{f}{\ZEROvect}(\REALS^{N^2})\right) = N^2$,
and we can choose an $N^2$ equation subsystem:
\begin{equation}
  \label{eq_f_subsystem}
  \tilde{f}(R)\ \ =\ \ \ZEROvect
\end{equation}
of (\ref{eq_f_system}) such that the differential
$\DIFFERENTIAL{\tilde{f}}{\ZEROvect}$ at $\ZEROvect$ is of rank
$N^2$,
i.e.
$\dim\left(\DIFFERENTIAL{\tilde{f}}{\ZEROvect}(\REALS^{N^2})\right) =
N^2$, and thus $\tilde{f}$ satisfies The Inverse Function
Theorem. The theorem implies in our case that, in a neighbourhood of
$\ZEROvect$,\ \  $R = \ZEROvect$ is the only solution to
(\ref{eq_f_subsystem}), as well as the only solution to
(\ref{eq_f_system}).
\bigskip

Also in this case, let us consider the differential, at $\ZEROvect$,
of the partial function vector $f^{\mathcal{U}}$, given by
(\ref{eq_f_Re_unitarity},\ref{eq_f_Im_unitarity}). This differential
value at $R$, $\DIFFERENTIAL{f^{\mathcal{U}}}{\ZEROvect}(R)$, is given
by the partial vector
(\ref{eq_Dfat0_Re_unitarity},\ref{eq_Dfat0_Im_unitarity}).
Since the 'remaining' differential, corresponding to the \DEPHASED\
'property condition' part of $f$, defined by 
(\ref{eq_Dfat0_1st_row_deph},\ref{eq_Dfat0_1st_col_deph}), is of rank
$(2N-1)$, the rank of $\DIFFERENTIAL{f^{\mathcal{U}}}{\ZEROvect}$ is
equal to $N^2 - (2N-1)$. Recall that from considering above the minimal
solution space of system (\ref{eq_rows_orthogonality_system}), it
cannot be greater than $N^2 - (2N-1)$. Were it smaller, the rank of
$\DIFFERENTIAL{f}{\ZEROvect}$ would be smaller than $N^2$, which
cannot be if the defect of $H$ is $0$, see above.

Then one can choose an $N^2 - (2N-1)$ equation subsystem
$\tilde{f}^{\mathcal{U}}(R)\ \ =\ \ \ZEROvect$\ \ of system 
$f^{\mathcal{U}}(R) = \ZEROvect$, with the full rank:
\begin{equation}
  \dim\left(  
    \DIFFERENTIAL{\tilde{f}^{\mathcal{U}}}{\ZEROvect}( \REALS^{N^2} )
  \right) = N^2 - (2N-1)
\end{equation}
thus defining a $(2N-1)$ dimensional manifold around $\ZEROvect$. This
manifold generates, by (\ref{eq_H_generated_Hadamard}), the $(2N-1)$
dimensional manifold containing all, not necessarily \DEPHASED,
complex Hadamard matrices in a neighbourhood of $H$. In fact, the
latter manifold is equal, around $H$, to the $(2N-1)$ dimensional
manifold of matrices obtained by left and right multiplication of $H$
by unitary diagonal matrices:
\begin{equation}
  \left\{\ \ \  
    \diag(\PHASE{\alpha_1},\ldots,\PHASE{\alpha_N}) \cdot
    H \cdot
    \diag(1,\PHASE{\beta_2},\ldots,\PHASE{\beta_N})\ \ \ 
  \right\}
\end{equation}

%
%

\subsection{Continuous orbits of Hadamard matrices}

The set of inequivalent Hadamard matrices is finite for $N=2$ and $N=3$,
but already for $N=4$ there exists a continuous, one parameter
family of equivalence classes.
 To characterize such orbits
we will introduce the notion of an
\AUF.


\begin{definition}
  \label{def_AUF}

  An \DEFINED{\AUF}\ $H(\AUFspace{R})$ stemming from a \DEPHASED\
  $N \times N$ complex Hadamard matrix $H$
is the set of matrices satisfying (\ref{Hadam11}),
 associated with a subspace
  $\AUFspace{R}$ of a space of all real $N \times N$ matrices with
  zeros in the first row and column, 
  \begin{equation}
    H(\AUFspace{R}) =
       \{H \HADprod \EXPentrywise(\Ii \cdot R)\ :\ R \in \AUFspace{R} \}
\ .
  \end{equation}
\label{affhadfam}
\end{definition}

The words 'family' and 'orbit' denote
submanifolds of $\mathbb{R}^{2n^2}$ consisting purely of \DEPHASED\ complex
Hadamard matrices.
We will often write $H(\alpha_1,\ldots,\alpha_m)$ if $\AUFspace{R}$ is
known to be an $m$-dimensional space with basis $R_1,\ldots,R_m$.
In this case, $H(\alpha_1,\ldots,\alpha_m)$ will also denote the
element of an \AUF:
\begin{eqnarray}
  H(\alpha_1,\ldots,\alpha_m) &  = & 
  H(R) \stackrel{\mbox{\tiny def}}{=}  H \HADprod \EXPentrywise(\Ii \cdot R)\\
    & \mbox{where} &  R  = \alpha_1 \cdot R_1 \ +\ \ldots\ +\ \alpha_m
    \cdot R_m  \nonumber
\end{eqnarray}
\medskip

  An \AUF\ $H(\AUFspace{R})$ stemming from a \DEPHASED\
  $N \times N$ complex Hadamard matrix $H$ is called a
  \DEFINED{\maxAUF}\ when it is not contained in any larger \AUF\
  $H(\AUFspace{R'})$ stemming from $H$:
  \begin{equation}
    \label{eq_maxAUF_space_cond}
    \AUFspace{R} \subset \AUFspace{R'}  \Longrightarrow
    \AUFspace{R} = \AUFspace{R'}
  \end{equation}
%
\medskip

Calculation of the defect of $H$, defined in the previous section, is a step
towards
determination of \AUFs\ stemming from $H$:


\begin{lemma}
  There are no \AUFs\ stemming from a \DEPHASED\
$N \times N$ complex Hadamard matrix $H$ if it is \ISOLATED, in
particular if the defect of $H$ is equal to $0$.
\end{lemma}


\begin{lemma}
  The dimension of 
a  continuous Hadamard orbit 
stemming from a \DEPHASED\ Hadamard matrix
  $H$ is not greater than the defect,  $d(H)$.
\end{lemma}
\smallskip 

A lower bound $d_c(N)$ for the maximal dimensionality 
of a continuous orbit of inequivalent Hadamard matrices of size $N$ 
was derived by Di{\c t}{\v a} \cite{Di04}.
Interestingly, for powers of a prime, $N=p^k$,
this bound coincides with the defect (\ref{powprime})
calculated at the Fourier matrix,
which provides an upper bound
for the dimension of an orbit stemming from $F_N$.
Thus in this very case 
the problem of determining $d_c(N)$
restricted to orbits including $F_N$ is solved 
and we know that the \maxAUF{} stemming from $F_N$ is not
contained in any non-affine orbit
of a larger dimension. 

\smallskip 

Finally, we introduce two notions of equivalence between \AUFs:


\begin{definition}
  \label{def_perm_equivalent_AUFs}

  Two \AUFs\ stemming from \DEPHASED\ $N \times N$ complex Hadamard matrices
  $H_1$ and $H_2$: $H_1(\AUFspace{R'})$ and $H_2(\AUFspace{R''})$,
  associated with real matrix spaces $\AUFspace{R'}$ and
  $\AUFspace{R''}$ of the same dimension, are called
  \DEFINED{\PERMUTATIONequivalent}\ if

  there exist two permutation matrices $P_r$ and $P_c$ such that
  \begin{equation}
    H_2(\AUFspace{R''})  =  P_r \cdot H_1(\AUFspace{R'}) \cdot P_c
  \end{equation}
  i.e. there is one to one correspondence, by row and column
  permutation, between the elements of $H_1(\AUFspace{R'})$ and
  $H_2(\AUFspace{R''})$.

\end{definition}

Note that permutation matrices considered in the above definition must
not shift the first row or column of a matrix.
\medskip

%
\begin{definition}
  \label{def_cognate_AUFs}

  Two \AUFs\ $H_1(\AUFspace{R'})$ and $H_2(\AUFspace{R''})$ stemming
  from \DEPHASED\ $N \times N$ complex Hadamard matrices $H_1$ and
  $H_2$, associated with real matrix spaces $\AUFspace{R'}$ and
  $\AUFspace{R''}$ of the same dimension, are called
  \DEFINED{\COGNATE}\ if
  \begin{eqnarray}
    \forall B \in H_2(\AUFspace{R''}) &   \exists A \in
    H_1(\AUFspace{R'})  &  \ \ \ \ B \RELofEQUI A^T \ , \\
    \forall A \in H_1(\AUFspace{R'}) &   \exists B \in
    H_2(\AUFspace{R''})  &  \ \ \ \ A \RELofEQUI B^T \ .
  \end{eqnarray}

The family 
  $H(\AUFspace{R})$ is called \DEFINED{\selfCOGNATE}\ if
  \begin{equation}
    \forall B \in H(\AUFspace{R}) \ \ \   \exists A \in
    H(\AUFspace{R})
    \ \ \ \ \ \ B \RELofEQUI A^T
  \end{equation}
  
\end{definition}

%
%
%

\section{Construction of Hadamard matrices}


\subsection{Same matrix size: reordering of entries and conjugation}

If H is a \DEPHASED\ Hadamard matrix, so are its transpose $H^T$, the
conjugated matrix $\CONJ{H}$ and Hermitian transpose
$H^{\dagger}$. It is not at all obvious, whether any pair of these is an
equivalent pair. However, in some special cases it is so. 

For example,
the \DEPHASED\ forms of a Hadamard circulant matrix $C$ and its
transpose $C^T$ are equivalent since $C$ and $C^T$ are always
permutation equivalent (see also the remark on the top of p.319 in
\cite{Ha96}):
\begin{equation}
  \label{eq_circulant_equivalence}
  C^T = P^T \cdot C \cdot P
\end{equation}
where $C$ is an $N \times N$ circulant matrix 
$C_{i,j} = x_{i-j \mod  N}$\ \ for some $x \in {\mathbbm C}^N$, and 
$P = [\STbasis{1},\ \STbasis{N},\ \STbasis{N-1},\ \ldots\
,\STbasis{2}]$, where $\STbasis{i}$ are the standard basis column vectors.
   
On the other hand, there are infinitely many
examples of equivalent and inequivalent pairs of circulant Hadamard matrices
$C,\ \CONJ{C}$, the same applying to their \DEPHASED\ forms.

Apart from transposition and conjugation, for certain dimensions
there exist other matrix reorderings that preserve
the Hadamard structure. Such operations that switch substructures of real
Hadamard matrices to generate inequivalent matrices have
recently been discussed in \cite{Or05}. It is likely that these methods
may  be useful to get inequivalent complex Hadamard matrices.
In this way only a finite number of inequivalent matrices
can be obtained.

%
%

\subsection{Same matrix size: linear variation of phases}

Starting from a given Hadamard matrix $H$ in the \DEPHASED\ form
one may investigate, whether it is possible to perform
infinitesimal changes of some of $(N-1)^2$ phases
 of the core of $H$
to preserve unitarity.
Assuming that all these phases 
($\Phi_{kl}$,  from Eq. (\ref{logHad}) $k,l=2,\dots,N$) 
 vary linearly with free parameters
one can find analytical form of such orbits, i.e. \AUFs,  stemming from 
e.g. Fourier matrices of  composite dimensions \cite{Ta06}.
\medskip

To obtain \AUFs\ stemming from $H$, one has to consider all pairs of
rows of $H$. Now, taking the inner product of the rows in the $i,j$-th
pair ($1 \leq i < j \leq N$), one gets zero as the sum of the
corresponding values in the sequence:
\begin{equation}
  \label{eq_i_j_chain}
  \left(\ \ \ \  
    H_{i,1} \cdot {\CONJ{H}}_{j,1}\ \ ,\ \ \ \ 
    H_{i,2} \cdot {\CONJ{H}}_{j,2}\ \ ,\ \ \ \ 
    \ldots\ \ ,\ \ \ \ 
    H_{i,N} \cdot {\CONJ{H}}_{j,N}\ \ \ \  
  \right)
\end{equation}
Such sequences will further be called \DEFINED{chains}, their
subsequences -- \DEFINED{subchains}. Thus (\ref{eq_i_j_chain}) features the
$i,j$-th chain of $H$. A chain (subchain) is \DEFINED{closed} if its
elements add up to zero. As\ \ $(1/\sqrt{N}) \cdot H$\ \ is unitary, all its
chains are closed. It is not obvious, however, that any of these
chains contain closed subchains.
\medskip

Let us now construct a \DEFINED{closed subchain pattern} for $H$. For
each chain of $H$, let us split it, disjointly, into closed subchains: 
\begin{equation}
  \label{eq_subchain}
  \left(\ \ \ \ 
    H_{i,k_1^{(s)}} \cdot {\CONJ{H}}_{j,k_1^{(s)}}\ \ ,\ \ \ \
    H_{i,k_2^{(s)}} \cdot {\CONJ{H}}_{j,k_2^{(s)}}\ \ ,\ \ \ \
    \ \ldots\ \ ,\ \ \ \
    H_{i,k_{p^{(s)}}^{(s)}} \cdot {\CONJ{H}}_{j,k_{p^{(s)}}^{(s)}}\ \ \ \
  \right)
\end{equation}
where
\begin{description}
  \item[] $s \in \{1,2,\ldots,r_{i,j}\}$\ \ designates subchains,\ \ \ $r_{i,j}$
        being the number of closed subchains the $i,j$-th chain is split
        into

  \item[] $\bigcup_{s=1}^{r_{i,j}} \{k_1^{(s)},\ldots,
        k_{p^{(s)}}^{(s)}\} = \{1,2,\ldots,N\}$,\ \ \ $p^{(s)}$ being the
        length of the $s$-th subchain

  \item [] $\{k_1^{(s_1)},\ldots, k_{p^{(s_1)}}^{(s_1)}\} \cap 
           \{k_1^{(s_2)},\ldots, k_{p^{(s_2)}}^{(s_2)}\} = \emptyset$
           \ \ \ if\ \ \ $s_1 \neq s_2$
\end{description}
and splitting of $\{1,2,\ldots,N\}$ into $\{k_1^{(s)},\ldots,
k_{p^{(s)}}^{(s)}\}$ is done independently for each chain of
$H$, that is the $k$-values above in fact depend also on $i,j$.

A pattern according to which all the chains of $H$ are split, in the
above way, into closed subchains, will be called a 
\DEFINED{closed subchain pattern}.
\medskip

A closed subchain pattern may give rise to the \AUF, stemming from $H$,
corresponding to this pattern. The space $\AUFspace{R}$, generating
this $H(\AUFspace{R})$ family (see Definition \ref{def_AUF}), is
defined by the equations:
\begin{eqnarray}
 R_{1,j} & = & 0\ \ \ \ \ \ j \in \{2,\ldots,N\}\ ,  \label{eq_1st_row_deph} \\
 R_{i,1} & = & 0\ \ \ \ \ \ i \in \{1,\ldots,N\}\ ,  \label{eq_1st_col_deph}
\end{eqnarray}
(as $H(\AUFspace{R})$ is made of \DEPHASED\ Hadamard matrices) and,
for all $1 \leq i < j \leq N$ and $s \in \{1,2,\ldots,r_{i,j}\}$ ,
\begin{equation}
  \label{eq_R_diff_equalities}
  \left( R_{i,k_1^{(s)}} - R_{j,k_1^{(s)}} \right)\ \  =\ \ \ldots\ \ =\ \ 
  \left( R_{i,k_{p^{(s)}}^{(s)}} - R_{j,k_{p^{(s)}}^{(s)}} \right) \ ,
\end{equation}
where the sets of indices $\{k_1^{(s)},\ldots, k_{p^{(s)}}^{(s)}\}$
correspond to the considered pattern splitting of the $i,j$-th chain
into closed subchains (\ref{eq_subchain}). Recall that the sets of
$k$'s depend on $i,j$'s, ommited for simplicity of notation.

If the system\ \ 
(\ref{eq_1st_row_deph},\ref{eq_1st_col_deph},\ref{eq_R_diff_equalities})\
\ yields a nonzero space $\AUFspace{R}$,
then $H(\AUFspace{R})$ is the \AUF\ corresponding to the chosen closed
subchain pattern. Actually, any \AUF\ stemming from $H$ is generated
by some space $\AUFspace{R}$ contained in a (probably larger) space
$\AUFspace{R}'$, corresponding to some closed subchain pattern for
$H$. The respective theorem and its proof will be published in
\cite{Ta06}.
\medskip

It may happen that the system\ \ 
(\ref{eq_1st_row_deph},\ref{eq_1st_col_deph},\ref{eq_R_diff_equalities}),\
\ for pattern $\CSP{P_1}{H}$, defines
space $\AUFspace{R}$, which is also obtained as the solution to\ \ 
(\ref{eq_1st_row_deph},\ref{eq_1st_col_deph},\ref{eq_R_diff_equalities})\
\ system shaped by another pattern
$\CSP{P_2}{H}$, imposing stronger conditions of type
(\ref{eq_R_diff_equalities}), as a result of there being longer
subchains in $\CSP{P_2}{H}$
composed of more than one subchain of $\CSP{P_1}{H}$, for a given pair
$i,j$. If this is not the case, we say that $\AUFspace{R}$ is \DEFINED{strictly
associated} with pattern $\CSP{P_1}{H}$. Then the
subchains of the $i,j$-th chain of 
$H(R) = H \HADprod \EXPentrywise(\Ii \cdot R),\ \ R \in \AUFspace{R}$, distinguished
according to pattern $\CSP{P_1}{H}$:
\begin{eqnarray}
  (\ \  &
  H_{i,k_1^{(s)}} {\CONJ{H}}_{j,k_1^{(s)}} 
  \exp
  \left(
    \Ii \cdot (\ R_{i,k_1^{(s)}} - R_{j,k_1^{(s)}}\ )
  \right)\ \ , 
  & \nonumber \\
  & H_{i,k_2^{(s)}} {\CONJ{H}}_{j,k_2^{(s)}}
  \exp
  \left( 
    \Ii \cdot (\ R_{i,k_2^{(s)}} - R_{j,k_2^{(s)}}\ )
  \right)\ \ , 
  & \nonumber \\
  & \ldots\ \ , & \nonumber \\
  & H_{i,k_{p^{(s)}}^{(s)}}  {\CONJ{H}}_{j,k_{p^{(s)}}^{(s)}}
  \exp
  \left(
    \Ii \cdot (\ R_{i,k_{p^{(s)}}^{(s)}} - R_{j,k_{p^{(s)}}^{(s)}}\ )
  \right) 
  & \ \ )  \label{eq_subchain_rotation}
\end{eqnarray}
rotate independently as $R$ runs along $\AUFspace{R}$ satisfying
(\ref{eq_R_diff_equalities}).
\medskip

Maximal \AUFs\ stemming from $H$ are generated by maximal, in the sense of
(\ref{eq_maxAUF_space_cond}), solutions to
(\ref{eq_1st_row_deph},\ref{eq_1st_col_deph},\ref{eq_R_diff_equalities})
systems shaped by some
specific closed subchain patterns, which we also call \DEFINED{maximal}.
We have been able to find all of
these for almost every complex Hadamard matrix considered in our catalogue. To
our understanding, however, it becomes a serious combinatorial problem
already for $N=12$. For example, for a real $12 \times 12$ Hadamard
matrix, each chain can be split in $(6!)^2$ ways into two element
closed subchains only.

Fortunately, as far as Fourier matrices are
concerned, the allowed maximal subchain patterns are
especially regular. We are thus able to obtain
all \maxAUFs\ stemming from $F_N$ for an arbitrary $N$ \cite{Ta06}, as
we have done for $N \leq 16$.
\bigskip

An alternative method of constructing \AUFs, developed by 
Di{\c t}{\v a} \cite{Di04}, is presented in section
\ref{subsec_Ditas_method}.
We also refer the reader to the article by Nicoara \cite{Ni04},
in which conditions are given for the existence of one parameter
families of commuting squares of finite dimensional von Neumann
algebras. These conditions can be used to establish the existence of
one parameter families of complex Hadamard matrices, stemming from some $H$, which
are not assumed to be affine.

%
%

\subsection{Duplication of the matrix size}

Certain ways of construction of real Hadamard  matrices
of an extended size work also in the complex case. If  $A$ and $B$
belong to $\HADAMARDS{N}$ then
\begin{equation}
  \label{Had2N}
  H=
  \left[ 
    \begin{array}{ll}
      A & {\ \ }B   \\
      A &      -B   \\
    \end{array} 
  \right] 
  \in \HADAMARDS{2N}
\end{equation}
Furthermore, if $A$ and $B$ are taken
to be in the \DEPHASED\ form, so is $H$.
This method, originally due to Hadamard,
can be generalized by realizing that $B$ can be
multiplied at left by an arbitrary diagonal unitary matrix,
$E=\diag( 1,\PHASE{d_1},\dots,\PHASE{d_{N-1}})$.
If $A$ and $B$ depend on $a$ and $b$ free parameters, respectively, then
\begin{equation}
  \label{Had2Nfas}
  H'=
  \left[ 
    \begin{array}{ll}
      A & {\ \ \ }EB   \\
      A &        -EB   \\
    \end{array} 
  \right]
\end{equation}
represents an\ \ $(a+b+N-1)$--parameter
family of Hadamard matrices of size $2N$
in the \DEPHASED\ form.

%
%

\subsection{Quadruplication of the matrix size}

In analogy to Eq. (\ref{Had2N}) one may quadruple the matrix size,
in a construction similar to that derived by Williamson \cite{Wi44}
from quaternions.
If $A,B,C,D \in  \HADAMARDS{N}$
are in the \DEPHASED\ form then 
\begin{equation}
  \label{Williamson}
  \left[
    \begin{array}{rrrr}
      A  &  B  &  C  &  D \\
      A  & -B  &  C  & -D \\
      A  &  B  & -C  & -D \\
      A  & -B  & -C  &  D
  \end{array} 
  \right] 
  \in \HADAMARDS{4N}
\end{equation}
and it has the \DEPHASED\ form.
 This form is preserved, if the
blocks $B,C$ and $D$ are multiplied by diagonal unitary matrices
$E_1,E_2$ and $E_3$ respectively, 
each containing unity and $N-1$ free phases.
Therefore (\ref{Williamson}) describes an\ \ $[a+b+c+d+3(N-1)]$--dimensional
family of Hadamard matrices \cite{Di04}, where $a,b,c,d$
denote the number of  free parameters
contained in $A,B,C,D$, respectively.

%
%

\subsection{Generalized method related to tensor product}
\label{subsec_Ditas_method}

It is not difficult to design a similar method which
increases the size of a Hadamard matrix by eight,
but more generally, we can use tensor product\footnote{
Tensor products of Hadamard matrices of the Butson type were
investigated in \cite{Ho05}.}
to increase the size of the matrix $K$ times.
For any two Hadamard matrices,
$A \in \HADAMARDS{K}$ and $B \in \HADAMARDS{M}$,
their tensor product $A \otimes B \in \HADAMARDS{KM}$.
A more general construction by Di{\c t}{\v a}
allows to use entire set of  $K$ (possibly different) Hadamard matrices
$\{B_1,\dots ,B_K \}$ of size $M$.
Then the matrix
\begin{equation}
  \label{eq_Ditas_method}
  H =  
  \left[
    \begin{array}{ccccc}
      A_{11} B_1  & A_{12}E_2 B_2   &  .  &  .  & A_{1K} E_K B_K  \\
           .      &      .          &  .  &  .  &      .          \\
           .      &      .          &  .  &  .  &      .          \\
      A_{K1} B_1  & A_{K2} E_2 B_2  &  .  &  .  & A_{KK} E_K B_K
    \end{array}
  \right]
\end{equation}
of size $N=KM$ is Hadamard \cite{Di04}. As before we introduce
additional 
free phases by using $K-1$ diagonal unitary matrices $E_k$. Each
matrix depends on $M-1$ phases, since the constraint
$\ELEMENTof{E_k}{1}{1}=1$ for $k=2,\dots, K$ is necessary to preserve the
\DEPHASED\ form of $H$. Thus this orbit of (mostly) not equivalent Hadamard
matrices depends on
\begin{equation}
  \label{Ditpar}
  d\ \ =\ \ a\ +\  \sum_{j=1}^K b_j\  +\  (K-1)(M-1)
\end{equation}
free parameters. Here $a$ denotes the number of
free parameters in $A$, while $b_j$
denotes the number of free parameters in $B_j$.
 A similar construction
giving at least $(K-1)(M-1)$ free parameters
was given by Haagerup \cite{Ha96}.
In the simplest case $K \cdot M = 2 \cdot 2$
these methods give the standard
$1$--parameter $N=4$ family (\ref{F4_maxAUF_explicit}),
while for $K \cdot M = 2 \cdot 3$
one arrives with $(2-1)(3-1)=2$ parameter
family (\ref{F6_maxAUF})
of (mostly) inequivalent $N=6$ Hadamard matrices.

The tensor product construction
can work only for composite $N$,
so it was conjectured \cite{Po83}
that for a prime dimension $N$ there 
exist only finitely many inequivalent complex Hadamard matrices.
However, this occurred to be false
after a discovery by Petrescu \cite{Pe97},
who found continuous families of complex Hadamard matrices
for certain prime dimensions.
We are going to present  his solution for $N=7$ and $N=13$,
while a similar construction \cite{Pe97}
works also for $N=19,31$ and $79$.
For all primes $N\ge 7$
there exist at least three isolated complex Hadamard matrices,
see \cite{Ha96}, Thm. 3.10, p.320.

%
%
%

\section{Catalogue  of complex Hadamard matrices}
\label{sec_catalogue}

In this section we list complex Hadamard matrices known to us. To
save space we do not describe their construction but present a
short characterization of each case.

 Each entry $H$,  provided in  a \DEPHASED\ form,
represents a continuous $(2N-1)$--dimensional family of matrices
$\bf H$ obtained by multiplication by diagonal unitary matrices:
\begin{equation}
  \label{diagonal}
  {\bf H}(\alpha_1,...,\alpha_N,\beta_2,...,\beta_N)=
  D_1(\alpha_1,\ldots,\alpha_N) \cdot H \cdot D_2(\beta_2,\ldots,\beta_N)
\end{equation}
where
\begin{eqnarray}
  D_1(\alpha_1,\ldots,\alpha_N) &  = &  \diag(\PHASE{\alpha_1},
  \ldots,\PHASE{\alpha_N}) \nonumber \\
  D_2(\beta_2,\ldots,\beta_N)   &  = &  \diag(1,
  \PHASE{\beta_2},\ldots,\PHASE{\beta_N}) \nonumber
\end{eqnarray}

Furthermore, each $H$ in the list represents matrices obtained by
discrete permutations of its rows and columns, and equivalent in
the sense of (\ref{equival}). For a given size $N$ we enumerate
families by capital letters associated with a given construction.
The superscript in brackets denotes the dimension of an orbit. For
instance, $F_6^{(2)}$ represents the two--parameter family of
$N=6$ complex Hadamard matrices stemming from the Fourier matrix
$F_6$. Displaying continuous families of Hadamard matrices we shall use the symbol
$\bO$
to denote zeros in phase variation matrices.
For completeness we shall start with a trivial case.

%
%

\subsection {$N=1$}
\quad \quad 
$\AUFfamily{F}{1}{0} = F_1=[1]$.
\ENDofSUBSECTION

%
%

\subsection {$N=2$}
\label{subsec__N_eq_2}

All complex $2 \times 2$ Hadamard matrices are equivalent to the Fourier matrix
$F_2$ (\cite{Ha96}, Prop.2.1, p.298):
\begin{equation}
  F_2 = \AUFfamily{F}{2}{0} = 
  \left[ 
    \begin{array}{cc} 
      1 & 1   \\ 
      1 & -1  \\ 
    \end{array} 
  \right] \ .
\end{equation}

The complex Hadamard matrix 
$F_2$ is \ISOLATED . 
\bigskip

The set of 
inequivalent Hadamard matrices of size $2$ contains one element only, 
$\HADreprs{2} = \{ F_2 \}$.
\ENDofSUBSECTION

%
%

\subsection {$N=3$}
\label{subsec__N_eq_3}

All complex $3 \times 3$ Hadamard matrices are equivalent to the Fourier matrix
$F_3$ (\cite{Ha96}, Prop.2.1, p.298): 
\begin{equation}
  F_3 = \AUFfamily{F}{3}{0} = 
  \left[ 
    \begin{array}{ccc} 
      1 & 1  & 1 \\ 
      1 & w & w^2 \\
      1 & w^2 & w 
    \end{array} 
  \right] \ ,
\end{equation}
where $w=\exp(\Ii \cdot 2\pi/3)$, so $w^3 = 1$.

$F_3$ is an \ISOLATED\ complex Hadamard matrix.
\bigskip

The set of \DEPHASED\ representatives can be taken as $\HADreprs{3} = \{
F_3 \}$.
\ENDofSUBSECTION

%
%

\subsection {$N=4$}
\label{subsec__N_eq_4}

Every $4 \times 4$ complex Hadamard matrix  is equivalent
to a matrix belonging to the only \maxAUF\ $\AUFfamily{F}{4}{1}(a)$ stemming
from
$F_4$ \cite{Ha96}. The $\AUFfamily{F}{4}{1}(a)$ family is given by the formula:
\begin{equation}
\label{F4_maxAUF}
  \AUFfamily{F}{4}{1}(a) = F_4 \HADprod \EXPentrywise (\Ii \cdot
\AUFphases{F}{4}{1}(a))
\end{equation}
where
\begin{equation}
  F_4 = 
%
  \left[
    \begin{array}{cccc}
      1 & 1 & 1 & 1\\
      1 & w & w^2 & w^3\\
      1 & w^2 & 1 & w^2\\
      1 & w^3 & w^2 & w
    \end{array}
  \right] =
%
  \left[
    \begin{array}{rrrr}
      1 &    1 &  1 &    1\\
      1 &  \Ii & -1 & -\Ii\\
      1 &   -1 &  1 &   -1\\
      1 & -\Ii & -1 &  \Ii
    \end{array}
  \right]
\end{equation}
where $w = \exp(\Ii \cdot 2\pi/4)= \Ii$, so $w^4 = 1,\ w^2 = -1$,
and
\begin{equation}
  \AUFphases{F}{4}{1}(a) =
  \left[
  \begin{array}{rr|rr}
    \bO & \bO & \bO & \bO\\
    \bO & a & \bO & a\\
    \hline
    \bO & \bO & \bO & \bO\\
    \bO & a & \bO & a
  \end{array}
  \right] \ .
\end{equation}
Thus the orbit stemming from $F_4$ reads:
\begin{equation}
  \label{F4_maxAUF_explicit}
  \AUFfamily{F}{4}{1}(a) =
  \left[
    \begin{array}{cccc}
      1 &                     1 &        1 &                     1 \\
      1 &  \Ii \cdot \PHASE{a} &       -1 & -\Ii \cdot \PHASE{a} \\
      1 &                    -1 &        1 &                    -1 \\
      1 & -\Ii \cdot \PHASE{a} &       -1 &  \Ii \cdot \PHASE{a}
    \end{array}
  \right] \ .
\end{equation}
\bigskip

The above orbit is \PERMUTATIONequivalent\ to:
\begin{equation}
\label{F4fam_Dita}
  \AUFfamily{\tilde{F}}{4}{1}(\alpha) = 
  \left[ \begin{array}{c|c}
            \ELEMENTof{F_2}{1}{1} \cdot F_2 & \ELEMENTof{F_2}{1}{2} \cdot
            D(\alpha) \cdot F_2 \\
            \hline
            \ELEMENTof{F_2}{2}{1} \cdot F_2 & \ELEMENTof{F_2}{2}{2} \cdot
            D(\alpha) \cdot F_2
         \end{array} \right] \ ,
\end{equation}
where $D(\alpha)$ is the $2 \times 2$ diagonal matrix
$\diag(1,\PHASE{\alpha})$.

This orbit is constructed with the Di{\c t}{\v a}'s method \cite{Di04}, by setting
$K=M=2,\ \ A = B_1 = B_2 = F_2,\ \ E_2 = D(\alpha)$ in (\ref{eq_Ditas_method}).
\medskip

It passes through a permuted $F_4$:
\begin{equation}
  F_4 = \AUFfamily{\tilde{F}}{4}{1}(\pi/2) \cdot [\STbasis{1}, \STbasis{3},
\STbasis{2},
  \STbasis{4}]^T
\end{equation} 
where $\STbasis{i}$ is the $i$-th standard basis column vector, and
through $F_2 \otimes F_2 = \AUFfamily{\tilde{F}}{4}{1}(0)$. Note that $F_4$ and
$F_2
\otimes F_2$ are, according to \cite{Ta05}, inequivalent.
\bigskip

The $\AUFfamily{F}{4}{1}$ orbit of (\ref{F4_maxAUF}) is symmetric, so it is
\selfCOGNATE. 
Replacing $a$ by $a+\pi$ yields $\AUFfamily{F}{4}{1}(a)$ with the $2$-nd and
$4$-th
column exchanged, so $\AUFfamily{F}{4}{1}(a) \RELofEQUI
\AUFfamily{F}{4}{1}(a+\pi)$.
\bigskip

The set of \DEPHASED\ representatives can be taken as 
$\HADreprs{4} = \{ \AUFfamily{F}{4}{1}(a):\ a \in [0,\pi) \}$
\ENDofSUBSECTION

%
%

\subsection {$N=5$}
\label{subsec__N_eq_5}

As shown by Haagerup in \cite{Ha96} (Th. 2.2, p. 298) 
for $N=5$
all complex Hadamard matrices are equivalent
to the Fourier matrix $F_5$:
\begin{equation}
  F_5 = \AUFfamily{F}{5}{0} = 
  \left[
    \begin{array}{ccccc}
      1 & 1 & 1 & 1 & 1\\
      1 & w & w^2 & w^3 & w^4\\
      1 & w^2 & w^4 & w & w^3\\
      1 & w^3 & w & w^4 & w^2\\
      1 & w^4 & w^3 & w^2 & w
    \end{array}
  \right] \ ,
\end{equation}
where $w = \exp(\Ii \cdot 2\pi/5)$ so $w^5 = 1$. 

The above matrix $F_5$ is \ISOLATED.
\bigskip

The set of \DEPHASED\ representatives can be taken as 
$\HADreprs{5} = \{ F_5 \}$. 
\ENDofSUBSECTION

%
%

\subsection{$N=6$}
\label{subsec__N_eq_6}


\subsubsection{Orbits stemming from $F_6$}

The only \maxAUFs\ stemming from $F_6$ are:
\begin{eqnarray}
  \AUFfamily{F}{6}{2}(a,b)  &  = & F_6 \HADprod \EXPentrywise
  \left(\Ii \cdot \AUFphases{F}{6}{2}(a,b)\right) \label{F6_maxAUF}\\
  \TRANSPOSE{\AUFfamily{F}{6}{2}(a,b)}  &  = & F_6 \HADprod
  \EXPentrywise \left(\Ii \cdot \TRANSPOSE{\AUFphases{F}{6}{2}(a,b)}
  \right)   \label{F6_maxAUF_transposed}
\end{eqnarray}
where
\begin{equation}
  F_6 =
  \left[
    \begin{array}{cccccc}
      1 & 1 & 1 & 1 & 1 & 1\\
      1 & w & w^2 & w^3 & w^4 & w^5\\
      1 & w^2 & w^4 & 1 & w^2 & w^4\\
      1 & w^3 & 1 & w^3 & 1 & w^3\\
      1 & w^4 & w^2 & 1 & w^4 & w^2\\
      1 & w^5 & w^4 & w^3 & w^2 & w
    \end{array}
  \right]
\end{equation}
where $w = \exp(\Ii \cdot 2\pi/6)$ so $w^6 = 1,\ \ w^3 = -1$, and
\begin{equation}
  \AUFphases{F}{6}{2}(a,b) = 
  \left[
    \begin{array}{ccc|ccc}
      \bO & \bO & \bO & \bO & \bO & \bO\\
      \bO & a & b & \bO & a & b\\
      \hline
      \bO & \bO & \bO & \bO & \bO & \bO\\
      \bO & a & b & \bO & a & b\\
      \hline
      \bO & \bO & \bO & \bO & \bO & \bO\\
      \bO & a & b & \bO & a & b
    \end{array}
  \right]
\end{equation}
\bigskip
The above families (\ref{F6_maxAUF}) and (\ref{F6_maxAUF_transposed})
are \COGNATE. The defect (see Def. \label{def_defect}) reads $d(F_6)=4$.

At least one of the above orbits is \PERMUTATIONequivalent\ to one of the
orbits obtained using the Di{\c t}{\v a}'s method, either
\begin{equation}
\label{F6fam_Dita_1}
  \AUFfamily[A]{\tilde{F}}{6}{2}(\alpha_1,\alpha_2) = 
  \left[ \begin{array}{c|c|c}
            \ELEMENTof{F_3}{1}{1} \cdot F_2 & \ELEMENTof{F_3}{1}{2} \cdot
            D(\alpha_1) \cdot F_2 & \ELEMENTof{F_3}{1}{3} \cdot
            D(\alpha_2) \cdot F_2 \\
            \hline
            \ELEMENTof{F_3}{2}{1} \cdot F_2 & \ELEMENTof{F_3}{2}{2} \cdot
            D(\alpha_1) \cdot F_2 & \ELEMENTof{F_3}{2}{3} \cdot
            D(\alpha_2) \cdot F_2 \\
            \hline
            \ELEMENTof{F_3}{3}{1} \cdot F_2 & \ELEMENTof{F_3}{3}{2} \cdot
            D(\alpha_1) \cdot F_2 & \ELEMENTof{F_3}{3}{3} \cdot
            D(\alpha_2) \cdot F_2 \\
         \end{array} \right]
\end{equation}
where $D(\alpha)$ is the $2 \times 2$ diagonal
matrix $\diag(1,\PHASE{\alpha})$, by setting $K=3\ ,M=2,\ \ A=F_3,\ B_1
= B_2 = B_3 = F_2,\ \ E_2 = D(\alpha_1),\ E_3 = D(\alpha_2)$ in
(\ref{eq_Ditas_method}), 

or
\begin{equation}
\label{F6fam_Dita_2}
  \AUFfamily[B]{\tilde{F}}{6}{2}(\beta_1,\beta_2) = 
  \left[ \begin{array}{c|c}
            \ELEMENTof{F_2}{1}{1} \cdot F_3 & \ELEMENTof{F_2}{1}{2} \cdot
            D(\beta_1,\beta_2) \cdot F_3 \\
            \hline
            \ELEMENTof{F_2}{2}{1} \cdot F_3 & \ELEMENTof{F_2}{2}{2} \cdot
            D(\beta_1,\beta_2) \cdot F_3
         \end{array} \right]
\end{equation}
where $D(\beta_1,\beta_2)$ is the $3 \times 3$ diagonal
matrix $\diag(1,\PHASE{\beta_1},\PHASE{\beta_2})$, by setting $K =
2,\ M = 3,\ \ A = F_2,\ B_1 = B_2 = F_3,\ \ E_2 = D(\beta_1,\beta_2)$
in (\ref{eq_Ditas_method}).
\medskip

The above statement is true because the orbits (\ref{F6fam_Dita_1}) and
(\ref{F6fam_Dita_2}) pass
through $\AUFfamily[A]{\tilde{F}}{6}{2}(0,0) = F_3 \otimes F_2$ and
$\AUFfamily[B]{\tilde{F}}{6}{2}(0,0) = F_2 \otimes F_3$, both of which are,
according to \cite{Ta05}, permutation equivalent to $F_6$. Thus
(\ref{F6fam_Dita_1}) and (\ref{F6fam_Dita_2}) are \maxAUFs\ stemming
from permuted $F_6$'s. These orbits were constructed in the work of Haagerup
\cite{Ha96}.  
\bigskip

A change of the phase $a$ by $\pi$ 
in the family $\AUFfamily{F}{6}{2}$ 
corresponds to the exchange of the $2$-nd and
$5$-th column of $\AUFfamily{F}{6}{2}(a,b)$,
while a change of $b$ by $\pi$ is equivalent
to the exchange of the $3$-rd and $6$-th column of $\AUFfamily{F}{6}{2}(a,b)$.
This
implies (permutation) equivalence relation:
\begin{equation} 
  \AUFfamily{F}{6}{2}(a,b)  \RELofEQUI   \AUFfamily{F}{6}{2}(a+\pi,b)
  \RELofEQUI  \AUFfamily{F}{6}{2}(a,b+\pi)  \RELofEQUI 
\AUFfamily{F}{6}{2}(a+\pi,b+\pi) \ .
\end{equation}


\subsubsection{$1$-parameter orbits}

There are precisely five \PERMUTATIONequivalent\ $1$-parameter \maxAUFs\
stemming from the symmetric matrix $D_6$:
\begin{equation}
\label{U1par}
  D_6 = 
  \left[
  \begin{array}{cccccc}
    1 & 1 & 1 & 1 & 1 & 1\\
    1 & -1 & \Ii & -\Ii & -\Ii & \Ii \\
    1 & \Ii & -1 & \Ii & -\Ii & -\Ii \\
    1 & -\Ii & \Ii & -1 & \Ii & -\Ii \\
    1 & -\Ii & -\Ii & \Ii & -1 & \Ii \\
    1 & \Ii & -\Ii & -\Ii & \Ii & -1
  \end{array}
  \right]
\end{equation}
namely 
\begin{equation}
\label{U6_1par_orbits}
 \AUFfamily{D}{6}{1}(c) = D_6 \HADprod \EXPentrywise \left( \Ii \cdot
\AUFphases{D}{6}{1}(c) \right)
\end{equation}
where
\begin{equation}
  \AUFphases{D}{6}{1}(c) = 
  \left[
    \begin{array}{cccccc}
      \bO & \bO &  \bO & \bO & \bO &  \bO\\
      \bO & \bO &  \bO & \bO & \bO &  \bO\\
      \bO & \bO &  \bO & c & c &  \bO\\
      \bO & \bO & -c & \bO & \bO & -c\\
      \bO & \bO & -c & \bO & \bO & -c\\
      \bO & \bO &  \bO & c & c &  \bO
    \end{array}
  \right]
\end{equation}
and
\begin{eqnarray}
    P_{1} \cdot \AUFfamily{D}{6}{1}(c) \cdot P_{1}^T & =  &
    D_6 \HADprod \EXPentrywise \left( \Ii \cdot P_{1} \AUFphases{D}{6}{1}(c)
P_{1}^T \right) 
   \\
    P_{2} \cdot \AUFfamily{D}{6}{1}(c) \cdot P_{2}^T & =  & 
    D_6 \HADprod \EXPentrywise \left( \Ii \cdot P_{2} \AUFphases{D}{6}{1}(c)
P_{2}^T \right)
  \\
    P_{3} \cdot \AUFfamily{D}{6}{1}(c) \cdot P_{3}^T & =  &
    D_6 \HADprod \EXPentrywise \left( \Ii \cdot P_{3} \AUFphases{D}{6}{1}(c)
P_{3}^T \right)
   \\
    P_{4} \cdot \AUFfamily{D}{6}{1}(c) \cdot P_{4}^T & =  & 
    D_6 \HADprod \EXPentrywise \left( \Ii \cdot P_{4} \AUFphases{D}{6}{1}(c)
P_{4}^T \right)
\end{eqnarray}
where
\begin{eqnarray}
P_{1} & = & [ \STbasis{1}, \STbasis{3}, \STbasis{2},
  \STbasis{6}, \STbasis{5}, \STbasis{4} ] \\
P_{2} & = & [ \STbasis{1}, \STbasis{4}, \STbasis{3},
  \STbasis{2}, \STbasis{6}, \STbasis{5} ] \\
P_{3} & = & [ \STbasis{1}, \STbasis{6}, \STbasis{2},
  \STbasis{3}, \STbasis{4}, \STbasis{5} ] \\
P_{4} & = & [ \STbasis{1}, \STbasis{5}, \STbasis{4},
  \STbasis{3}, \STbasis{2}, \STbasis{6} ]
\end{eqnarray}
and $\STbasis{i}$ denotes the $i$-th standard basis column
vector. 
\bigskip

We have the permutation
equivalence
\begin{equation}
  \AUFfamily{D}{6}{1}(c+\pi) = P_{-}^T \cdot \AUFfamily{D}{6}{1}(c) \cdot
P_{-}\  \ \ \mbox{for}\ \ \
  P_{-} = 
  [ \STbasis{1}, \STbasis{2}, \STbasis{3}, \STbasis{5}, \STbasis{4},
  \STbasis{6} ]
\end{equation}

Also
\begin{equation}
  \AUFfamily{D}{6}{1}(-c) = \TRANSPOSE{\AUFfamily{D}{6}{1}(c)}
\end{equation}
thus $\{ \AUFfamily{D}{6}{1}(c):\ c \in [0,2\pi)\}$ and $\{
\TRANSPOSE{\AUFfamily{D}{6}{1}(c)} :\ c \in [0,2\pi)\}$
are equal sets, that is $\AUFfamily{D}{6}{1}$ is a \selfCOGNATE\ family

None of the matrices of $\AUFfamily{F}{6}{2}$ and
$\TRANSPOSE{\AUFfamily{F}{6}{2}}$  of (\ref{F6_maxAUF}) and
(\ref{F6_maxAUF_transposed}) are
equivalent to any of the matrices of $\AUFfamily{D}{6}{1}$.

Obviously, the above remarks apply to the remaining orbits stemming
from $D_6$.

\bigskip
The $\AUFfamily{D}{6}{1}$ orbit was presented in \cite{Di04} in the
Introduction,
and its 'starting point' matrix $D_6$ of (\ref{U1par}) even in the earlier
work \cite{Ha96} p.307 (not \DEPHASED).


\subsubsection{The 'cyclic $6$--roots' matrix}

There exists another, inequivalent to any of the above $6
\times 6$ matrices, complex Hadamard matrix derived in \cite{Ha96}
from the results of \cite{BF94} on so called cyclic $6$--roots.

The matrix $\AUFfamily{\tilde{C}}{6}{0}$ below is circulant, i.e. it has the
structure
$\ELEMENTof{\AUFfamily{\tilde{C}}{6}{0}}{i}{j} =  x_{(i-j \mod 6)+1}$, where
\begin{equation}
  x = [ 1,\ \Ii/d,\ -1/d,\ -\Ii,\ -d,\ \Ii d ]^T
\end{equation}
and
\begin{equation}
  d = \frac{1-\sqrt{3}}{2} + \Ii \cdot {\left( \frac{\sqrt{3}}{2}
    \right)}^{\frac{1}{2}}
\end{equation}
is a root of the equation
$d^2 - (1-\sqrt{3})d + 1=0$.
\bigskip

The matrix $\AUFfamily{\tilde{C}}{6}{0}$ and its \DEPHASED\ form
$\AUFfamily{C}{6}{0}$ read:
\begin{eqnarray}
  \AUFfamily{\tilde{C}}{6}{0} & = &
  \left[
    \begin{array}{cccccc}
      1 & \Ii\, d & - d & -\Ii & -{d}^{\SSS -1} & {\Ii}{d}^{\SSS -1} \\
      {\Ii}{d}^{\SSS -1} & 1 & \Ii\, d & - d & -\Ii & -{d}^{\SSS -1} \\
      -{d}^{\SSS -1} & {\Ii}{d}^{\SSS -1} & 1 & \Ii\, d & - d & -\Ii \\
      -\Ii & -{d}^{\SSS -1} & {\Ii}{d}^{\SSS -1} & 1 & \Ii\, d & - d \\
      - d & -\Ii & -{d}^{\SSS -1} & {\Ii}{d}^{\SSS -1} & 1 & \Ii\, d \\
      \Ii\, d & - d & -\Ii & -{d}^{\SSS -1} & {\Ii}{d}^{\SSS -1} & 1
    \end{array}
  \right]
\\
  \AUFfamily{C}{6}{0} & = & 
  \left[
    \begin{array}{cccccc}
      1 & 1 & 1 & 1 & 1 & 1\\
      1 & -1 & - d & - d^{\SSS 2} & d^{\SSS 2} & d\\
      1 & -{d}^{\SSS -1} & 1 & d^{\SSS 2} & - d^{\SSS 3} & d^{\SSS 2}\\
      1 & -{d}^{\SSS -2} & {d}^{\SSS -2} & -1 & d^{\SSS 2} & - d^{\SSS 2}\\
      1 & {d}^{\SSS -2} & -{d}^{\SSS -3} & {d}^{\SSS -2} & 1 & - d\\
      1 & {d}^{\SSS -1} & {d}^{\SSS -2} & -{d}^{\SSS -2} & -{d}^{\SSS -1} & -1
     \end{array}
    \right]
\end{eqnarray}
\bigskip

The circulant structure of $\AUFfamily{\tilde{C}}{6}{0}$ implies
that it is equivalent to
$\TRANSPOSE{\AUFfamily{\tilde{C}}{6}{0}}$, so
$\AUFfamily{C}{6}{0}  \RELofEQUI \TRANSPOSE{\AUFfamily{C}{6}{0}}$
(see (\ref{eq_circulant_equivalence})).  Thus $\AUFfamily{C}{6}{0} \RELofEQUI
\CONJ{\AUFfamily{C}{6}{0}}$ (
$\Longleftrightarrow \AUFfamily{\tilde{C}}{6}{0} \RELofEQUI
\CONJ{\AUFfamily{\tilde{C}}{6}{0}}$).

No \AUF\ stems from $\AUFfamily{C}{6}{0}$.
However, 
we do not know if the cyclic $6$--roots matrix is isolated
since the defect $d(C_6^{(0)} )=4$ and
we cannot exclude existence of some other orbit.


\subsubsection{The 'spectral set' $6 \times 6$ matrix}

Another complex Hadamard matrix found by Tao
\cite{Ta04} plays an important role in
investigation of spectral sets and disproving
the Fuglede's conjecture \cite{Ma04,KM04}.
It is a symmetric matrix $S_6^{(0)}$, which belongs
to the Butson class $H(3,6)$,
so its entries depend on
the third root of unity $\omega = \exp(\Ii \cdot 2\pi/3)$,
\begin{equation}
  \label{S6}
  \AUFfamily{S}{6}{0} =
  \left[
    \begin{array}{cccccc}
      1 & 1 & 1 &  1 & 1 &  1\\
      1 & 1 & \omega &  \omega & \omega^2 &  \omega^2\\
      1 & \omega & 1 &  \omega^2 & \omega^2 &  \omega\\
      1 & \omega & \omega^2 &  1 & \omega &  \omega^2\\
      1 & \omega^2 & \omega^2 & \omega  & 1 &  \omega\\
      1 & \omega^2 & \omega &  \omega^2 & \omega &  1
    \end{array}
  \right] \ .
\end{equation}

Thus the corresponding log-Hadamard matrix reads
\begin{equation}
  \label{eq_logHad_S6}
\Phi_{S_6} \ = \ 
  \frac{2 \pi}{3} 
  \left[
    \begin{array}{cccccc}
      0 & 0 & 0 &  0 & 0 &  0\\
      0 & 0 & 1 &  1 & 2 &  2\\
      0 & 1 & 0 &  2 & 2 &  1\\
      0 & 1 & 2 &  0 & 1 &  2\\
      0 & 2 & 2 &  1 & 0 &  1\\
      0 & 2 & 1 &  2 & 1 &  0
    \end{array}
  \right] \ .
\end{equation}

Its defect is equal to zero, $d(S_6)=0$,
 hence the matrix (\ref{S6}) is isolated.
Spectral sets allow to construct certain Hadamard matrices
for other composite dimensions (see prop. 2.2. in \cite{KM04a}),
but it is not yet established, in which cases
this method yields new solutions.

\ENDofSUBSECTION

%
%

\subsection {$N=7$}
\label{subsec__N_eq_7}


\subsubsection{Orbits stemming from $F_7$}

$F_7$ is an \ISOLATED\ $7 \times 7$ complex Hadamard matrix:
\begin{equation}
  F_7 = \AUFfamily{F}{7}{0} =
  \left[
    \begin{array}{ccccccc}
      1 & 1 & 1 & 1 & 1 & 1 & 1\\
      1 & w & w^2 & w^3 & w^4 & w^5 & w^6\\
      1 & w^2 & w^4 & w^6 & w & w^3 & w^5\\
      1 & w^3 & w^6 & w^2 & w^5 & w & w^4\\
      1 & w^4 & w & w^5 & w^2 & w^6 & w^3\\
      1 & w^5 & w^3 & w & w^6 & w^4 & w^2\\
      1 & w^6 & w^5 & w^4 & w^3 & w^2 & w
    \end{array}
  \right] \ ,
\end{equation}
where $w = \exp(\Ii \cdot 2\pi/7)$, so $w^7 = 1$.


\subsubsection{$1$-parameter orbits}

There are precisely three \PERMUTATIONequivalent\ $1$-parameter \maxAUFs\
stemming from the symmetric matrix being a permuted 'starting point' for the
$1$-parameter orbit found by Petrescu \cite{Pe97}:
\begin{equation}
\label{U_Petr7}
  P_7 = 
  \left[
    \begin{array}{ccccccc}
      1 & 1 & 1 & 1 & 1 & 1 & 1\\
      1 & w & w^4 & w^5 & w^3 & w^3 & w\\
      1 & w^4 & w & w^3 & w^5 & w^3 & w\\
      1 & w^5 & w^3 & w & w^4 & w & w^3\\
      1 & w^3 & w^5 & w^4 & w & w & w^3\\
      1 & w^3 & w^3 & w & w & w^4 & w^5\\
      1 & w & w & w^3 & w^3 & w^5 & w^4
    \end{array}
  \right] \ ,
\end{equation}
where $w = \exp(\Ii \cdot 2\pi/6)$, so $w^6 = 1,\ w^3 = -1$. 

They are: 
\begin{equation}
\label{U_Petr7_orbits}
 \AUFfamily{P}{7}{1}(a) = P_7 \HADprod \EXPentrywise \left( \Ii \cdot
\AUFphases{P}{7}{1}(a) \right)
\end{equation}
where
\begin{equation}
  \AUFphases{P}{7}{1}(a) = 
  \left[
    \begin{array}{ccccccc}
      \bO &  \bO &  \bO &  \bO &  \bO &  \bO &  \bO \\
      \bO &  a &  a &  \bO &  \bO &  \bO &  \bO \\
      \bO &  a &  a &  \bO &  \bO &  \bO &  \bO \\
      \bO &  \bO &  \bO & -a & -a &  \bO &  \bO \\
      \bO &  \bO &  \bO & -a & -a &  \bO &  \bO \\
      \bO &  \bO &  \bO &  \bO &  \bO &  \bO &  \bO \\
      \bO &  \bO &  \bO &  \bO &  \bO &  \bO &  \bO
    \end{array}
  \right]
\end{equation}
and
\begin{eqnarray}
  P_1 \cdot \AUFfamily{P}{7}{1}(a) \cdot P_2^T  & = &
  P_7 \HADprod \EXPentrywise \left( \Ii \cdot P_1 \AUFphases{P}{7}{1}(a) P_2^T
\right)  \\
  P_2 \cdot \AUFfamily{P}{7}{1}(a)  \cdot P_1^T  & = & 
  P_7 \HADprod \EXPentrywise \left( \Ii \cdot P_2 \AUFphases{P}{7}{1}(a) P_1^T
\right)
\end{eqnarray}
where
\begin{equation}
  P_1 = [ \STbasis{1},\ \STbasis{7},\ \STbasis{3},\ \STbasis{4},\ \STbasis{6},\
  \STbasis{5},\ \STbasis{2} ]
  \ \ \ \ \ \ \
  P_2 = [ \STbasis{1},\ \STbasis{2},\ \STbasis{7},\ \STbasis{6},\ \STbasis{5},\
  \STbasis{4},\ \STbasis{3} ]
\end{equation}
and $\STbasis{i}$ denotes the $i$-th standard basis column
vector.
\bigskip

The above orbits
are \PERMUTATIONequivalent\ to the $7 \times 7$ family found
by Petrescu \cite{Pe97}. They all are \COGNATE, and the
$\AUFfamily{P}{7}{1}$ orbit is even \selfCOGNATE\ since
\begin{equation}
  \AUFfamily{P}{7}{1}(a)  = \TRANSPOSE{\AUFfamily{P}{7}{1}(a)} \ .
\end{equation}
\bigskip

Also
\begin{equation}
  \AUFfamily{P}{7}{1}(-a) = P^T \cdot \AUFfamily{P}{7}{1}(a) \cdot P
  \ \ \ \ \mbox{for}\ \ \ \ 
  P = [ \STbasis{1},\ \STbasis{4},\ \STbasis{5},\ \STbasis{2},\ \STbasis{3},\
  \STbasis{7},\ \STbasis{6} ]
\end{equation}
so $\AUFfamily{P}{7}{1}(-a) \RELofEQUI \AUFfamily{P}{7}{1}(a)$
\bigskip

Due to some freedom in the construction of family components 
the method of Petrescu allows one to build other
families of Hadamard matrices similar to $\AUFfamily{P}{7}{1}$. 
Not knowing if they are inequivalent
we are not going to consider them here.


\subsubsection{The 'cyclic $7$--roots' matrices}

There exist only four inequivalent $7 \times 7$
complex Hadamard matrices, inequivalent to $F_7$ and the
$1$-parameter family found by Petrescu (see previous subsection),
associated with nonclassical cyclic $7$--roots.
This result was obtained
in \cite{Ha96} and is based on the catalogue of all cyclic $7$--roots
presented in \cite{BF94}.

\bigskip

The two matrices $\AUFfamily[A]{\tilde{C}}{7}{0},\
\AUFfamily[B]{\tilde{C}}{7}{0}$
correspond to the so--called 'index 2' solutions to the cyclic $7$--roots
problem. They have the circulant structure $\ELEMENTof{U}{i}{j} = x_{(i-j \mod
7)+1}$, where
\begin{eqnarray}
  x = [ 1,\  1,\  1,\       d,\    1,\       d,\        d  ] & for &
\AUFfamily[A]{\tilde{C}}{7}{0} \\
  x = [ 1,\  1,\  1,\ \CONJ{d},\   1,\ \CONJ{d},\ \CONJ{d} ] & for &
\AUFfamily[B]{\tilde{C}}{7}{0}
\end{eqnarray}
and
\begin{equation}
  d = \frac{-3+\Ii \sqrt{7}}{4}\ \ \ \  \mbox{such that}\ \ \ \  d^2 +
  \frac{3}{2}d + 1 = 0 \ \ \ \Longrightarrow d \cdot \CONJ{d} = 1
\end{equation}

The corresponding \DEPHASED\ matrices are denoted as $\AUFfamily[A]{C}{7}{0}$
and
$\AUFfamily[B]{C}{7}{0}$:

\begin{equation}
  \AUFfamily[A]{\tilde{C}}{7}{0} = 
  \left[
    \begin{array}{ccccccc}
      1 & d & d & 1 & d & 1 & 1\\
      1 & 1 & d & d & 1 & d & 1\\
      1 & 1 & 1 & d & d & 1 & d\\
      d & 1 & 1 & 1 & d & d & 1\\
      1 & d & 1 & 1 & 1 & d & d\\
      d & 1 & d & 1 & 1 & 1 & d\\
      d & d & 1 & d & 1 & 1 & 1
    \end{array}
  \right]
\ \ \ \ 
 \AUFfamily[A]{C}{7}{0} = 
  \left[
    \begin{array}{ccccccc}
      1 & 1 & 1 & 1 & 1 & 1 & 1\\
      1 & d^{\SSS -1} & 1 & d & d^{\SSS -1} & d & 1\\
      1 & d^{\SSS -1} & d^{\SSS -1} & d & 1 & 1 & d\\
      1 & d^{\SSS -2} & d^{\SSS -2} & d^{\SSS -1} & d^{\SSS -1} & 1 & d^{\SSS
-1}\\
      1 & 1 & d^{\SSS -1} & 1 & d^{\SSS -1} & d & d\\
      1 & d^{\SSS -2} & d^{\SSS -1} & d^{\SSS -1} & d^{\SSS -2} & d^{\SSS -1} &
1\\
      1 & d^{\SSS -1} & d^{\SSS -2} & 1 & d^{\SSS -2} & d^{\SSS -1} &
      d^{\SSS -1}
    \end{array}
  \right]
\end{equation}

\begin{equation}
  \AUFfamily[B]{\tilde{C}}{7}{0} =
  \left[
    \begin{array}{ccccccc}
      1 & d^{\SSS -1} & d^{\SSS -1} & 1 & d^{\SSS -1} & 1 & 1\\
      1 & 1 & d^{\SSS -1} & d^{\SSS -1} & 1 & d^{\SSS -1} & 1\\
      1 & 1 & 1 & d^{\SSS -1} & d^{\SSS -1} & 1 & d^{\SSS -1}\\
      d^{\SSS -1} & 1 & 1 & 1 & d^{\SSS -1} & d^{\SSS -1} & 1\\
      1 & d^{\SSS -1} & 1 & 1 & 1 & d^{\SSS -1} & d^{\SSS -1}\\
      d^{\SSS -1} & 1 & d^{\SSS -1} & 1 & 1 & 1 & d^{\SSS -1}\\
      d^{\SSS -1} & d^{\SSS -1} & 1 & d^{\SSS -1} & 1 & 1 & 1
    \end{array}
  \right]
\ \ \ \
  \AUFfamily[B]{C}{7}{0} =
  \left[
    \begin{array}{ccccccc}
      1 & 1 & 1 & 1 & 1 & 1 & 1\\
      1 & d & 1 & d^{\SSS -1} & d & d^{\SSS -1} & 1\\
      1 & d & d & d^{\SSS -1} & 1 & 1 & d^{\SSS -1}\\
      1 & d^2 & d^2 & d & d & 1 & d\\
      1 & 1 & d & 1 & d & d^{\SSS -1} & d^{\SSS -1}\\
      1 & d^2 & d & d & d^2 & d & 1\\
      1 & d & d^2 & 1 & d^2 & d & d
    \end{array}
  \right]
\end{equation}

There holds $\AUFfamily[B]{C}{7}{0} = \CONJ{\AUFfamily[A]{C}{7}{0}}$\ \ \ 
(\ \ $\Longleftrightarrow \AUFfamily[B]{\tilde{C}}{7}{0} =
\CONJ{\AUFfamily[A]{\tilde{C}}{7}{0}}$\ \ ) 
and the matrices
%
$\AUFfamily[A]{C}{7}{0}$, $\AUFfamily[A]{\tilde{C}}{7}{0}$ and 
$\AUFfamily[B]{C}{7}{0}$, $\AUFfamily[B]{\tilde{C}}{7}{0}$
are equivalent to 
$\TRANSPOSE{\AUFfamily[A]{C}{7}{0}}$, 
$\TRANSPOSE{\AUFfamily[A]{\tilde{C}}{7}{0}}$ 
and 
$\TRANSPOSE{\AUFfamily[B]{C}{7}{0}}$, 
$\TRANSPOSE{\AUFfamily[B]{\tilde{C}}{7}{0}}$ 
respectively (see (\ref{eq_circulant_equivalence})).
\bigskip
\bigskip

The structure of $\AUFfamily[C]{\tilde{C}}{7}{0},
\AUFfamily[D]{\tilde{C}}{7}{0}$
related to 'index 3' solutions to the cyclic $7$--roots problem is again
$\ELEMENTof{U}{i}{j} = x_{(i-j \mod 7)+1}$, where $x$ is a bit more
complicated. 

We put:
\begin{eqnarray}
  x = [ 1,\ A,\ B,\ C,\ C,\ B,\ A ]^T  &  \mbox{for}  &
  \AUFfamily[C]{\tilde{C}}{7}{0} \\
  x = [ 1,\ \CONJ{A},\ \CONJ{B},\ \CONJ{C},\ \CONJ{C},\ \CONJ{B},\ \CONJ{A} ]^T 
&  \mbox{for}  &
  \AUFfamily[D]{\tilde{C}}{7}{0}
\end{eqnarray}
where $A=a,\ \ B=ab,\ \ C=abc$\ \  are products of algebraic numbers $a,\
b,\ c$\ \  of modulus equal to $1$ (see the final remark 3.11 of
\cite{Ha96}). Their numerical approximations are given by
\begin{eqnarray}
  a & \approx & \exp(\Ii \cdot 4.312839) \\ 
  b & \approx & \exp(\Ii \cdot 1.356228) \\
  c & \approx & \exp(\Ii \cdot 1.900668) \\
\end{eqnarray}
and then
\begin{eqnarray}
  A & = & a   \approx (-0.389004)  + \Ii \cdot (-0.921236) \\
  B & = & ab  \approx ( 0.817282)  + \Ii \cdot (-0.576238) \\
  C & = & abc \approx ( 0.280434)  + \Ii \cdot ( 0.959873)
\end{eqnarray}

%
%

Again, $\AUFfamily[C]{C}{7}{0},\ \AUFfamily[D]{C}{7}{0}$ denote the \DEPHASED\
versions of
 $\AUFfamily[C]{\tilde{C}}{7}{0},\ \AUFfamily[D]{\tilde{C}}{7}{0}$, which read:
\begin{equation}
  \AUFfamily[C]{\tilde{C}}{7}{0} =
  \left[
    \begin{array}{ccccccc}
      1 & A & B & C & C & B & A\\
      A & 1 & A & B & C & C & B\\
      B & A & 1 & A & B & C & C\\
      C & B & A & 1 & A & B & C\\
      C & C & B & A & 1 & A & B\\
      B & C & C & B & A & 1 & A\\
      A & B & C & C & B & A & 1
    \end{array}
  \right]
\ \ \ \
 \AUFfamily[D]{\tilde{C}}{7}{0} =
  \left[
    \begin{array}{ccccccc}
      1 & A^{\SSS -1} & B^{\SSS -1} & C^{\SSS -1} & C^{\SSS -1} & B^{\SSS -1} &
A^{\SSS -1}\\
      A^{\SSS -1} & 1 & A^{\SSS -1} & B^{\SSS -1} & C^{\SSS -1} & C^{\SSS -1} &
B^{\SSS -1}\\
      B^{\SSS -1} & A^{\SSS -1} & 1 & A^{\SSS -1} & B^{\SSS -1} & C^{\SSS -1} &
C^{\SSS -1}\\
      C^{\SSS -1} & B^{\SSS -1} & A^{\SSS -1} & 1 & A^{\SSS -1} & B^{\SSS -1} &
C^{\SSS -1}\\
      C^{\SSS -1} & C^{\SSS -1} & B^{\SSS -1} & A^{\SSS -1} & 1 & A^{\SSS -1} &
B^{\SSS -1}\\
      B^{\SSS -1} & C^{\SSS -1} & C^{\SSS -1} & B^{\SSS -1} & A^{\SSS -1} & 1 &
A^{\SSS -1}\\
      A^{\SSS -1} & B^{\SSS -1} & C^{\SSS -1} & C^{\SSS -1} & B^{\SSS -1} &
      A^{\SSS -1} & 1
    \end{array}
  \right]
\end{equation}
\bigskip

Then
\begin{eqnarray}
  \AUFfamily[C]{C}{7}{0} &  =  &
  \left[
    \begin{array}{c|c|c|c|c|c|c}
      1 & 1 & 1 & 1 & 1 & 1 & 1\\
      \hline
      1 & A^{\SSS -2} & B^{\SSS -1} & A^{\SSS -1} B C^{\SSS -1} &
      A^{\SSS -1} & A^{\SSS -1} B^{\SSS -1} C & A^{\SSS -2} B\\
      \hline
      1 & B^{\SSS -1} & B^{\SSS -2} & A B^{\SSS -1} C^{\SSS -1} &
      C^{\SSS -1} & B^{\SSS -2} C & A^{\SSS -1} B^{\SSS -1} C\\
      \hline
      1 & A^{\SSS -1} B C^{\SSS -1} & A B^{\SSS -1} C^{\SSS -1} &
      C^{\SSS -2} & A C^{\SSS -2} & C^{\SSS -1} & A^{\SSS -1}\\
      \hline
      1 & A^{\SSS -1} & C^{\SSS -1} & A C^{\SSS -2} & C^{\SSS -2} & A
      B^{\SSS -1} C^{\SSS -1} & A^{\SSS -1} B C^{\SSS -1}\\
      \hline
      1 & A^{\SSS -1} B^{\SSS -1} C & B^{\SSS -2} C & C^{\SSS -1} & A
      B^{\SSS -1} C^{\SSS -1} & B^{\SSS -2} & B^{\SSS -1}\\
      \hline
      1 & A^{\SSS -2} B & A^{\SSS -1} B^{\SSS -1} C & A^{\SSS -1} & A^{\SSS -1}
B C^{\SSS -1} &
      B^{\SSS -1} & A^{\SSS -2}
    \end{array}
  \right]
  \nonumber
\\
               &  =  & 
  \left[
    \begin{array}{c|c|c|c|c|c|c}
      1 & 1 & 1 & 1 & 1 & 1 & 1\\
      \hline
      1 & a^{\SSS -2} & a^{\SSS -1} b^{\SSS -1} & a^{\SSS -1} c^{\SSS
        -1} & a^{\SSS -1} & a^{\SSS -1} c & a^{\SSS -1} b\\
      \hline
      1 & a^{\SSS -1} b^{\SSS -1} & a^{\SSS -2} b^{\SSS -2} & a^{\SSS
        -1} b^{\SSS -2} c^{\SSS -1} & a^{\SSS -1} b^{\SSS -1} c^{\SSS
        -1} & a^{\SSS -1} b^{\SSS -1} c & a^{\SSS -1} c\\
      \hline
      1 & a^{\SSS -1} c^{\SSS -1} & a^{\SSS -1} b^{\SSS -2} c^{\SSS
        -1} & a^{\SSS -2} b^{\SSS -2} c^{\SSS -2} & a^{\SSS -1}
      b^{\SSS -2} c^{\SSS -2} & a^{\SSS -1} b^{\SSS -1} c^{\SSS -1} &
      a^{\SSS -1}\\
      \hline
      1 & a^{\SSS -1} & a^{\SSS -1} b^{\SSS -1} c^{\SSS -1} & a^{\SSS
        -1} b^{\SSS -2} c^{\SSS -2} & a^{\SSS -2} b^{\SSS -2} c^{\SSS
        -2} & a^{\SSS -1} b^{\SSS -2} c^{\SSS -1} & a^{\SSS -1}
      c^{\SSS -1}\\
      \hline
      1 & a^{\SSS -1} c & a^{\SSS -1} b^{\SSS -1} c & a^{\SSS -1}
      b^{\SSS -1} c^{\SSS -1} & a^{\SSS -1} b^{\SSS -2} c^{\SSS -1} &
      a^{\SSS -2} b^{\SSS -2} & a^{\SSS -1} b^{\SSS -1}\\
      \hline
      1 & a^{\SSS -1} b & a^{\SSS -1} c & a^{\SSS -1} & a^{\SSS -1} c^{\SSS -1}
&
      a^{\SSS -1} b^{\SSS -1} & a^{\SSS -2}
    \end{array}
  \right]
\end{eqnarray}
and
\begin{eqnarray}
 \AUFfamily[D]{C}{7}{0} &  =  & 
  \left[
    \begin{array}{c|c|c|c|c|c|c}
      1 & 1 & 1 & 1 & 1 & 1 & 1\\
      \hline
      1 & A^2 & B & A B^{\SSS -1} C & A & A B C^{\SSS -1} & A^{\SSS 2}
      B^{\SSS -1}\\
      \hline
      1 & B & B^2 & A^{\SSS -1} B C & C & B^{\SSS 2} C^{\SSS -1} & A B
      C^{\SSS -1}\\
      \hline
      1 & A B^{\SSS -1} C & A^{\SSS -1} B C & C^2 &  A^{\SSS -1}
      C^{\SSS 2} & C & A\\
      \hline
      1 & A & C &  A^{\SSS -1} C^{\SSS 2} & C^2 & A^{\SSS -1} B C & A
      B^{\SSS -1} C\\
      \hline
      1 & A B C^{\SSS -1} & B^{\SSS 2} C^{\SSS -1} & C & A^{\SSS -1} B
      C & B^2 & B\\
      \hline
      1 & A^{\SSS 2} B^{\SSS -1} & A B C^{\SSS -1} & A & A B^{\SSS -1} C & B &
A^2
    \end{array}
  \right]
  \nonumber
\\
               &  =  &
  \left[
    \begin{array}{c|c|c|c|c|c|c}
      1 & 1 & 1 & 1 & 1 & 1 & 1\\
      \hline
      1 & a^2 & a b & a c & a & a c^{\SSS -1} & a b^{\SSS -1}\\
      \hline
      1 & a b & a^2 b^2 & a b^2 c & a b c & a b c^{\SSS -1} & a
      c^{\SSS -1}\\
      \hline
      1 & a c & a b^2 c & a^2 b^2 c^2 & a b^2 c^2 & a b c & a\\
      \hline
      1 & a & a b c & a b^2 c^2 & a^2 b^2 c^2 & a b^2 c & a c\\
      \hline
      1 & a c^{\SSS -1} & a b c^{\SSS -1} & a b c & a b^2 c & a^2 b^2
      & a b\\
      \hline
      1 & a b^{\SSS -1} & a c^{\SSS -1} & a & a c & a b & a^2
    \end{array}
  \right]
\end{eqnarray}


The matrices 
$\AUFfamily[C]{C}{7}{0}$ and $\AUFfamily[D]{C}{7}{0}$
are symmetric 
and are related by complex conjugation,
$\AUFfamily[D]{C}{7}{0} = \CONJ{\AUFfamily[C]{C}{7}{0}}$.
All four cyclic $7$--roots Hadamard matrices
are isolated \cite{Ni04}.

\ENDofSUBSECTION

%
%
 
\subsection {$N=8$}
\label{subsec__N_eq_8}


\subsubsection{Orbits stemming from $F_{8}$}

The only \maxAUF\ stemming from $F_8$ is the 
$5$-parameter orbit:
\begin{equation}
\label{F8_maxAUF}
  \AUFfamily{F}{8}{5}(a,b,c,d,e) = F_8 \HADprod \EXPentrywise (\Ii \cdot
\AUFphases{F}{8}{5}(a,b,c,d,e))
\end{equation}
where
\begin{equation}
  F_8 =
  \left[
    \begin{array}{cccccccc}
      1 & 1 & 1 & 1 & 1 & 1 & 1 & 1\\
      1 & w & w^2 & w^3 & w^4 & w^5 & w^6 & w^7\\
      1 & w^2 & w^4 & w^6 & 1 & w^2 & w^4 & w^6\\
      1 & w^3 & w^6 & w & w^4 & w^7 & w^2 & w^5\\
      1 & w^4 & 1 & w^4 & 1 & w^4 & 1 & w^4\\
      1 & w^5 & w^2 & w^7 & w^4 & w & w^6 & w^3\\
      1 & w^6 & w^4 & w^2 & 1 & w^6 & w^4 & w^2\\
      1 & w^7 & w^6 & w^5 & w^4 & w^3 & w^2 & w
    \end{array}
  \right] \ ,
\end{equation}
where $w = \exp(\Ii \cdot 2\pi/8)$, so $w^8=1,\ w^4=-1,\ w^2=\Ii$, and
\begin{equation}
   \AUFphases{F}{8}{5}(a,b,c,d,e) = 
   \left[
     \begin{array}{cccc|cccc}
       \bO & \bO & \bO & \bO & \bO & \bO & \bO & \bO\\
       \bO & a & b & c & \bO & a & b & c\\
       \bO & d & \bO & d & \bO & d & \bO & d\\
       \bO & e & b & c - a + e & \bO & e & b & c - a + e\\
       \hline
       \bO & \bO & \bO & \bO & \bO & \bO & \bO & \bO\\
       \bO & a & b & c & \bO & a & b & c\\
       \bO & d & \bO & d & \bO & d & \bO & d\\
       \bO & e & b & c - a + e & \bO & e & b & c - a + e
     \end{array}
   \right] \ .
\end{equation}
The family $\AUFfamily{F}{8}{5}$ is \selfCOGNATE.
\bigskip

The above orbit is \PERMUTATIONequivalent\ to the orbit constructed
with  the Di{\c t}{\v a}'s method \cite{Di04}:
\begin{equation}
\label{F8fam_Dita}
  \AUFfamily{\tilde{F}}{8}{5}(\alpha_1,\ldots,\alpha_5) = 
  \left[ \begin{array}{c|c}
            \ELEMENTof{F_2}{1}{1} \cdot \AUFfamily{F}{4}{1}(\alpha_1) &
\ELEMENTof{F_2}{1}{2} \cdot
            D(\alpha_3,\alpha_4,\alpha_5) \cdot \AUFfamily{F}{4}{1}(\alpha_2)
\\
            \hline
            \ELEMENTof{F_2}{2}{1} \cdot \AUFfamily{F}{4}{1}(\alpha_1) &
\ELEMENTof{F_2}{2}{2} \cdot
            D(\alpha_3,\alpha_4,\alpha_5) \cdot \AUFfamily{F}{4}{1}(\alpha_2)
         \end{array} \right]
\end{equation}
where
\begin{equation}
\label{F4_mAUF}
  \AUFfamily{F}{4}{1}(\alpha) = 
  \left[ 
    \begin{array}{cccc} 
      1 & 1 &  1 & 1  \\ 
      1 & \Ii \PHASE{\alpha} & -1 & -\Ii \PHASE{\alpha}  \\ 
      1 & -1 & 1 & -1  \\ 
      1 & -\Ii \PHASE{\alpha} & -1 & \Ii \PHASE{\alpha}  
    \end{array} \right]
\end{equation}
is the only \maxAUF\ stemming from $F_4$ and
$D(\alpha_3,\alpha_4,\alpha_5)$ is the $4 \times 4$ diagonal matrix
$\diag(1,\PHASE{\alpha_3},\PHASE{\alpha_4},\PHASE{\alpha_5})$. 
\medskip

Eq. 
(\ref{F8fam_Dita}) leads to $F_2
\otimes F_4$ for $\alpha_1 = \ldots = \alpha_5 = 0$, which is not
equivalent to $F_8$, see \cite{Ta05}. However,
$\AUFfamily{\tilde{F}}{8}{5}(0,0,(1/8)2\pi,(2/8)2\pi,(3/8)2\pi)$ is permutation
equivalent to $F_8$, since 
\begin{equation}
  F_8 = \AUFfamily{\tilde{F}}{8}{5}(0,0,(1/8)2\pi,(2/8)2\pi,(3/8)2\pi) \cdot        
[\STbasis{1},\STbasis{3},\STbasis{5},\STbasis{7},\STbasis{2},\STbasis{4},\STbasis{6},\STbasis{8}]^{T}
\ ,
\end{equation}
where $\STbasis{i}$ denotes the $i$-th column vector of the standard
basis of $\mathbf{C}^8$.
Thus Eq. (\ref{F8fam_Dita}) generates the only \maxAUF\ stemming from  permuted
$F_8$. 

The matrix  
$\AUFfamily{\tilde{F}}{8}{5}(\pi/2,\pi/2,0,0,0)$ yields the only
real $8 \times 8$
Hadamard matrix, up to permutations
and multiplying rows and columns by $-1$.
 It is \DEPHASED, so it is permutation equivalent to $F_2 \otimes F_2
\otimes F_2$.

Therefore all appropriately permuted tensor products
of Fourier matrices,
 $F_2 \otimes F_2
\otimes F_2$, $F_2 \otimes F_4$ and $F_8$, 
although inequivalent \cite{Ta05},
 are connected by the orbit (\ref{F8fam_Dita}).
\ENDofSUBSECTION

%
%

\subsection {$N=9$}
\label{subsec__N_eq_9}


\subsubsection{Orbits stemming from $F_{9}$}

The only \maxAUF\ stemming from $F_9$ is the 
$4$-parameter orbit:
\begin{equation}
\label{F9_maxAUF}
  \AUFfamily{F}{9}{4}(a,b,c,d) = F_9 \HADprod \EXPentrywise (\Ii \cdot
\AUFphases{F}{9}{4}(a,b,c,d)) \ ,
\end{equation}
where
\begin{equation}
  F_9 =
  \left[
    \begin{array}{ccccccccc}
      1 & 1 & 1 & 1 & 1 & 1 & 1 & 1 & 1\\
      1 & w & w^2 & w^3 & w^4 & w^5 & w^6 & w^7 & w^8\\
      1 & w^2 & w^4 & w^6 & w^8 & w & w^3 & w^5 & w^7\\
      1 & w^3 & w^6 & 1 & w^3 & w^6 & 1 & w^3 & w^6\\
      1 & w^4 & w^8 & w^3 & w^7 & w^2 & w^6 & w & w^5\\
      1 & w^5 & w & w^6 & w^2 & w^7 & w^3 & w^8 & w^4\\
      1 & w^6 & w^3 & 1 & w^6 & w^3 & 1 & w^6 & w^3\\
      1 & w^7 & w^5 & w^3 & w & w^8 & w^6 & w^4 & w^2\\
      1 & w^8 & w^7 & w^6 & w^5 & w^4 & w^3 & w^2 & w
    \end{array}
  \right] \ ,
\end{equation}
with  $w = \exp(\Ii \cdot 2\pi/9)$, so $w^9 = 1$, and
\begin{equation}
  \AUFphases{F}{9}{4}(a,b,c,d) =
  \left[
    \begin{array}{ccc|ccc|ccc}
      \bO & \bO & \bO & \bO & \bO & \bO & \bO & \bO & \bO\\
      \bO & a & b & \bO & a & b & \bO & a & b\\
      \bO & c & d & \bO & c & d & \bO & c & d\\
      \hline
      \bO & \bO & \bO & \bO & \bO & \bO & \bO & \bO & \bO\\
      \bO & a & b & \bO & a & b & \bO & a & b\\
      \bO & c & d & \bO & c & d & \bO & c & d\\
      \hline
      \bO & \bO & \bO & \bO & \bO & \bO & \bO & \bO & \bO\\
      \bO & a & b & \bO & a & b & \bO & a & b\\
      \bO & c & d & \bO & c & d & \bO & c & d
    \end{array}
  \right] \ .
\end{equation}
The orbit 
$\AUFfamily{F}{9}{4}$ is \selfCOGNATE.
Observe that its dimension is equal to the defect, 
$d(F_{9})=4$, which follows from Eq. (\ref{powprime}).
\bigskip

\bigskip

It is \PERMUTATIONequivalent\ to the $4$-dimensional orbit passing
through a permuted $F_9$, constructed using  the Di{\c t}{\v a}'s method:
\begin{equation}
\label{F9fam_Dita}
  \AUFfamily{\tilde{F}}{9}{4}(\alpha_1,\ldots,\alpha_4) = 
  \left[ \begin{array}{c|c|c}
            \ELEMENTof{F_3}{1}{1} \cdot F_3 & \ELEMENTof{F_3}{1}{2} \cdot
            D(\alpha_1,\alpha_2) \cdot F_3 & \ELEMENTof{F_3}{1}{3} \cdot
            D(\alpha_3,\alpha_4) \cdot F_3 \\
            \hline
            \ELEMENTof{F_3}{2}{1} \cdot F_3 & \ELEMENTof{F_3}{2}{2} \cdot
            D(\alpha_1,\alpha_2) \cdot F_3 & \ELEMENTof{F_3}{2}{3} \cdot
            D(\alpha_3,\alpha_4) \cdot F_3 \\
            \hline
            \ELEMENTof{F_3}{3}{1} \cdot F_3 & \ELEMENTof{F_3}{3}{2} \cdot
            D(\alpha_1,\alpha_2) \cdot F_3 & \ELEMENTof{F_3}{3}{3} \cdot
            D(\alpha_3,\alpha_4) \cdot F_3
         \end{array} \right]
\end{equation}
where $D(\alpha,\beta)$ is the $3 \times 3$ diagonal matrix
$\diag(1,\PHASE{\alpha},\PHASE{\beta})$.
\medskip

The matrix  
$\AUFfamily{\tilde{F}}{9}{4}(0,0,0,0) = F_3 \otimes F_3$ is not  equivalent to
$F_9$ \cite{Ta05}, but
\begin{equation}
  F_9 = \AUFfamily{\tilde{F}}{9}{4}((1/9)2\pi,(2/9)2\pi,(2/9)2\pi,(4/9)2\pi)
\cdot         [\STbasis{1},\STbasis{4},\STbasis{7},\STbasis{2},\STbasis{5},
\STbasis{8},\STbasis{3},\STbasis{6},\STbasis{9}]^{T} \ ,
\end{equation}
where $\STbasis{i}$ are the standard basis column vectors.

Thus both inequivalent permuted  matrices, $F_3 \otimes F_3$ and
$F_9$, are connected by the orbit (\ref{F9fam_Dita}).
\ENDofSUBSECTION

%
%

\subsection {$N=10$}
\label{subsec__N_eq_10}


\subsubsection{Orbits stemming from $F_{10}$}

The only \maxAUFs\  stemming from $F_{10}$ are:
\begin{eqnarray}
  \AUFfamily{F}{10}{4}(a,b,c,d)  &  = & F_{10} \HADprod \EXPentrywise
  \left(\Ii \cdot \AUFphases{F}{10}{4}(a,b,c,d)\right)  \label{F10_maxAUF}\\
  \TRANSPOSE{\AUFfamily{F}{10}{4}(a,b,c,d)}  &  = & F_{10} \HADprod
\EXPentrywise
  \left(\Ii \cdot \TRANSPOSE{\AUFphases{F}{10}{4}(a,b,c,d)} \right) \ ,
\label{F10_maxAUF_transposed} 
\end{eqnarray}
where
\begin{equation}
  F_{10} =
  \left[
    \begin{array}{cccccccccc}
      1 & 1 & 1 & 1 & 1 & 1 & 1 & 1 & 1 & 1\\
      1 & w & w^2 & w^3 & w^4 & w^5 & w^6 & w^7 & w^8 & w^9\\
      1 & w^2 & w^4 & w^6 & w^8 & 1 & w^2 & w^4 & w^6 & w^8\\
      1 & w^3 & w^6 & w^9 & w^2 & w^5 & w^8 & w & w^4 & w^7\\
      1 & w^4 & w^8 & w^2 & w^6 & 1 & w^4 & w^8 & w^2 & w^6\\
      1 & w^5 & 1 & w^5 & 1 & w^5 & 1 & w^5 & 1 & w^5\\
      1 & w^6 & w^2 & w^8 & w^4 & 1 & w^6 & w^2 & w^8 & w^4\\
      1 & w^7 & w^4 & w & w^8 & w^5 & w^2 & w^9 & w^6 & w^3\\
      1 & w^8 & w^6 & w^4 & w^2 & 1 & w^8 & w^6 & w^4 & w^2\\
      1 & w^9 & w^8 & w^7 & w^6 & w^5 & w^4 & w^3 & w^2 & w
    \end{array}
  \right]
\end{equation}
with $w = \exp(\Ii \cdot 2\pi/10)$, so $w^{10} = 1,\ w^5 = -1$, and
\begin{equation}
 \AUFphases{F}{10}{4}(a,b,c,d) = 
  \left[
    \begin{array}{ccccc|ccccc}
      \bO & \bO & \bO & \bO & \bO & \bO & \bO & \bO & \bO & \bO\\
      \bO & a & b & c & d & \bO & a & b & c & d\\
      \hline
      \bO & \bO & \bO & \bO & \bO & \bO & \bO & \bO & \bO & \bO\\
      \bO & a & b & c & d & \bO & a & b & c & d\\
      \hline
      \bO & \bO & \bO & \bO & \bO & \bO & \bO & \bO & \bO & \bO\\
      \bO & a & b & c & d & \bO & a & b & c & d\\
      \hline
      \bO & \bO & \bO & \bO & \bO & \bO & \bO & \bO & \bO & \bO\\
      \bO & a & b & c & d & \bO & a & b & c & d\\
      \hline
      \bO & \bO & \bO & \bO & \bO & \bO & \bO & \bO & \bO & \bO\\
      \bO & a & b & c & d & \bO & a & b & c & d
    \end{array}
  \right]
\end{equation}
The affine Hadamard families 
$\AUFfamily{F}{10}{4}$ and $\TRANSPOSE{\AUFfamily{F}{10}{4}}$ are
cognate. 
\bigskip

At least one of them must be \PERMUTATIONequivalent\ to an orbit
constructed using  the Di{\c t}{\v a}'s method, either:
\begin{equation}
\label{F10fam_Dita_1}
  \AUFfamily[A]{\tilde{F}}{10}{4}(\alpha_1,\ldots,\alpha_4) = 
  \left[ \begin{array}{c|c}
            \ELEMENTof{F_2}{1}{1} \cdot F_5 & \ELEMENTof{F_2}{1}{2} \cdot
            D(\alpha_1,\ldots,\alpha_4) \cdot F_5 \\
            \hline
            \ELEMENTof{F_2}{2}{1} \cdot F_5 & \ELEMENTof{F_2}{2}{2} \cdot
            D(\alpha_1,\ldots,\alpha_4) \cdot F_5
         \end{array} \right]
\end{equation}
where $D(\alpha_1,\ldots,\alpha_4)$ is the $5 \times 5$ diagonal
matrix $\diag(1,\PHASE{\alpha_1},\ldots,\PHASE{\alpha_4})$,
or
\begin{equation}
\label{F10fam_Dita_2}
  \AUFfamily[B]{\tilde{F}}{10}{4}(\beta_1,\ldots,\beta_4)
\end{equation}
such that its\ \  $i,j$-th $2 \times 2$ block is equal to
$\ELEMENTof{F_5}{i}{j}
\cdot D(\alpha) \cdot F_2$, where $i,j \in \{1 \ldots 5\}$,
$D(\alpha)$ is the diagonal matrix $\diag(1,\PHASE{\alpha})$ and
$\alpha = 0,\beta_1,\ldots,\beta_4$ for $j = 1,2,\ldots,5$
respectively.
\medskip

This is because $\AUFfamily[A]{\tilde{F}}{10}{4}(\ZEROvect) = F_2 \otimes F_5$
and
$\AUFfamily[B]{\tilde{F}}{10}{4}(\ZEROvect) = F_5 \otimes F_2$ are, according
to
\cite{Ta05}, permutation equivalent to $F_{10}$, so both Di{\c t}{\v a}'s
orbits are \maxAUFs\ stemming from
permuted $F_{10}$'s.
\ENDofSUBSECTION

%
%

\subsection {$N=11$}
\label{subsec__N_eq_11}


\subsubsection{Orbits stemming from $F_{11}$}

The Fourier matrix 
$F_{11}$ is an \ISOLATED\ $11 \times 11$ complex Hadamard matrix:
\begin{equation}
  F_{11} = \AUFfamily{F}{11}{0} = 
  \left[
    \begin{array}{ccccccccccc}
      1 & 1 & 1 & 1 & 1 & 1 & 1 & 1 & 1 & 1 & 1\\
      1 & w & w^2 & w^3 & w^4 & w^5 & w^6 & w^7 & w^8 & w^9 & w^{10}\\
      1 & w^2 & w^4 & w^6 & w^8 & w^{10} & w & w^3 & w^5 & w^7 & w^9\\
      1 & w^3 & w^6 & w^9 & w & w^4 & w^7 & w^{10} & w^2 & w^5 & w^8\\
      1 & w^4 & w^8 & w & w^5 & w^9 & w^2 & w^6 & w^{10} & w^3 & w^7\\
      1 & w^5 & w^{10} & w^4 & w^9 & w^3 & w^8 & w^2 & w^7 & w & w^6\\
      1 & w^6 & w & w^7 & w^2 & w^8 & w^3 & w^9 & w^4 & w^{10} & w^5\\
      1 & w^7 & w^3 & w^{10} & w^6 & w^2 & w^9 & w^5 & w & w^8 & w^4\\
      1 & w^8 & w^5 & w^2 & w^{10} & w^7 & w^4 & w & w^9 & w^6 & w^3\\
      1 & w^9 & w^7 & w^5 & w^3 & w & w^{10} & w^8 & w^6 & w^4 & w^2\\
      1 & w^{10} & w^9 & w^8 & w^7 & w^6 & w^5 & w^4 & w^3 & w^2 & w
    \end{array}
  \right]
\end{equation}
where $w = \exp(\Ii \cdot 2\pi/11)$, so $w^{11} = 1$.


\subsubsection{'Cyclic $11$--roots' matrices}

There are precisely two $\RELofEQUI$ equivalence classes of complex Hadamard
matrices, inequivalent to
$F_{11}$, associated with the so--called nonclassical 'index 2' cyclic $11$--roots.
This
result is drawn in \cite{Ha96} p.319.

The classes are represented by the matrices $\AUFfamily[A]{\tilde{C}}{11}{0},\
\AUFfamily[B]{\tilde{C}}{11}{0}$ below, their respective \DEPHASED\ forms are
denoted by $\AUFfamily[A]{C}{11}{0},\ \AUFfamily[B]{C}{11}{0}$. Both matrices
have the
circulant structure  $\ELEMENTof{U}{i}{j} = x_{(i-j \mod 11)+1}$, where
\begin{eqnarray}
  x = [ 1,\  1,\  e,\ 1,\ 1,\ 1,\ e,\ e,\ e,\ 1,\ e ] & \mbox{for} &
\AUFfamily[A]{\tilde{C}}{11}{0} \\
  x = [  1,\  1,\  \CONJ{e},\ 1,\ 1,\ 1,\ \CONJ{e},\ \CONJ{e},\
  \CONJ{e},\ 1,\ \CONJ{e} ] & \mbox{for} & \AUFfamily[B]{\tilde{C}}{11}{0} 
\end{eqnarray}
and
\begin{equation}
  e = -\frac{5}{6}  +  \Ii \cdot \frac{\sqrt{11}}{6}
\end{equation}.

There holds
\begin{equation}
  \AUFfamily[B]{\tilde{C}}{11}{0}  = \CONJ{\AUFfamily[A]{\tilde{C}}{11}{0}}
  \ \ \ \ \mbox{so}\ \ \ \ 
  \AUFfamily[B]{C}{11}{0} =  \CONJ{\AUFfamily[A]{C}{11}{0}}
\end{equation}

Applying transposition to $\AUFfamily[A]{\tilde{C}}{11}{0},\
\AUFfamily[B]{\tilde{C}}{11}{0}$ -- equivalently to $\AUFfamily[A]{C}{11}{0},\
\AUFfamily[B]{C}{11}{0}$ -- yields a matrix equivalent to the
original one (see (\ref{eq_circulant_equivalence})).

The matrices are given by:

\begin{equation}
  \AUFfamily[A]{\tilde{C}}{11}{0} =
  \left[
    \begin{array}{ccccccccccc}
      1 & e & 1 & e & e & e & 1 & 1 & 1 & e & 1\\
      1 & 1 & e & 1 & e & e & e & 1 & 1 & 1 & e\\
      e & 1 & 1 & e & 1 & e & e & e & 1 & 1 & 1\\
      1 & e & 1 & 1 & e & 1 & e & e & e & 1 & 1\\
      1 & 1 & e & 1 & 1 & e & 1 & e & e & e & 1\\
      1 & 1 & 1 & e & 1 & 1 & e & 1 & e & e & e\\
      e & 1 & 1 & 1 & e & 1 & 1 & e & 1 & e & e\\
      e & e & 1 & 1 & 1 & e & 1 & 1 & e & 1 & e\\
      e & e & e & 1 & 1 & 1 & e & 1 & 1 & e & 1\\
      1 & e & e & e & 1 & 1 & 1 & e & 1 & 1 & e\\
      e & 1 & e & e & e & 1 & 1 & 1 & e & 1 & 1
    \end{array}
  \right] \ ,
\end{equation}

\begin{equation}
  \AUFfamily[B]{\tilde{C}}{11}{0} =
  \left[
    \begin{array}{ccccccccccc}
      1 & e^{\SSS -1} & 1 & e^{\SSS -1} & e^{\SSS -1} & e^{\SSS -1} & 1 & 1 & 1
& e^{\SSS -1} & 1\\
      1 & 1 & e^{\SSS -1} & 1 & e^{\SSS -1} & e^{\SSS -1} & e^{\SSS -1} & 1 & 1
& 1 & e^{\SSS -1}\\
      e^{\SSS -1} & 1 & 1 & e^{\SSS -1} & 1 & e^{\SSS -1} & e^{\SSS -1} &
e^{\SSS -1} & 1 & 1 & 1\\
      1 & e^{\SSS -1} & 1 & 1 & e^{\SSS -1} & 1 & e^{\SSS -1} & e^{\SSS -1} &
e^{\SSS -1} & 1 & 1\\
      1 & 1 & e^{\SSS -1} & 1 & 1 & e^{\SSS -1} & 1 & e^{\SSS -1} & e^{\SSS -1}
& e^{\SSS -1} & 1\\
      1 & 1 & 1 & e^{\SSS -1} & 1 & 1 & e^{\SSS -1} & 1 & e^{\SSS -1} & e^{\SSS
-1} & e^{\SSS -1}\\
      e^{\SSS -1} & 1 & 1 & 1 & e^{\SSS -1} & 1 & 1 & e^{\SSS -1} & 1 & e^{\SSS
-1} & e^{\SSS -1}\\
      e^{\SSS -1} & e^{\SSS -1} & 1 & 1 & 1 & e^{\SSS -1} & 1 & 1 & e^{\SSS -1}
& 1 & e^{\SSS -1}\\
      e^{\SSS -1} & e^{\SSS -1} & e^{\SSS -1} & 1 & 1 & 1 & e^{\SSS -1} & 1 & 1
& e^{\SSS -1} & 1\\
      1 & e^{\SSS -1} & e^{\SSS -1} & e^{\SSS -1} & 1 & 1 & 1 & e^{\SSS -1} & 1
& 1 & e^{\SSS -1}\\
      e^{\SSS -1} & 1 & e^{\SSS -1} & e^{\SSS -1} & e^{\SSS -1} & 1 & 1 & 1
      & e^{\SSS -1} & 1 & 1
    \end{array}
  \right]
\end{equation}

\begin{equation}
  \AUFfamily[A]{C}{11}{0} = 
  \left[
    \begin{array}{ccccccccccc}
      1 & 1 & 1 & 1 & 1 & 1 & 1 & 1 & 1 & 1 & 1\\
      1 & e^{\SSS -1} & e & e^{\SSS -1} & 1 & 1 & e & 1 & 1 & e^{\SSS -1} & e\\
      1 & e^{\SSS -2} & e^{\SSS -1} & e^{\SSS -1} & e^{\SSS -2} & e^{\SSS -1} &
1 & 1 & e^{\SSS -1} & e^{\SSS -2} & e^{\SSS -1}\\
      1 & 1 & 1 & e^{\SSS -1} & 1 & e^{\SSS -1} & e & e & e & e^{\SSS -1} & 1\\
      1 & e^{\SSS -1} & e & e^{\SSS -1} & e^{\SSS -1} & 1 & 1 & e & e & 1 & 1\\
      1 & e^{\SSS -1} & 1 & 1 & e^{\SSS -1} & e^{\SSS -1} & e & 1 & e & 1 & e\\
      1 & e^{\SSS -2} & e^{\SSS -1} & e^{\SSS -2} & e^{\SSS -1} & e^{\SSS -2} &
e^{\SSS -1} & 1 & e^{\SSS -1} & e^{\SSS -1} & 1\\
      1 & e^{\SSS -1} & e^{\SSS -1} & e^{\SSS -2} & e^{\SSS -2} & e^{\SSS -1} &
e^{\SSS -1} & e^{\SSS -1} & 1 & e^{\SSS -2} & 1\\
      1 & e^{\SSS -1} & 1 & e^{\SSS -2} & e^{\SSS -2} & e^{\SSS -2} & 1 &
e^{\SSS -1} & e^{\SSS -1} & e^{\SSS -1} & e^{\SSS -1}\\
      1 & 1 & e & 1 & e^{\SSS -1} & e^{\SSS -1} & 1 & e & 1 & e^{\SSS -1} & e\\
      1 & e^{\SSS -2} & 1 & e^{\SSS -1} & e^{\SSS -1} & e^{\SSS -2} &
      e^{\SSS -1} & e^{\SSS -1} & 1 & e^{\SSS -2} & e^{\SSS -1}
    \end{array}
  \right]
\end{equation}

\begin{equation}
  \AUFfamily[B]{C}{11}{0} =
  \left[
    \begin{array}{ccccccccccc}
      1 & 1 & 1 & 1 & 1 & 1 & 1 & 1 & 1 & 1 & 1\\
      1 & e & e^{\SSS -1} & e & 1 & 1 & e^{\SSS -1} & 1 & 1 & e & e^{\SSS -1}\\
      1 & e^2 & e & e & e^2 & e & 1 & 1 & e & e^2 & e\\
      1 & 1 & 1 & e & 1 & e & e^{\SSS -1} & e^{\SSS -1} & e^{\SSS -1} & e & 1\\
      1 & e & e^{\SSS -1} & e & e & 1 & 1 & e^{\SSS -1} & e^{\SSS -1} & 1 & 1\\
      1 & e & 1 & 1 & e & e & e^{\SSS -1} & 1 & e^{\SSS -1} & 1 & e^{\SSS -1}\\
      1 & e^2 & e & e^2 & e & e^2 & e & 1 & e & e & 1\\
      1 & e & e & e^2 & e^2 & e & e & e & 1 & e^2 & 1\\
      1 & e & 1 & e^2 & e^2 & e^2 & 1 & e & e & e & e\\
      1 & 1 & e^{\SSS -1} & 1 & e & e & 1 & e^{\SSS -1} & 1 & e & e^{\SSS -1}\\
      1 & e^2 & 1 & e & e & e^2 & e & e & 1 & e^2 & e
    \end{array}
  \right]
\end{equation}


\subsubsection{Nicoara's $11 \times 11$ complex Hadamard matrix}

Another equivalence class of $11 \times 11$ Hadamard matrices is
represented by the matrix communicated to the authors by
Nicoara:
\begin{equation}
  \AUFfamily{N}{11}{0} =
  \left[
    \begin{array}{ccccccccccc}
      1   &   1   &   1   &   1   &   1   &   1   &   1   &   1   &   1   &   1  
&   1 \\
      1  &  a  &  -a  &  -a  &  -a  &  -1  &  -1  &  -a^{\SSS -1}  &  -a^{\SSS
-1}  &  -1  &  -1 \\
      1  &  -a  &  a  &  -a  &  -a  &  -1  &  -1  &  -1  &  -1  &  -a^{\SSS -1} 
&  -a^{\SSS -1} \\
      1  &  -a  &  -a  &  -1  &  a  &  -1  &  -a  &  -a^{\SSS -1}  &  -1  & 
-a^{\SSS -1}  &  -1 \\
      1  &  -a  &  -a  &  a  &  -1  &  -a  &  -1  &  -1  &  -a^{\SSS -1}  &  -1 
&  -a^{\SSS -1} \\
      1  &  -1  &  -1  &  -1  &  -a  &  -a  &  1  &  -a^{\SSS -1}  &  -1  &  -1 
&  -a^{\SSS -1} \\
      1  &  -1  &  -1  &  -a  &  -1  &  1  &  -a  &  -1  &  -a^{\SSS -1}  & 
-a^{\SSS -1}  &  -1 \\
      1  &  -a^{\SSS -1}  &  -1  &  -a^{\SSS -1}  &  -1  &  -a^{\SSS -1}  &  -1 
&  -a^{\SSS -1}  &  a^{\SSS -1}  &  -a^{\SSS -2}  &  -a^{\SSS -2} \\
      1  &  -a^{\SSS -1}  &  -1  &  -1  &  -a^{\SSS -1}  &  -1  &  -a^{\SSS -1} 
&  a^{\SSS -1}  &  -a^{\SSS -1}  &  -a^{\SSS -2}  &  -a^{\SSS -2} \\
      1  &  -1  &  -a^{\SSS -1}  &  -a^{\SSS -1}  &  -1  &  -1  &  -a^{\SSS -1} 
&  -a^{\SSS -2}  &  -a^{\SSS -2}  &  -a^{\SSS -2}  &  a^{\SSS -2} \\
      1  &  -1  &  -a^{\SSS -1}  &  -1  &  -a^{\SSS -1}  &  -a^{\SSS -1}  &  -1 
&  -a^{\SSS -2}  &  -a^{\SSS -2}  &  a^{\SSS -2}  &  -a^{\SSS -2} 
    \end{array}
  \right]
\end{equation}
where
\begin{equation}
  a= -\frac{3}{4} - \Ii \cdot \frac{\sqrt{7}}{4}
\end{equation}

The above matrix is isolated\footnote{R.Nicoara, private communication}
since its defect $d(N_{11}^{(0)})=0$.
\ENDofSUBSECTION

%
%

\subsection {$N=12$}
\label{subsec__N_eq_12}


\subsubsection{Orbits stemming from $F_{12}$}

The only \maxAUFs\ stemming from $F_{12}$ are:
\begin{eqnarray}
  \AUFfamily[A]{F}{12}{9}(a,b,c,d,e,f,g,h,i)  &  = & F_{12} \HADprod
\EXPentrywise
  \left(\Ii \cdot \AUFphases[A]{F}{12}{9}(a,\ldots,i)\right)
  \label{F10_maxAUF_D1A2}  \\
  \AUFfamily[B]{F}{12}{9}(a,b,c,d,e,f,g,h,i)  &  = & F_{12} \HADprod
\EXPentrywise
  \left(\Ii \cdot \AUFphases[B]{F}{12}{9}(a,\ldots,i)\right)
  \label{F10_maxAUF_D1D2}  \\
  \AUFfamily[C]{F}{12}{9}(a,b,c,d,e,f,g,h,i)  &  = & F_{12} \HADprod
\EXPentrywise
  \left(\Ii \cdot \AUFphases[C]{F}{12}{9}(a,\ldots,i)\right)
  \label{F10_maxAUF_B1A2}  \\
  \AUFfamily[D]{F}{12}{9}(a,b,c,d,e,f,g,h,i)  &  = & F_{12} \HADprod
\EXPentrywise
  \left(\Ii \cdot \AUFphases[D]{F}{12}{9}(a,\ldots,i)\right)
  \label{F10_maxAUF_A1A2}  \\
  \TRANSPOSE{\AUFfamily[B]{F}{12}{9}(a,b,c,d,e,f,g,h,i)}  &  = & F_{12}
\HADprod \EXPentrywise
  \left(\Ii \cdot \TRANSPOSE{\AUFphases[B]{F}{12}{9}(a,\ldots,i)} \right)
  \label{F10_maxAUF_C1A2}  \nonumber \\
  \TRANSPOSE{\AUFfamily[C]{F}{12}{9}(a,b,c,d,e,f,g,h,i)}  &  = & F_{12}
\HADprod \EXPentrywise
  \left(\Ii \cdot \TRANSPOSE{\AUFphases[C]{F}{12}{9}(a,\ldots,i)} \right)
  \label{F10_maxAUF_D1C2}  \nonumber \\
  \TRANSPOSE{\AUFfamily[D]{F}{12}{9}(a,b,c,d,e,f,g,h,i)}  &  = & F_{12}
\HADprod \EXPentrywise
  \left(\Ii \cdot \TRANSPOSE{\AUFphases[D]{F}{12}{9}(a,\ldots,i)} \right)
  \label{F10_maxAUF_D1B2}  \nonumber
\end{eqnarray}
where
\begin{equation}
  F_{12} =
  \left[
    \begin{array}{cccccccccccc}
      1 & 1 & 1 & 1 & 1 & 1 & 1 & 1 & 1 & 1 & 1 & 1\\
      1 & w & w^2 & w^3 & w^4 & w^5 & w^6 & w^7 & w^8 & w^9 & w^{10} & w^{11}\\
      1 & w^2 & w^4 & w^6 & w^8 & w^{10} & 1 & w^2 & w^4 & w^6 & w^8 & w^{10}\\
      1 & w^3 & w^6 & w^9 & 1 & w^3 & w^6 & w^9 & 1 & w^3 & w^6 & w^9\\
      1 & w^4 & w^8 & 1 & w^4 & w^8 & 1 & w^4 & w^8 & 1 & w^4 & w^8\\
      1 & w^5 & w^{10} & w^3 & w^8 & w & w^6 & w^{11} & w^4 & w^9 & w^2 & w^7\\
      1 & w^6 & 1 & w^6 & 1 & w^6 & 1 & w^6 & 1 & w^6 & 1 & w^6\\
      1 & w^7 & w^2 & w^9 & w^4 & w^{11} & w^6 & w & w^8 & w^3 & w^{10} & w^5\\
      1 & w^8 & w^4 & 1 & w^8 & w^4 & 1 & w^8 & w^4 & 1 & w^8 & w^4\\
      1 & w^9 & w^6 & w^3 & 1 & w^9 & w^6 & w^3 & 1 & w^9 & w^6 & w^3\\
      1 & w^{10} & w^8 & w^6 & w^4 & w^2 & 1 & w^{10} & w^8 & w^6 & w^4 & w^2\\
      1 & w^{11} & w^{10} & w^9 & w^8 & w^7 & w^6 & w^5 & w^4 & w^3 & w^2 & w
    \end{array}
  \right]
\end{equation}
where $w = \exp(\Ii \cdot 2\pi/12)$, so $w^{12}=1,\ w^6 = -1,\ w^3 =
\Ii$, and
\begin{equation} 
   \AUFphases[A]{F}{12}{9}(a,\ldots,i) = 
   \left[
     \begin{array}{cccccc|cccccc}
       \bO & \bO & \bO & \bO & \bO & \bO & \bO & \bO & \bO & \bO & \bO & \bO\\
       \bO & a & b & c & d & e & \bO & a & b & c & d & e\\
       \bO & f & \bO & f & \bO & f & \bO & f & \bO & f & \bO & f\\
       \bO & g & b & c - a + g & d & e - a + g & \bO & g & b & c - a + g & d &
e - a + g\\
       \bO & h & \bO & h & \bO & h & \bO & h & \bO & h & \bO & h\\
       \bO & i & b & c - a + i & d & e - a + i & \bO & i & b & c - a + i &
       d & e - a + i\\
       \hline
       \bO & \bO & \bO & \bO & \bO & \bO & \bO & \bO & \bO & \bO & \bO & \bO\\
       \bO & a & b & c & d & e & \bO & a & b & c & d & e\\
       \bO & f & \bO & f & \bO & f & \bO & f & \bO & f & \bO & f\\
       \bO & g & b & c - a + g & d & e - a + g & \bO & g & b & c - a + g & d &
e - a + g\\
       \bO & h & \bO & h & \bO & h & \bO & h & \bO & h & \bO & h\\
       \bO & i & b & c - a + i & d & e - a + i & \bO & i & b & c - a + i & d &
e - a + i
     \end{array}
   \right]
\end{equation}

\begin{equation} 
  \AUFphases[B]{F}{12}{9}(a,\ldots,i) =
  \left[
    \begin{array}{cccccc|cccccc}
      \bO & \bO & \bO & \bO & \bO & \bO & \bO & \bO & \bO & \bO & \bO & \bO\\
      \bO & a & b & c & d & e & \bO & a & b & c & d & e\\
      \bO & f & g & \bO & f & g & \bO & f & g & \bO & f & g\\
      \bO & h & i & c & d - a + h & e - b + i & \bO & h & i & c & d - a +
      h & e - b + i\\
      \hline
      \bO & \bO & \bO & \bO & \bO & \bO & \bO & \bO & \bO & \bO & \bO & \bO\\
      \bO & a & b & c & d & e & \bO & a & b & c & d & e\\
      \bO & f & g & \bO & f & g & \bO & f & g & \bO & f & g\\
      \bO & h & i & c & d - a + h & e - b + i & \bO & h & i & c & d - a +
      h & e - b + i\\
      \hline
      \bO & \bO & \bO & \bO & \bO & \bO & \bO & \bO & \bO & \bO & \bO & \bO\\
      \bO & a & b & c & d & e & \bO & a & b & c & d & e\\
      \bO & f & g & \bO & f & g & \bO & f & g & \bO & f & g\\
      \bO & h & i & c & d - a + h & e - b + i & \bO & h & i & c & d - a + h & e
- b + i
    \end{array}
  \right]
\end{equation}

\begin{equation} 
  \AUFphases[C]{F}{12}{9}(a,\ldots,i) =
  \left[
    \begin{array}{cccccccccccc}
      \bO & \bO & \bO & \bO & \bO & \bO & \bO & \bO & \bO & \bO & \bO & \bO\\
      \bO & a & b & c & \bO & d & b & a & \bO & c & b & d\\
      \bO & e & f & e & \bO & e & f & e & \bO & e & f & e\\
      \bO & g & \bO & c - a + g & \bO & d - a + g & \bO & g & \bO & c - a + g &
\bO & d - a + g\\
      \bO & h & b & h & \bO & h & b & h & \bO & h & b & h\\
      \bO & i & f & c - a + i & \bO & d - a + i & f & i & \bO & c - a + i &
      f & d - a + i\\
      \hline
      \bO & \bO & \bO & \bO & \bO & \bO & \bO & \bO & \bO & \bO & \bO & \bO\\
      \bO & a & b & c & \bO & d & b & a & \bO & c & b & d\\
      \bO & e & f & e & \bO & e & f & e & \bO & e & f & e\\
      \bO & g & \bO & c - a + g & \bO & d - a + g & \bO & g & \bO & c - a + g &
\bO & d - a + g\\
      \bO & h & b & h & \bO & h & b & h & \bO & h & b & h\\
      \bO & i & f & c - a + i & \bO & d - a + i & f & i & \bO & c - a + i & f &
d - a + i
    \end{array}
  \right]
\end{equation}

\begin{equation} 
  \AUFphases[D]{F}{12}{9}(a,\ldots,i) =
  \left[
    \begin{array}{cccccccccccc}
      \bO & \bO & \bO & \bO & \bO & \bO & \bO & \bO & \bO & \bO & \bO & \bO\\
      \bO & a & b & c & d & a & \bO & c & b & a & d & c\\
      \bO & e & \bO & f & \bO & e & \bO & f & \bO & e & \bO & f\\
      \bO & g & b & g & d & g & \bO & g & b & g & d & g\\
      \bO & h & \bO & c - a + h & \bO & h & \bO & c - a + h & \bO & h & \bO & c
- a + h\\
      \bO & i & b & f - e + i & d & i & \bO & f - e + i & b & i & d & f -
      e + i\\
      \hline
      \bO & \bO & \bO & \bO & \bO & \bO & \bO & \bO & \bO & \bO & \bO & \bO\\
      \bO & a & b & c & d & a & \bO & c & b & a & d & c\\
      \bO & e & \bO & f & \bO & e & \bO & f & \bO & e & \bO & f\\
      \bO & g & b & g & d & g & \bO & g & b & g & d & g\\
      \bO & h & \bO & c - a + h & \bO & h & \bO & c - a + h & \bO & h & \bO & c
- a + h\\
      \bO & i & b & f - e + i & d & i & \bO & f - e + i & b & i & d & f - e + i
    \end{array}
  \right]
\end{equation}

Thus we have three pairs of \COGNATE\ families, and the
$\AUFfamily[A]{F}{12}{9}$ family is \selfCOGNATE.
\bigskip

At least one of the above orbits is \PERMUTATIONequivalent\ to the
orbit constructed using  the Di{\c t}{\v a}'s method:
\begin{eqnarray}
\label{F12fam_Dita}
  \lefteqn{\AUFfamily{\tilde{F}}{12}{9}(\alpha_1,\ldots,\alpha_9) =}\\
  & & 
  \left[ 
    \begin{array}{c|c|c}
      \ELEMENTof{F_3}{1}{1} \cdot \AUFfamily{F}{4}{1}(\alpha_1) &
\ELEMENTof{F_3}{1}{2} \cdot
      D(\alpha_4,\alpha_5,\alpha_6) \cdot \AUFfamily{F}{4}{1}(\alpha_2) &
\ELEMENTof{F_3}{1}{3} \cdot
      D(\alpha_7,\alpha_8,\alpha_9) \cdot \AUFfamily{F}{4}{1}(\alpha_3) \\
      \hline
      \ELEMENTof{F_3}{2}{1} \cdot \AUFfamily{F}{4}{1}(\alpha_1) &
\ELEMENTof{F_3}{2}{2} \cdot
      D(\alpha_4,\alpha_5,\alpha_6) \cdot \AUFfamily{F}{4}{1}(\alpha_2) &
\ELEMENTof{F_3}{2}{3} \cdot
      D(\alpha_7,\alpha_8,\alpha_9) \cdot \AUFfamily{F}{4}{1}(\alpha_3) \\
      \hline
      \ELEMENTof{F_3}{3}{1} \cdot \AUFfamily{F}{4}{1}(\alpha_1) &
\ELEMENTof{F_3}{3}{2} \cdot
      D(\alpha_4,\alpha_5,\alpha_6) \cdot \AUFfamily{F}{4}{1}(\alpha_2) &
\ELEMENTof{F_3}{3}{3} \cdot
      D(\alpha_7,\alpha_8,\alpha_9) \cdot \AUFfamily{F}{4}{1}(\alpha_3)
    \end{array} 
  \right] \nonumber
\end{eqnarray}
where $D(\alpha,\beta,\gamma)$ denotes the $4 \times 4$ diagonal
matrix $\diag(1,\PHASE{\alpha},\PHASE{\beta},\PHASE{\gamma})$ and
$\AUFfamily{F}{4}{1}(\alpha)$ is given by (\ref{F4_maxAUF}).
\medskip

It passes through $\AUFfamily{\tilde{F}}{12}{9}(\ZEROvect)
= F_3 \otimes F_4$, which is, according to \cite{Ta05}, permutation
equivalent to $F_{12}$, so $\AUFfamily{\tilde{F}}{12}{9}$
is a \maxAUF\ stemming from a permuted $F_{12}$.

It also passes through
$\AUFfamily{\tilde{F}}{12}{9}(\pi/2,\pi/2,\pi/2,\ZEROvect)$, which is
\DEPHASED,
so it is a permuted $F_3 \otimes F_2 \otimes F_2$. 
Thus permuted and
inequivalent $F_{12}$ and $F_3 \otimes F_2 \otimes F_2$ (see
\cite{Ta05}) are connected by
the orbit of (\ref{F12fam_Dita}).
\medskip

Note also that a similar construction using the Di{\c t}{\v a}'s method
with the role of $F_3$ and $\AUFfamily{F}{4}{1}$ exchanged yields a
$7$-dimensional orbit, also passing through a permuted $F_{12}$: $F_4
\otimes F_3$, which is a suborbit of one of the existing
$9$-dimensional \maxAUFs\ stemming from $F_4
\otimes F_3$.


\subsubsection{Other $12 \times 12$ orbits}

Other \DEPHASED\ $12 \times 12$ orbits can be obtained, using the Di{\c t}{\v a}'s
method, from
$F_2$ and \DEPHASED\ $6 \times 6$ complex Hadamard matrices
from section \ref{subsec__N_eq_6}, for example:
\begin{description}
  \item[]
        \begin{equation}
          \AUFfamily{FD}{12}{8}(\alpha_1,\ldots,\alpha_8)\ \ \  =\ \ \   
          \DITAtwoBYtwo{F_2}
          {\AUFfamily{F}{6}{2}(\alpha_1,\alpha_2)}
          {D(\alpha_4,\ldots,\alpha_8) \cdot
            \AUFfamily{D}{6}{1}(\alpha_3)}
        \end{equation}

  \item
        \begin{equation}
          \AUFfamily{FC}{12}{7}(\alpha_1,\ldots,\alpha_7)\ \ \ =\ \ \  
          \DITAtwoBYtwo{F_2}
          {\AUFfamily{F}{6}{2}(\alpha_1,\alpha_2)}
          {D(\alpha_3,\ldots,\alpha_7) \cdot
            \AUFfamily{C}{6}{0}}
        \end{equation}

  \item
      \begin{equation}
        \AUFfamily{FS}{12}{7}(\alpha_1,\ldots,\alpha_7)\ \ \ =\ \ \    
        \DITAtwoBYtwo{F_2}
        {\AUFfamily{F}{6}{2}(\alpha_1,\alpha_2)}
        {D(\alpha_3,\ldots,\alpha_7) \cdot
          \AUFfamily{S}{6}{0}}
      \end{equation}

  \item
        \begin{equation}
          \AUFfamily{DD}{12}{7}(\alpha_1,\ldots,\alpha_7)\ \ \ =\ \ \ 
          \DITAtwoBYtwo{F_2}
          {\AUFfamily{D}{6}{1}(\alpha_1)}
          {D(\alpha_3,\ldots,\alpha_7) \cdot
            \AUFfamily{D}{6}{1}(\alpha_2)}
        \end{equation}
        
  \item
        \begin{equation}
          \AUFfamily{DC}{12}{6}(\alpha_1,\ldots,\alpha_6)\ \ \ =\ \ \   
          \DITAtwoBYtwo{F_2}
          {\AUFfamily{D}{6}{1}(\alpha_1)}
          {D(\alpha_2,\ldots,\alpha_6) \cdot
            \AUFfamily{C}{6}{0}}
        \end{equation}
        
  \item
        \begin{equation}
          \AUFfamily{DS}{12}{6}(\alpha_1,\ldots,\alpha_6)\ \ \ =\ \ \   
          \DITAtwoBYtwo{F_2}
          {\AUFfamily{D}{6}{1}(\alpha_1)}
          {D(\alpha_2,\ldots,\alpha_6) \cdot
            \AUFfamily{S}{6}{0}}
        \end{equation}

  \item
        \begin{equation}
          \AUFfamily{CC}{12}{5}(\alpha_1,\ldots,\alpha_5)\ \ \ =\ \ \ 
          \DITAtwoBYtwo{F_2}
          {\AUFfamily{C}{6}{0}}
          {D(\alpha_1,\ldots,\alpha_5) \cdot
            \AUFfamily{C}{6}{0}}
        \end{equation}

  \item
        \begin{equation}
          \AUFfamily{CS}{12}{5}(\alpha_1,\ldots,\alpha_5)\ \ \ =\ \ \   
          \DITAtwoBYtwo{F_2}
          {\AUFfamily{C}{6}{0}}
          {D(\alpha_1,\ldots,\alpha_5) \cdot
            \AUFfamily{S}{6}{0}}
        \end{equation}

  \item
        \begin{equation}
          \AUFfamily{SS}{12}{5}(\alpha_1,\ldots,\alpha_5)\ \ \ =\ \ \ 
          \DITAtwoBYtwo{F_2}
          {\AUFfamily{S}{6}{0}}
          {D(\alpha_1,\ldots,\alpha_5) \cdot
            \AUFfamily{S}{6}{0}}
        \end{equation}
\end{description}
where $D(\beta_1,\ldots,\beta_5)$ denotes the $6 \times 6$ diagonal
matrix $\diag(1,\PHASE{\beta_1},\ldots,\PHASE{\beta_5})$.
\ENDofSUBSECTION

%
%
\subsection {$N=13$}
\label{subsec__N_eq_13}


\subsubsection{Orbits stemming from $F_{13}$}

The Fourier matrix 
$F_{13}$ is an \ISOLATED\ $13 \times 13$ complex Hadamard matrix:
\begin{equation}
  F_{13} = \AUFfamily{F}{13}{0} = 
  \left[
    \begin{array}{ccccccccccccc}
      1 & 1 & 1 & 1 & 1 & 1 & 1 & 1 & 1 & 1 & 1 & 1 & 1\\
      1 & w & w^2 & w^3 & w^4 & w^5 & w^6 & w^7 & w^8 & w^9 & w^{10} & w^{11} &
w^{12}\\
      1 & w^2 & w^4 & w^6 & w^8 & w^{10} & w^{12} & w & w^3 & w^5 & w^7 & w^9 &
w^{11}\\
      1 & w^3 & w^6 & w^9 & w^{12} & w^2 & w^5 & w^8 & w^{11} & w & w^4 & w^7 &
w^{10}\\
      1 & w^4 & w^8 & w^{12} & w^3 & w^7 & w^{11} & w^2 & w^6 & w^{10} & w &
w^5 & w^9\\
      1 & w^5 & w^{10} & w^2 & w^7 & w^{12} & w^4 & w^9 & w & w^6 & w^{11} &
w^3 & w^8\\
      1 & w^6 & w^{12} & w^5 & w^{11} & w^4 & w^{10} & w^3 & w^9 & w^2 & w^8 &
w & w^7\\
      1 & w^7 & w & w^8 & w^2 & w^9 & w^3 & w^{10} & w^4 & w^{11} & w^5 &
w^{12} & w^6\\
      1 & w^8 & w^3 & w^{11} & w^6 & w & w^9 & w^4 & w^{12} & w^7 & w^2 &
w^{10} & w^5\\
      1 & w^9 & w^5 & w & w^{10} & w^6 & w^2 & w^{11} & w^7 & w^3 & w^{12} &
w^8 & w^4\\
      1 & w^{10} & w^7 & w^4 & w & w^{11} & w^8 & w^5 & w^2 & w^{12} & w^9 &
w^6 & w^3\\
      1 & w^{11} & w^9 & w^7 & w^5 & w^3 & w & w^{12} & w^{10} & w^8 & w^6 &
w^4 & w^2\\
      1 & w^{12} & w^{11} & w^{10} & w^9 & w^8 & w^7 & w^6 & w^5 & w^4 & w^3
      & w^2 & w
    \end{array}
  \right]  
\end{equation}
where $w = \exp(\Ii \cdot 2\pi/13)$, so $w^{13} = 1$.

\subsubsection{Petrescu $13 \times 13$ orbit}
There exists a continuous $2$-parameter orbit 
$\AUFfamily{P}{13}{2}$
of $13 \times 13$ complex Hadamard matrices
 found by Petrescu \cite{Pe97}.
\begin{equation}
  \AUFfamily{P}{13}{2}(e,f)   =  P_{13} \HADprod \EXPentrywise
  \left(\Ii \cdot \AUFphases{P}{13}{2}(e,f)\right)
\end{equation}
where
\begin{equation}
 P_{13} = 
  \left[
    \begin{array}{c|cccc|cccc|cccc}
      1 & 1 &	 1 & 1 & 1 & 1 & 1 & 1 & 1 & 1 & 1 & 1 & 1\\
      \hline
      1 & -1 & t^{10}	 & - t^5 & t^5 & \Ii t^5 & -\Ii t^5 & \Ii t^{15} & -\Ii
t^{15} & t^{16} & t^4 & t^{22} & t^{28}\\
      1 & t^{10} & -1 & t^5 & - t^5 & -\Ii t^5 & \Ii t^5 & -\Ii t^{15} & \Ii
t^{15} & t^{16} & t^4 & t^{22} & t^{28}\\
      1 & - t^5 & t^5 & -1 & t^{10} & \Ii t^{15} & -\Ii t^{15} & \Ii t^5 & -\Ii
t^5 & t^4 & t^{16} & t^{28} & t^{22}\\
      1 & t^5 & - t^5 & t^{10} & -1 & -\Ii t^{15} & \Ii t^{15} & -\Ii
      t^5 & \Ii t^5 & t^4 & t^{16} & t^{28} & t^{22}\\
      \hline
      1 & \Ii t^5 & -\Ii t^5 & \Ii t^{25} & -\Ii t^{25} & -1 & t^{10} & - t^5 &
t^5 & t^{22} & t^{28} & t^4 & t^{16}\\
      1 & -\Ii t^5 & \Ii t^5 & -\Ii t^{25} & \Ii t^{25} & t^{10} & -1 & t^5 & -
t^5 & t^{22} & t^{28} & t^4 & t^{16}\\
      1 & \Ii t^{25} & -\Ii t^{25} & \Ii t^5 & -\Ii t^5 & - t^5 & t^5 & -1 &
t^{10} & t^{28} & t^{22} & t^{16} & t^4\\
      1 & -\Ii t^{25} & \Ii t^{25} & -\Ii t^5 & \Ii t^5 & t^5 & - t^5
      & t^{10} & -1 & t^{28} & t^{22} & t^{16} & t^4\\
      \hline
      1 & t^4 & t^4 & t^{16} & t^{16} & t^{28} & t^{28} & t^{22} & t^{22} &
t^{20} & t^{10} & t^{10} & t^{10}\\
      1 & t^{16} & t^{16} & t^4 & t^4 & t^{22} & t^{22} & t^{28} & t^{28} &
t^{10} & t^{20} & t^{10} & t^{10}\\
      1 & t^{28} & t^{28} & t^{22} & t^{22} & t^{16} & t^{16} & t^4 & t^4 &
t^{10} & t^{10} & t^{20} & t^{10}\\
      1 & t^{22} & t^{22} & t^{28} & t^{28} & t^4 & t^4 & t^{16} & t^{16} &
      t^{10} & t^{10} & t^{10} & t^{20}
    \end{array}
  \right]
\end{equation}
where $t = \exp( \Ii \cdot 2\pi/{30} )$, so $t^{30} = 1,\ t^{15} =
-1$, and
{ 
\newcommand{\manySPACES}
{\ \ \ \ \ \ \ \ \ \ \ \ \ \ \ \ \ \ \ \ \ \ \ \ \ \ \ \ \ \ \ \ \ \ \
\ \ \ \ \ \ \ \ \ \ \ \ \ \ \ \ \ \ }
\begin{eqnarray}
  \AUFphases{P}{13}{2}(e,f) = \manySPACES & & \\
   {\tiny 
     \left[
       \begin{array}{c|cccc|cccc|cccc}
         \bO & \bO & \bO & \bO & \bO & \bO & \bO & \bO & \bO & \bO &
         \bO & \bO & \bO\\
         \hline
         \bO & \bO & \bO & f & f & e & e & e + G\left(f\right) & e +
G\left(f\right) & \bO & \bO & \bO & \bO\\
         \bO & \bO & \bO & f & f & e & e & e + G\left(f\right) & e +
G\left(f\right) & \bO & \bO & \bO & \bO\\
         \bO & f & f & \bO & \bO & e + G\left(f\right) & e + G\left(f\right) &
e & e & \bO & \bO & \bO & \bO\\
         \bO & f & f & \bO & \bO & e + G\left(f\right) & e +
         G\left(f\right) & e & e & \bO & \bO & \bO & \bO\\
         \hline
         \bO & - e & - e &  - e - G\left(f\right) &  - e -
         G\left(f\right) & \bO & \bO & - f & - f & \bO & \bO & \bO &
         \bO\\ 
         \bO & - e & - e &  - e - G\left(f\right) &  - e -
         G\left(f\right) & \bO & \bO & - f & - f & \bO & \bO & \bO &
         \bO\\ 
         \bO &  - e - G\left(f\right) &  - e - G\left(f\right) & - e &
         - e & - f & - f & \bO & \bO & \bO & \bO & \bO & \bO\\ 
         \bO &  - e - G\left(f\right) &  - e - G\left(f\right) & - e &
         - e & - f & - f & \bO & \bO & \bO & \bO & \bO & \bO\\
         \hline
         \bO & \bO & \bO & \bO & \bO & \bO & \bO & \bO & \bO & \bO & \bO & \bO
& \bO\\
         \bO & \bO & \bO & \bO & \bO & \bO & \bO & \bO & \bO & \bO & \bO & \bO
& \bO\\
         \bO & \bO & \bO & \bO & \bO & \bO & \bO & \bO & \bO & \bO & \bO & \bO
& \bO\\
         \bO & \bO & \bO & \bO & \bO & \bO & \bO & \bO & \bO & \bO & \bO & \bO
& \bO
       \end{array}
     \right]
   } & & \nonumber
\end{eqnarray}
}  
where
\begin{equation}
  G\left(f\right) = \arg\left( -\frac{\cos(f)}{2}\ \ +\ \ \Ii \cdot
    \frac{\sqrt{2}}{4} \sqrt{7 - \cos(2f)} \right)\ \ \ -\ \ \ \frac{2\pi}{3}
\end{equation}

Since the function $G(f)$ is nonlinear,
the above family is not an \AUF, but it is not clear
  whether it could be contained in any
\AUF{} of a larger dimension.
\medskip

Due to some freedom in the construction of family components 
the method of Petrescu allows one to build other
similar families of Hadamard matrices. 
Not knowing whether they are inequivalent
we are not going to consider them here.

\subsubsection{'Cyclic $13$--roots' matrices}

There are precisely two $\RELofEQUI$ equivalence classes of complex Hadamard
matrices, inequivalent to
$F_{13}$, associated with the so--called 'index 2' cyclic $13$--roots. This
result is drawn in \cite{Ha96} p.319.

The classes are represented by the matrices $\AUFfamily[A]{\tilde{C}}{13}{0},\
\AUFfamily[B]{\tilde{C}}{13}{0}$ below, their respective \DEPHASED\ forms are
denoted by $\AUFfamily[A]{C}{13}{0},\ \AUFfamily[B]{C}{13}{0}$. Both matrices
have the
circulant structure  $\ELEMENTof{U}{i}{j} = x_{(i-j \mod 13)+1}$, where
\begin{eqnarray}
  x = [ 1,\  c,\  \CONJ{c},\ c,\ c,\ \CONJ{c},\ \CONJ{c},\ \CONJ{c},\
  \CONJ{c},\ c,\ c,\ \CONJ{c},\ c ] & \mbox{for} &
\AUFfamily[A]{\tilde{C}}{13}{0} \\
  x = [ 1,\  d,\  \CONJ{d},\ d,\ d,\ \CONJ{d},\ \CONJ{d},\ \CONJ{d},\
  \CONJ{d},\ d,\ d,\ \CONJ{d},\ d ] & \mbox{for} &
\AUFfamily[B]{\tilde{C}}{13}{0} 
\end{eqnarray}
and
\begin{eqnarray}
  c & = & \left( \frac{-1+\sqrt{13}}{12} \right)   +   \Ii \cdot
  \left( \frac{\sqrt{130 + 2\sqrt{13}}}{12} \right) \\
  d & = & \left( \frac{-1-\sqrt{13}}{12} \right)   +   \Ii \cdot
  \left( \frac{\sqrt{130 - 2\sqrt{13}}}{12} \right)
\end{eqnarray}.

Conjugating $\AUFfamily[k]{C}{13}{0},\ \AUFfamily[k]{\tilde{C}}{13}{0},\
k=A,B$ yields a matrix
equivalent to the original one \cite{Ha96}.
\bigskip

The matrices 
$\AUFfamily[A]{C}{13}{0}$, $\AUFfamily[A]{\tilde{C}}{13}{0}$ and 
$\AUFfamily[B]{C}{13}{0}$, $\AUFfamily[B]{\tilde{C}}{13}{0}$
are symmetric. They read

\begin{equation}
  \AUFfamily[A]{\tilde{C}}{13}{0} = 
  \left[
    \begin{array}{ccccccccccccc}
      1 & c & c^{\SSS -1} & c & c & c^{\SSS -1} & c^{\SSS -1} & c^{\SSS -1} &
c^{\SSS -1} & c & c & c^{\SSS -1} & c\\
      c & 1 & c & c^{\SSS -1} & c & c & c^{\SSS -1} & c^{\SSS -1} & c^{\SSS -1}
& c^{\SSS -1} & c & c & c^{\SSS -1}\\
      c^{\SSS -1} & c & 1 & c & c^{\SSS -1} & c & c & c^{\SSS -1} & c^{\SSS -1}
& c^{\SSS -1} & c^{\SSS -1} & c & c\\
      c & c^{\SSS -1} & c & 1 & c & c^{\SSS -1} & c & c & c^{\SSS -1} & c^{\SSS
-1} & c^{\SSS -1} & c^{\SSS -1} & c\\
      c & c & c^{\SSS -1} & c & 1 & c & c^{\SSS -1} & c & c & c^{\SSS -1} &
c^{\SSS -1} & c^{\SSS -1} & c^{\SSS -1}\\
      c^{\SSS -1} & c & c & c^{\SSS -1} & c & 1 & c & c^{\SSS -1} & c & c &
c^{\SSS -1} & c^{\SSS -1} & c^{\SSS -1}\\
      c^{\SSS -1} & c^{\SSS -1} & c & c & c^{\SSS -1} & c & 1 & c & c^{\SSS -1}
& c & c & c^{\SSS -1} & c^{\SSS -1}\\
      c^{\SSS -1} & c^{\SSS -1} & c^{\SSS -1} & c & c & c^{\SSS -1} & c & 1 & c
& c^{\SSS -1} & c & c & c^{\SSS -1}\\
      c^{\SSS -1} & c^{\SSS -1} & c^{\SSS -1} & c^{\SSS -1} & c & c & c^{\SSS
-1} & c & 1 & c & c^{\SSS -1} & c & c\\
      c & c^{\SSS -1} & c^{\SSS -1} & c^{\SSS -1} & c^{\SSS -1} & c & c &
c^{\SSS -1} & c & 1 & c & c^{\SSS -1} & c\\
      c & c & c^{\SSS -1} & c^{\SSS -1} & c^{\SSS -1} & c^{\SSS -1} & c & c &
c^{\SSS -1} & c & 1 & c & c^{\SSS -1}\\
      c^{\SSS -1} & c & c & c^{\SSS -1} & c^{\SSS -1} & c^{\SSS -1} & c^{\SSS
-1} & c & c & c^{\SSS -1} & c & 1 & c\\
      c & c^{\SSS -1} & c & c & c^{\SSS -1} & c^{\SSS -1} & c^{\SSS -1} &
      c^{\SSS -1} & c & c & c^{\SSS -1} & c & 1
    \end{array}
  \right]
\end{equation}

\begin{equation}
  \AUFfamily[B]{\tilde{C}}{13}{0} = 
  \left[
    \begin{array}{ccccccccccccc}
      1 & d & d^{\SSS -1} & d & d & d^{\SSS -1} & d^{\SSS -1} & d^{\SSS -1} &
d^{\SSS -1} & d & d & d^{\SSS -1} & d\\
      d & 1 & d & d^{\SSS -1} & d & d & d^{\SSS -1} & d^{\SSS -1} & d^{\SSS -1}
& d^{\SSS -1} & d & d & d^{\SSS -1}\\
      d^{\SSS -1} & d & 1 & d & d^{\SSS -1} & d & d & d^{\SSS -1} & d^{\SSS -1}
& d^{\SSS -1} & d^{\SSS -1} & d & d\\
      d & d^{\SSS -1} & d & 1 & d & d^{\SSS -1} & d & d & d^{\SSS -1} & d^{\SSS
-1} & d^{\SSS -1} & d^{\SSS -1} & d\\
      d & d & d^{\SSS -1} & d & 1 & d & d^{\SSS -1} & d & d & d^{\SSS -1} &
d^{\SSS -1} & d^{\SSS -1} & d^{\SSS -1}\\
      d^{\SSS -1} & d & d & d^{\SSS -1} & d & 1 & d & d^{\SSS -1} & d & d &
d^{\SSS -1} & d^{\SSS -1} & d^{\SSS -1}\\
      d^{\SSS -1} & d^{\SSS -1} & d & d & d^{\SSS -1} & d & 1 & d & d^{\SSS -1}
& d & d & d^{\SSS -1} & d^{\SSS -1}\\
      d^{\SSS -1} & d^{\SSS -1} & d^{\SSS -1} & d & d & d^{\SSS -1} & d & 1 & d
& d^{\SSS -1} & d & d & d^{\SSS -1}\\
      d^{\SSS -1} & d^{\SSS -1} & d^{\SSS -1} & d^{\SSS -1} & d & d & d^{\SSS
-1} & d & 1 & d & d^{\SSS -1} & d & d\\
      d & d^{\SSS -1} & d^{\SSS -1} & d^{\SSS -1} & d^{\SSS -1} & d & d &
d^{\SSS -1} & d & 1 & d & d^{\SSS -1} & d\\
      d & d & d^{\SSS -1} & d^{\SSS -1} & d^{\SSS -1} & d^{\SSS -1} & d & d &
d^{\SSS -1} & d & 1 & d & d^{\SSS -1}\\
      d^{\SSS -1} & d & d & d^{\SSS -1} & d^{\SSS -1} & d^{\SSS -1} & d^{\SSS
-1} & d & d & d^{\SSS -1} & d & 1 & d\\
      d & d^{\SSS -1} & d & d & d^{\SSS -1} & d^{\SSS -1} & d^{\SSS -1} &
      d^{\SSS -1} & d & d & d^{\SSS -1} & d & 1
    \end{array}
  \right]
\end{equation}

\begin{equation}
  \AUFfamily[A]{C}{13}{0} = 
  \left[
    \begin{array}{ccccccccccccc}
      1 & 1 & 1 & 1 & 1 & 1 & 1 & 1 & 1 & 1 & 1 & 1 & 1\\
      1 & c^{\SSS -2} & c & c^{\SSS -3} & c^{\SSS -1} & c & c^{\SSS -1} &
c^{\SSS -1} & c^{\SSS -1} & c^{\SSS -3} & c^{\SSS -1} & c & c^{\SSS -3}\\
      1 & c & c^2 & c & c^{\SSS -1} & c^3 & c^3 & c & c & c^{\SSS -1} & c^{\SSS
-1} & c^3 & c\\
      1 & c^{\SSS -3} & c & c^{\SSS -2} & c^{\SSS -1} & c^{\SSS -1} & c & c &
c^{\SSS -1} & c^{\SSS -3} & c^{\SSS -3} & c^{\SSS -1} & c^{\SSS -1}\\
      1 & c^{\SSS -1} & c^{\SSS -1} & c^{\SSS -1} & c^{\SSS -2} & c & c^{\SSS
-1} & c & c & c^{\SSS -3} & c^{\SSS -3} & c^{\SSS -1} & c^{\SSS -3}\\
      1 & c & c^3 & c^{\SSS -1} & c & c^2 & c^3 & c & c^3 & c & c^{\SSS -1} & c
& c^{\SSS -1}\\
      1 & c^{\SSS -1} & c^3 & c & c^{\SSS -1} & c^3 & c^2 & c^3 & c & c & c & c
& c^{\SSS -1}\\
      1 & c^{\SSS -1} & c & c & c & c & c^3 & c^2 & c^3 & c^{\SSS -1} & c & c^3
& c^{\SSS -1}\\
      1 & c^{\SSS -1} & c & c^{\SSS -1} & c & c^3 & c & c^3 & c^2 & c & c^{\SSS
-1} & c^3 & c\\
      1 & c^{\SSS -3} & c^{\SSS -1} & c^{\SSS -3} & c^{\SSS -3} & c & c &
c^{\SSS -1} & c & c^{\SSS -2} & c^{\SSS -1} & c^{\SSS -1} & c^{\SSS -1}\\
      1 & c^{\SSS -1} & c^{\SSS -1} & c^{\SSS -3} & c^{\SSS -3} & c^{\SSS -1} &
c & c & c^{\SSS -1} & c^{\SSS -1} & c^{\SSS -2} & c & c^{\SSS -3}\\
      1 & c & c^3 & c^{\SSS -1} & c^{\SSS -1} & c & c & c^3 & c^3 & c^{\SSS -1}
& c & c^2 & c\\
      1 & c^{\SSS -3} & c & c^{\SSS -1} & c^{\SSS -3} & c^{\SSS -1} &
      c^{\SSS -1} & c^{\SSS -1} & c & c^{\SSS -1} & c^{\SSS -3} & c &
      c^{\SSS -2}
    \end{array}
  \right]
\end{equation}

\begin{equation}
  \AUFfamily[B]{C}{13}{0} = 
  \left[
    \begin{array}{ccccccccccccc}
      1 & 1 & 1 & 1 & 1 & 1 & 1 & 1 & 1 & 1 & 1 & 1 & 1\\
      1 & d^{\SSS -2} & d & d^{\SSS -3} & d^{\SSS -1} & d & d^{\SSS -1} &
d^{\SSS -1} & d^{\SSS -1} & d^{\SSS -3} & d^{\SSS -1} & d & d^{\SSS -3}\\
      1 & d & d^2 & d & d^{\SSS -1} & d^3 & d^3 & d & d & d^{\SSS -1} & d^{\SSS
-1} & d^3 & d\\
      1 & d^{\SSS -3} & d & d^{\SSS -2} & d^{\SSS -1} & d^{\SSS -1} & d & d &
d^{\SSS -1} & d^{\SSS -3} & d^{\SSS -3} & d^{\SSS -1} & d^{\SSS -1}\\
      1 & d^{\SSS -1} & d^{\SSS -1} & d^{\SSS -1} & d^{\SSS -2} & d & d^{\SSS
-1} & d & d & d^{\SSS -3} & d^{\SSS -3} & d^{\SSS -1} & d^{\SSS -3}\\
      1 & d & d^3 & d^{\SSS -1} & d & d^2 & d^3 & d & d^3 & d & d^{\SSS -1} & d
& d^{\SSS -1}\\
      1 & d^{\SSS -1} & d^3 & d & d^{\SSS -1} & d^3 & d^2 & d^3 & d & d & d & d
& d^{\SSS -1}\\
      1 & d^{\SSS -1} & d & d & d & d & d^3 & d^2 & d^3 & d^{\SSS -1} & d & d^3
& d^{\SSS -1}\\
      1 & d^{\SSS -1} & d & d^{\SSS -1} & d & d^3 & d & d^3 & d^2 & d & d^{\SSS
-1} & d^3 & d\\
      1 & d^{\SSS -3} & d^{\SSS -1} & d^{\SSS -3} & d^{\SSS -3} & d & d &
d^{\SSS -1} & d & d^{\SSS -2} & d^{\SSS -1} & d^{\SSS -1} & d^{\SSS -1}\\
      1 & d^{\SSS -1} & d^{\SSS -1} & d^{\SSS -3} & d^{\SSS -3} & d^{\SSS -1} &
d & d & d^{\SSS -1} & d^{\SSS -1} & d^{\SSS -2} & d & d^{\SSS -3}\\
      1 & d & d^3 & d^{\SSS -1} & d^{\SSS -1} & d & d & d^3 & d^3 & d^{\SSS -1}
& d & d^2 & d\\
      1 & d^{\SSS -3} & d & d^{\SSS -1} & d^{\SSS -3} & d^{\SSS -1} &
      d^{\SSS -1} & d^{\SSS -1} & d & d^{\SSS -1} & d^{\SSS -3} & d &
      d^{\SSS -2}
    \end{array}
  \right]
\end{equation}
\ENDofSUBSECTION

%
%

\subsection {$N=14$}
\label{subsec__N_eq_14}


\subsubsection{Orbits stemming from $F_{14}$}

The only \maxAUFs\ stemming from $F_{14}$ are:
\begin{eqnarray}
  \AUFfamily{F}{14}{6}(a,b,c,d,e,f)  &  = & F_{14} \HADprod \EXPentrywise
  \left(\Ii \cdot \AUFphases{F}{14}{6}(a,b,c,d,e,f)\right) 
\label{F14_maxAUF}\\
  \TRANSPOSE{\AUFfamily{F}{14}{6}(a,b,c,d,e,f)}  &  = & F_{14} \HADprod
\EXPentrywise
  \left(\Ii \cdot \TRANSPOSE{\AUFphases{F}{14}{6}(a,b,c,d,e,f)} \right)
\nonumber 
\end{eqnarray}
where
\begin{equation}
  F_{14} =
  \left[
    \begin{array}{cccccccccccccc}
      1 & 1 & 1 & 1 & 1 & 1 & 1 & 1 & 1 & 1 & 1 & 1 & 1 & 1\\
      1 & w & w^2 & w^3 & w^4 & w^5 & w^6 & w^7 & w^8 & w^9 & w^{10} & w^{11} &
w^{12} & w^{13}\\
      1 & w^2 & w^4 & w^6 & w^8 & w^{10} & w^{12} & 1 & w^2 & w^4 & w^6 & w^8 &
w^{10} & w^{12}\\
      1 & w^3 & w^6 & w^9 & w^{12} & w & w^4 & w^7 & w^{10} & w^{13} & w^2 &
w^5 & w^8 & w^{11}\\
      1 & w^4 & w^8 & w^{12} & w^2 & w^6 & w^{10} & 1 & w^4 & w^8 & w^{12} &
w^2 & w^6 & w^{10}\\
      1 & w^5 & w^{10} & w & w^6 & w^{11} & w^2 & w^7 & w^{12} & w^3 & w^8 &
w^{13} & w^4 & w^9\\
      1 & w^6 & w^{12} & w^4 & w^{10} & w^2 & w^8 & 1 & w^6 & w^{12} & w^4 &
w^{10} & w^2 & w^8\\
      1 & w^7 & 1 & w^7 & 1 & w^7 & 1 & w^7 & 1 & w^7 & 1 & w^7 & 1 & w^7\\
      1 & w^8 & w^2 & w^{10} & w^4 & w^{12} & w^6 & 1 & w^8 & w^2 & w^{10} &
w^4 & w^{12} & w^6\\
      1 & w^9 & w^4 & w^{13} & w^8 & w^3 & w^{12} & w^7 & w^2 & w^{11} & w^6 &
w & w^{10} & w^5\\
      1 & w^{10} & w^6 & w^2 & w^{12} & w^8 & w^4 & 1 & w^{10} & w^6 & w^2 &
w^{12} & w^8 & w^4\\
      1 & w^{11} & w^8 & w^5 & w^2 & w^{13} & w^{10} & w^7 & w^4 & w & w^{12} &
w^9 & w^6 & w^3\\
      1 & w^{12} & w^{10} & w^8 & w^6 & w^4 & w^2 & 1 & w^{12} & w^{10} & w^8 &
w^6 & w^4 & w^2\\
      1 & w^{13} & w^{12} & w^{11} & w^{10} & w^9 & w^8 & w^7 & w^6 & w^5 &
      w^4 & w^3 & w^2 & w
    \end{array}
  \right]
\end{equation}
where $w = \exp(\Ii \cdot 2\pi/14)$, so $w^{14} = 1,\ w^7 = -1$, and
\begin{equation}
  \AUFphases{F}{14}{6}(a,b,c,d,e,f) = 
  \left[
  \begin{array}{ccccccc|ccccccc}
    \bO & \bO & \bO & \bO & \bO & \bO & \bO & \bO & \bO & \bO & \bO & \bO & \bO
& \bO\\
    \bO & a & b & c & d & e & f & \bO & a & b & c & d & e & f\\
    \hline
    \bO & \bO & \bO & \bO & \bO & \bO & \bO & \bO & \bO & \bO & \bO & \bO & \bO
& \bO\\
    \bO & a & b & c & d & e & f & \bO & a & b & c & d & e & f\\
    \hline
    \bO & \bO & \bO & \bO & \bO & \bO & \bO & \bO & \bO & \bO & \bO & \bO & \bO
& \bO\\
    \bO & a & b & c & d & e & f & \bO & a & b & c & d & e & f\\
    \hline
    \bO & \bO & \bO & \bO & \bO & \bO & \bO & \bO & \bO & \bO & \bO & \bO & \bO
& \bO\\
    \bO & a & b & c & d & e & f & \bO & a & b & c & d & e & f\\
    \hline
    \bO & \bO & \bO & \bO & \bO & \bO & \bO & \bO & \bO & \bO & \bO & \bO & \bO
& \bO\\
    \bO & a & b & c & d & e & f & \bO & a & b & c & d & e & f\\
    \hline
    \bO & \bO & \bO & \bO & \bO & \bO & \bO & \bO & \bO & \bO & \bO & \bO & \bO
& \bO\\
    \bO & a & b & c & d & e & f & \bO & a & b & c & d & e & f\\
    \hline
    \bO & \bO & \bO & \bO & \bO & \bO & \bO & \bO & \bO & \bO & \bO & \bO & \bO
& \bO\\
    \bO & a & b & c & d & e & f & \bO & a & b & c & d & e & f\\
  \end{array}
  \right]
\end{equation}
$\AUFfamily{F}{14}{6}$ and $\TRANSPOSE{\AUFfamily{F}{14}{6}}$ are a pair
of \COGNATE\ families.
\bigskip

At least one of them can be obtained by permuting one of Di{\c t}{\v a}'s
constructions, either
\begin{equation}
\label{F14fam_Dita_1}
  \AUFfamily[A]{\tilde{F}}{14}{6}(\alpha_1,\ldots,\alpha_6) = 
  \left[ \begin{array}{c|c}
            \ELEMENTof{F_2}{1}{1} \cdot F_7 & \ELEMENTof{F_2}{1}{2} \cdot
            D(\alpha_1,\ldots,\alpha_6) \cdot F_7 \\
            \hline
            \ELEMENTof{F_2}{2}{1} \cdot F_7 & \ELEMENTof{F_2}{2}{2} \cdot
            D(\alpha_1,\ldots,\alpha_6) \cdot F_7
         \end{array} \right]
\end{equation}
where $D(\alpha_1,\ldots,\alpha_6)$ is the $7 \times 7$ diagonal
matrix $\diag(1,\PHASE{\alpha_1},\ldots,\PHASE{\alpha_6})$,

or
\begin{equation}
\label{F14fam_Dita_2}
  \AUFfamily[B]{\tilde{F}}{14}{6}(\beta_1,\ldots,\beta_6)
\end{equation}
such that its\ \  $i,j$-th $2 \times 2$ block is equal to
$\ELEMENTof{F_7}{i}{j}
\cdot D(\alpha) \cdot F_2$, where $i,j \in \{1 \ldots 7\}$,
$D(\alpha)$ is the diagonal matrix $\diag(1,\PHASE{\alpha})$ and
$\alpha = 0,\beta_1,\ldots,\beta_6$ for $j = 1,2,\ldots,7$
respectively.
\medskip

This is because $\AUFfamily[A]{\tilde{F}}{14}{6}(\ZEROvect) = F_2 \otimes F_7$
and
$\AUFfamily[B]{\tilde{F}}{14}{6}(\ZEROvect) = F_7 \otimes F_2$ are, according
to
\cite{Ta05}, permutation equivalent to $F_{14}$, so both Di{\c t}{\v a}'s
orbits
are \maxAUFs\ stemming from
permuted $F_{14}$'s.


\subsubsection{Other $14 \times 14$ orbits}

Other \DEPHASED\ $14 \times 14$ orbits can be obtained, using the Di{\c t}{\v a}'s
method, from
$F_2$ and \DEPHASED\ $7 \times 7$ complex Hadamard matrices
from section \ref{subsec__N_eq_7}, for example:
\begin{description}
  \item
        \begin{equation}
          \AUFfamily{FP}{14}{7}(\alpha_1,\ldots,\alpha_7)\ \ \ =\ \ \   
          \DITAtwoBYtwo{F_2}
          {\AUFfamily{F}{7}{0}}
          {D(\alpha_2,\ldots,\alpha_7) \cdot
            \AUFfamily{P}{7}{1}(\alpha_1)}
        \end{equation}

  \item
        \begin{equation}
          \AUFfamily[k]{FC}{14}{6}(\alpha_1,\ldots,\alpha_6)\ \ \ =\ \ \ 
          \DITAtwoBYtwo{F_2}
          {\AUFfamily{F}{7}{0}}
          {D(\alpha_1,\ldots,\alpha_6) \cdot
            \AUFfamily[k]{C}{7}{0}}
        \end{equation}

  \item
        \begin{equation}
          \AUFfamily{PP}{14}{8}(\alpha_1,\ldots,\alpha_8)\ \ \ =\ \ \   
          \DITAtwoBYtwo{F_2}
          {\AUFfamily{P}{7}{1}(\alpha_1)}
          {D(\alpha_3,\ldots,\alpha_8) \cdot
            \AUFfamily{P}{7}{1}(\alpha_2)}
        \end{equation}

  \item
        \begin{equation}
          \AUFfamily[k]{PC}{14}{7}(\alpha_1,\ldots,\alpha_7)\ \ \ =\ \ \ 
          \DITAtwoBYtwo{F_2}
          {\AUFfamily{P}{7}{1}(\alpha_1)}
          {D(\alpha_2,\ldots,\alpha_7) \cdot
            \AUFfamily[k]{C}{7}{0}}
        \end{equation}

  \item
        \begin{equation}
          \AUFfamily[lm]{CC}{14}{6}(\alpha_1,\ldots,\alpha_6)\ \ \ =\ \ \   
          \DITAtwoBYtwo{F_2}
          {\AUFfamily[l]{C}{7}{0}}
          {D(\alpha_1,\ldots,\alpha_6) \cdot
            \AUFfamily[m]{C}{7}{0}}
        \end{equation}
\end{description}
where $k \in \{A,B,C,D\}$ and 
\begin{displaymath}
  (l,m) \in \{(A,A),(A,B),(A,C),(A,D),(B,B),(B,C),(B,D),(C,C),(C,D),(D,D)\}
\end{displaymath}
designate $7\times 7$
complex Hadamard matrices associated with cyclic $7$--roots,
and $D(\beta_1,\ldots,\beta_6)$ denotes the $7 \times 7$ diagonal
matrix $\diag(1,\PHASE{\beta_1},\ldots,\PHASE{\beta_6})$.
\ENDofSUBSECTION

%
%
 
\subsection {$N=15$}
\label{subsec__N_eq_15}


\subsubsection{Orbits stemming from $F_{15}$}

The only \maxAUFs\ stemming from $F_{15}$ are :
\begin{eqnarray}
  \AUFfamily{F}{15}{8}(a,b,c,d,e,f,g,h)  &  = & F_{15} \HADprod \EXPentrywise
  \left(\Ii \cdot \AUFphases{F}{15}{8}(a,b,c,d,e,f,g,h)\right) 
\label{F15_maxAUF}\\
  \TRANSPOSE{\AUFfamily{F}{15}{8}(a,b,c,d,e,f,g,h)}  &  = & F_{15} \HADprod
\EXPentrywise
  \left(\Ii \cdot \TRANSPOSE{\AUFphases{F}{15}{8}(a,b,c,d,e,f,g,h)} \right)
\nonumber
\end{eqnarray}
where
\begin{equation}
  F_{15} =
  \left[
    \begin{array}{ccccccccccccccc}
      1 & 1 & 1 & 1 & 1 & 1 & 1 & 1 & 1 & 1 & 1 & 1 & 1 & 1 & 1\\
      1 & w & w^2 & w^3 & w^4 & w^5 & w^6 & w^7 & w^8 & w^9 & w^{10} & w^{11} &
w^{12} & w^{13} & w^{14}\\
      1 & w^2 & w^4 & w^6 & w^8 & w^{10} & w^{12} & w^{14} & w & w^3 & w^5 &
w^7 & w^9 & w^{11} & w^{13}\\
      1 & w^3 & w^6 & w^9 & w^{12} & 1 & w^3 & w^6 & w^9 & w^{12} & 1 & w^3 &
w^6 & w^9 & w^{12}\\
      1 & w^4 & w^8 & w^{12} & w & w^5 & w^9 & w^{13} & w^2 & w^6 & w^{10} &
w^{14} & w^3 & w^7 & w^{11}\\
      1 & w^5 & w^{10} & 1 & w^5 & w^{10} & 1 & w^5 & w^{10} & 1 & w^5 & w^{10}
& 1 & w^5 & w^{10}\\
      1 & w^6 & w^{12} & w^3 & w^9 & 1 & w^6 & w^{12} & w^3 & w^9 & 1 & w^6 &
w^{12} & w^3 & w^9\\
      1 & w^7 & w^{14} & w^6 & w^{13} & w^5 & w^{12} & w^4 & w^{11} & w^3 &
w^{10} & w^2 & w^9 & w & w^8\\
      1 & w^8 & w & w^9 & w^2 & w^{10} & w^3 & w^{11} & w^4 & w^{12} & w^5 &
w^{13} & w^6 & w^{14} & w^7\\
      1 & w^9 & w^3 & w^{12} & w^6 & 1 & w^9 & w^3 & w^{12} & w^6 & 1 & w^9 &
w^3 & w^{12} & w^6\\
      1 & w^{10} & w^5 & 1 & w^{10} & w^5 & 1 & w^{10} & w^5 & 1 & w^{10} & w^5
& 1 & w^{10} & w^5\\
      1 & w^{11} & w^7 & w^3 & w^{14} & w^{10} & w^6 & w^2 & w^{13} & w^9 & w^5
& w & w^{12} & w^8 & w^4\\
      1 & w^{12} & w^9 & w^6 & w^3 & 1 & w^{12} & w^9 & w^6 & w^3 & 1 & w^{12}
& w^9 & w^6 & w^3\\
      1 & w^{13} & w^{11} & w^9 & w^7 & w^5 & w^3 & w & w^{14} & w^{12} &
w^{10} & w^8 & w^6 & w^4 & w^2\\
      1 & w^{14} & w^{13} & w^{12} & w^{11} & w^{10} & w^9 & w^8 & w^7 & w^6
      & w^5 & w^4 & w^3 & w^2 & w
    \end{array}
  \right]
\end{equation}
where $w = \exp(\Ii \cdot 2\pi/15)$, so $w^{15} = 1$, and
\begin{equation}
  \AUFphases{F}{15}{8}(a,b,c,d,e,f,g,h) = 
  \left[
    \begin{array}{ccccc|ccccc|ccccc}
      \bO & \bO & \bO & \bO & \bO & \bO & \bO & \bO & \bO & \bO & \bO & \bO &
\bO & \bO & \bO\\
      \bO & a & b & c & d & \bO & a & b & c & d & \bO & a & b & c & d\\
      \bO & e & f & g & h & \bO & e & f & g & h & \bO & e & f & g & h\\
      \hline
      \bO & \bO & \bO & \bO & \bO & \bO & \bO & \bO & \bO & \bO & \bO & \bO &
\bO & \bO & \bO\\
      \bO & a & b & c & d & \bO & a & b & c & d & \bO & a & b & c & d\\
      \bO & e & f & g & h & \bO & e & f & g & h & \bO & e & f & g & h\\
      \hline
      \bO & \bO & \bO & \bO & \bO & \bO & \bO & \bO & \bO & \bO & \bO & \bO &
\bO & \bO & \bO\\
      \bO & a & b & c & d & \bO & a & b & c & d & \bO & a & b & c & d\\
      \bO & e & f & g & h & \bO & e & f & g & h & \bO & e & f & g & h\\
      \hline
      \bO & \bO & \bO & \bO & \bO & \bO & \bO & \bO & \bO & \bO & \bO & \bO &
\bO & \bO & \bO\\
      \bO & a & b & c & d & \bO & a & b & c & d & \bO & a & b & c & d\\
      \bO & e & f & g & h & \bO & e & f & g & h & \bO & e & f & g & h\\
      \hline
      \bO & \bO & \bO & \bO & \bO & \bO & \bO & \bO & \bO & \bO & \bO & \bO &
\bO & \bO & \bO\\
      \bO & a & b & c & d & \bO & a & b & c & d & \bO & a & b & c & d\\
      \bO & e & f & g & h & \bO & e & f & g & h & \bO & e & f & g & h\\
    \end{array}
  \right] \ .
\end{equation}
The families 
$\AUFfamily{F}{15}{8}$ and $\TRANSPOSE{\AUFfamily{F}{15}{8}}$ are
cognate.
\bigskip

At least one of them can be obtained by permuting one of Di{\c t}{\v a}'s
constructions, either
\begin{equation}
\label{F15fam_Dita_1}
  \AUFfamily[A]{\tilde{F}}{15}{8}(\alpha_1,\ldots,\alpha_8) = 
  \left[ \begin{array}{c|c|c}
            \ELEMENTof{F_3}{1}{1} \cdot F_5 & \ELEMENTof{F_3}{1}{2} \cdot
            D(\alpha_1,\ldots,\alpha_4) \cdot F_5 & \ELEMENTof{F_3}{1}{3} \cdot
            D(\alpha_5,\ldots,\alpha_8) \cdot F_5 \\
            \hline
            \ELEMENTof{F_3}{2}{1} \cdot F_5 & \ELEMENTof{F_3}{2}{2} \cdot
            D(\alpha_1,\ldots,\alpha_4) \cdot F_5 & \ELEMENTof{F_3}{2}{3} \cdot
            D(\alpha_5,\ldots,\alpha_8) \cdot F_5 \\
            \hline
            \ELEMENTof{F_3}{3}{1} \cdot F_5 & \ELEMENTof{F_3}{3}{2} \cdot
            D(\alpha_1,\ldots,\alpha_4) \cdot F_5 & \ELEMENTof{F_3}{3}{3} \cdot
            D(\alpha_5,\ldots,\alpha_8) \cdot F_5 \\
         \end{array} \right]
\end{equation}
where $D(\alpha_1,\ldots,\alpha_4)$ is the $5 \times 5$ diagonal
matrix $\diag(1,\PHASE{\alpha_1},\ldots,\PHASE{\alpha_4})$,

or
\begin{equation}
\label{F15fam_Dita_2}
  \AUFfamily[B]{\tilde{F}}{15}{8}(\beta_1,\ldots,\beta_8)
\end{equation}
such that its\ \ $i,j$-th $3 \times 3$ block is equal to $\ELEMENTof{F_5}{i}{j}
\cdot D(\alpha,\beta) \cdot F_3$, where $i,j \in \{1 \ldots 5\}$,
$D(\alpha,\beta)$ is the diagonal matrix
$\diag(1,\PHASE{\alpha},\PHASE{\beta})$ and
\begin{displaymath}
  (\alpha,\beta)\  =\ 
  (0,0),\ (\beta_1,\beta_2),\ (\beta_3,\beta_4),\ (\beta_5,\beta_6),\
  (\beta_7,\beta_8)
\end{displaymath}
for $j = 1,2,\ldots,5$ respectively.
\medskip

This is because $\AUFfamily[A]{\tilde{F}}{15}{8}(\ZEROvect) = F_3 \otimes F_5$
and
$\AUFfamily[B]{\tilde{F}}{15}{8}(\ZEROvect) = F_5 \otimes F_3$ are, according
to
\cite{Ta05}, permutation equivalent to $F_{15}$, so both Di{\c t}{\v a}'s
orbits
are \maxAUFs\ stemming from
permuted $F_{15}$'s.
\ENDofSUBSECTION

%
%
 
\subsection {$N=16$}
\label{subsec__N_eq_16}


\subsubsection{Orbits stemming from $F_{16}$}

The only \maxAUF\ stemming from $F_{16}$ is the 
$17$-parameter orbit:
\begin{equation}
\label{F16_maxAUF}
  \AUFfamily{F}{16}{17}(a,b,c,d,e,f,g,h,i,j,k,l,m,n,o,p,r) = F_{16} \HADprod
  \EXPentrywise  (\Ii \cdot \AUFphases{F}{16}{17}(a,\ldots,r))
\end{equation}
where
\begin{equation}
  F_{16} =
  \left[
    \begin{array}{cccccccccccccccc}
      1 & 1 & 1 & 1 & 1 & 1 & 1 & 1 & 1 & 1 & 1 & 1 & 1 & 1 & 1 & 1\\
      1 & w & w^2 & w^3 & w^4 & w^5 & w^6 & w^7 & w^8 & w^9 & w^{10} & w^{11} &
w^{12} & w^{13} & w^{14} & w^{15}\\
      1 & w^2 & w^4 & w^6 & w^8 & w^{10} & w^{12} & w^{14} & 1 & w^2 & w^4 &
w^6 & w^8 & w^{10} & w^{12} & w^{14}\\
      1 & w^3 & w^6 & w^9 & w^{12} & w^{15} & w^2 & w^5 & w^8 & w^{11} & w^{14}
& w & w^4 & w^7 & w^{10} & w^{13}\\
      1 & w^4 & w^8 & w^{12} & 1 & w^4 & w^8 & w^{12} & 1 & w^4 & w^8 & w^{12}
& 1 & w^4 & w^8 & w^{12}\\
      1 & w^5 & w^{10} & w^{15} & w^4 & w^9 & w^{14} & w^3 & w^8 & w^{13} & w^2
& w^7 & w^{12} & w & w^6 & w^{11}\\
      1 & w^6 & w^{12} & w^2 & w^8 & w^{14} & w^4 & w^{10} & 1 & w^6 & w^{12} &
w^2 & w^8 & w^{14} & w^4 & w^{10}\\
      1 & w^7 & w^{14} & w^5 & w^{12} & w^3 & w^{10} & w & w^8 & w^{15} & w^6 &
w^{13} & w^4 & w^{11} & w^2 & w^9\\
      1 & w^8 & 1 & w^8 & 1 & w^8 & 1 & w^8 & 1 & w^8 & 1 & w^8 & 1 & w^8 & 1 &
w^8\\
      1 & w^9 & w^2 & w^{11} & w^4 & w^{13} & w^6 & w^{15} & w^8 & w & w^{10} &
w^3 & w^{12} & w^5 & w^{14} & w^7\\
      1 & w^{10} & w^4 & w^{14} & w^8 & w^2 & w^{12} & w^6 & 1 & w^{10} & w^4 &
w^{14} & w^8 & w^2 & w^{12} & w^6\\
      1 & w^{11} & w^6 & w & w^{12} & w^7 & w^2 & w^{13} & w^8 & w^3 & w^{14} &
w^9 & w^4 & w^{15} & w^{10} & w^5\\
      1 & w^{12} & w^8 & w^4 & 1 & w^{12} & w^8 & w^4 & 1 & w^{12} & w^8 & w^4
& 1 & w^{12} & w^8 & w^4\\
      1 & w^{13} & w^{10} & w^7 & w^4 & w & w^{14} & w^{11} & w^8 & w^5 & w^2 &
w^{15} & w^{12} & w^9 & w^6 & w^3\\
      1 & w^{14} & w^{12} & w^{10} & w^8 & w^6 & w^4 & w^2 & 1 & w^{14} &
w^{12} & w^{10} & w^8 & w^6 & w^4 & w^2\\
      1 & w^{15} & w^{14} & w^{13} & w^{12} & w^{11} & w^{10} & w^9 & w^8 &
      w^7 & w^6 & w^5 & w^4 & w^3 & w^2 & w
    \end{array}
  \right]
\end{equation}
where $w = \exp(\Ii \cdot 2\pi/16)$, so $w^{16} = 1,\ w^8 = -1,\ w^4 =
\Ii$, and
(where typographic purposes we denoted
$e-a+k$ by
\begin{displaymath} 
  \left(
    \begin{array}{c}
      e-a \\
      + k
    \end{array}
  \right),
\end{displaymath}
so this is not a binomial coefficient)

{
%
%
%
\newcommand{\SQEEZE}[3]
{
  \left(
    \begin{array}{@{}c@{}}
      #1 - #2 \\
      + #3
    \end{array}
  \right)
}
%
%
\newcommand{\manySPACES}
{\ \ \ \ \ \ \ \ \ \ \ \ \ \ \ \ \ \ \ \ \ \ \ \ \ \ \ \ \ \ \ \ \ \ \
\ \ \ \ \ \ \ \ \ \ \ \ \ \ \ \ \ \ }
\begin{eqnarray}
  \AUFphases{F}{16}{17}(a,\ldots,r) = \manySPACES & & \\
  {\tiny
  \left[
    \begin{array}{cccccccc|cccccccc}
      \bO & \bO & \bO & \bO & \bO & \bO & \bO & \bO & \bO & \bO & \bO & \bO &
\bO & \bO & \bO & \bO\\[2mm]
      \bO & a & b & c & d & e & f & g & \bO & a & b & c & d & e & f & g\\[2mm]
      \bO & h & i & j & \bO & h & i & j & \bO & h & i & j & \bO & h & i &
j\\[2mm]
      \bO & k & l & m & d & \SQEEZE{e}{a}{k} & \SQEEZE{f}{b}{l} &
\SQEEZE{g}{c}{m} & \bO & k & l
      & m & d & \SQEEZE{e}{a}{k} & \SQEEZE{f}{b}{l} & \SQEEZE{g}{c}{m}\\[2mm]
      \bO & n & \bO & n & \bO & n & \bO & n & \bO & n & \bO & n & \bO & n & \bO
& n\\[2mm]
      \bO & o & b & \SQEEZE{c}{a}{o} & d & \SQEEZE{e}{a}{o} & f &
\SQEEZE{g}{a}{o} & \bO & o & b
      & \SQEEZE{c}{a}{o} & d & \SQEEZE{e}{a}{o} & f & \SQEEZE{g}{a}{o}\\[2mm]
      \bO & p & i & \SQEEZE{j}{h}{p} & \bO & p & i & \SQEEZE{j}{h}{p}
      & \bO & p & i & \SQEEZE{j}{h}{p} & \bO & p & i &
      \SQEEZE{j}{h}{p} \\[2mm] 
      \bO & r & l & \SQEEZE{m}{k}{r} & d & \SQEEZE{e}{a}{r} & \SQEEZE{f}{b}{l}
&
      \SQEEZE{g}{c+m}{r-k} & \bO & r & l & \SQEEZE{m}{k}{r} & d &
      \SQEEZE{e}{a}{r} & \SQEEZE{f}{b}{l} & \SQEEZE{g}{c+m}{r-k} \\[2mm]  
      \hline
      \bO & \bO & \bO & \bO & \bO & \bO & \bO & \bO & \bO & \bO & \bO & \bO &
\bO & \bO & \bO & \bO\\[2mm]
      \bO & a & b & c & d & e & f & g & \bO & a & b & c & d & e & f & g\\[2mm]
      \bO & h & i & j & \bO & h & i & j & \bO & h & i & j & \bO & h & i &
j\\[2mm]
      \bO & k & l & m & d & \SQEEZE{e}{a}{k} & \SQEEZE{f}{b}{l} &
\SQEEZE{g}{c}{m} & \bO & k & l
      & m & d & \SQEEZE{e}{a}{k} & \SQEEZE{f}{b}{l} & \SQEEZE{g}{c}{m}\\[2mm]
      \bO & n & \bO & n & \bO & n & \bO & n & \bO & n & \bO & n & \bO & n & \bO
& n\\[2mm]
      \bO & o & b & \SQEEZE{c}{a}{o} & d & \SQEEZE{e}{a}{o} & f &
\SQEEZE{g}{a}{o} & \bO & o & b
      & \SQEEZE{c}{a}{o} & d & \SQEEZE{e}{a}{o} & f & \SQEEZE{g}{a}{o}\\[2mm]
      \bO & p & i & \SQEEZE{j}{h}{p} & \bO & p & i & \SQEEZE{j}{h}{p}
      & \bO & p & i & \SQEEZE{j}{h}{p} & \bO & p & i & \SQEEZE{j}{h}{p}\\[2mm]
      \bO & r & l & \SQEEZE{m}{k}{r} & d & \SQEEZE{e}{a}{r} & \SQEEZE{f}{b}{l}
&
      \SQEEZE{g}{c+m}{r-k}  & \bO & r & l & \SQEEZE{m}{k}{r} & d &
\SQEEZE{e}{a}{r}
      & \SQEEZE{f}{b}{l} & \SQEEZE{g}{c+m}{r-k}
    \end{array}
  \right] \ .
  }
  & & \nonumber
\end{eqnarray}
} 

The $17$--dimensional family $\AUFfamily{F}{16}{17}$
of $N=16$ complex Hadamard matrices is \selfCOGNATE.
Note that this dimensionality coincides with the defect, 
$d(F_{16})=17$, which follows from Eq. (\ref{powprime}).
\bigskip

The above orbit is \PERMUTATIONequivalent\ to the orbit constructed
with the Di{\c t}{\v a}'s method:
\begin{equation}
\label{F16fam_Dita}
  \AUFfamily{\tilde{F}}{16}{17}(\alpha_1,\ldots,\alpha_{17}) = 
  \left[ \begin{array}{c|c}
            \ELEMENTof{F_2}{1}{1} \cdot
\AUFfamily{F}{8}{5}(\alpha_1,\ldots,\alpha_5) & \ELEMENTof{F_2}{1}{2} \cdot
            D(\alpha_{11},\ldots,\alpha_{17}) \cdot
\AUFfamily{F}{8}{5}(\alpha_6,\ldots,\alpha_{10}) \\
            \hline
            \ELEMENTof{F_2}{2}{1} \cdot
\AUFfamily{F}{8}{5}(\alpha_1,\ldots,\alpha_5) & \ELEMENTof{F_2}{2}{2} \cdot
            D(\alpha_{11},\ldots,\alpha_{17}) \cdot
\AUFfamily{F}{8}{5}(\alpha_6,\ldots,\alpha_{10}) \\
         \end{array} \right]
\end{equation}
where the only \maxAUF\ stemming from
$F_8$: $\AUFfamily{F}{8}{5}$ such that $\AUFfamily{F}{8}{5}(\ZEROvect)=F_8$, is
given by
(\ref{F8_maxAUF}), and $D(\alpha_{11},\ldots,\alpha_{17})$ is the $8
\times 8$ diagonal matrix
$\diag(1,\PHASE{\alpha_{11}},\ldots,\PHASE{\alpha_{17}})$.
\medskip

This is because
\begin{eqnarray}
  F_{16} & = &
\AUFfamily{\tilde{F}}{16}{17}(\ZEROvect,(1/16)2\pi,(2/16)2\pi,\ldots,(7/16)2\pi) 
\cdot    \\
         &   &        
[\STbasis{1},\STbasis{3},\STbasis{5},\STbasis{7},\STbasis{9},\STbasis{11},\STbasis{13},\STbasis{15},\STbasis{2},\STbasis{4},\STbasis{6},\STbasis{8},\STbasis{10},\STbasis{12},\STbasis{14},\STbasis{16}]^{T}
\nonumber
\end{eqnarray}
where $\STbasis{i}$ denotes the $i$-th standard basis column vector, so indeed
(\ref{F16fam_Dita}) generates the
only \maxAUF\ stemming from a permuted
$F_{16}$.
\medskip

The above orbit also passes through $\AUFfamily{\tilde{F}}{16}{17}(\ZEROvect) =
F_2 \otimes F_8$, as well as through permuted $F_2 \otimes F_2 \otimes
F_4$ and $F_2 \otimes F_2 \otimes F_2 \otimes F_2$, since
$\AUFfamily{\tilde{F}}{8}{5}$ of (\ref{F8fam_Dita}), or the
\PERMUTATIONequivalent\ $\AUFfamily{F}{8}{5}$ of (\ref{F8_maxAUF}), both pass
through permuted $F_2 \otimes F_4$ and $F_2 \otimes F_2 \otimes
F_2$. Note that all the tensor products $F_{16}$, $F_2 \otimes F_8$, $F_2
\otimes F_2 \otimes
F_4$ and $F_2 \otimes F_2 \otimes F_2 \otimes F_2$ are inequivalent
\cite{Ta05}.
\ENDofSUBSECTION

%
%
%

%
%

\section {Closing remarks}

Let us summarize our work  by proposing a set
of \DEPHASED\ representatives $\HADreprs{N}$ of equivalence classes of
Hadamard matrices of size $N=2,3,4,5$, and by enumerating the sets from the
sum of
which one should be able to extract such a set of representatives for
$N=6,\dots,16$. 
The dots indicate that
the existence of other equivalence classes cannot be excluded. 
For instance, one could look for new inequivalent families 
for composite $N$ 
using the construction by  Di{\c t}{\v a} \cite{Di04}
with permuted {\sl some} of the component families 
of Hadamard matrices of smaller size.

We tend to believe that the compiled list is minimal
in the sense that each family is necessary,
since it contains at least some matrices not equivalent
to all others.
However, the presented orbits of Hadamard matrices may be
(partially) equivalent, and equivalences
may hold within families as well as between them.

In the list of complex Hadamard matrices presented below,
let $\{ \AUFfamily{X}{N}{d} \}$ denote the set of
elements of the family $\AUFfamily{X}{N}{d}$.

\begin{description}
  \newcommand{\SPACESlevelONE}{\ \ \ \ }
  \newcommand{\SPACESlevelTWO}{\ \ \ \ \ \ \ \ \ \ \ \ \ \ \ \ }
  \item[$N=2$]   
        \SPACESlevelONE $\HADreprs{2}\ =\ \{ \AUFfamily{F}{2}{0} \}$.
  \item
  \item[$N=3$]
        \SPACESlevelONE $\HADreprs{3}\ =\ \{ \AUFfamily{F}{3}{0} \}$.
  \item
  \item[$N=4$]
        \SPACESlevelONE $\HADreprs{4}\ =\ \{ \AUFfamily{F}{4}{1}(a); a\in
[0,\pi) \}$.
  \item
  \item[$N=5$]
        \SPACESlevelONE $\HADreprs{5}\ =\ \{ \AUFfamily{F}{5}{0} \}$.
  \item
  \item[$N=6$]
        \SPACESlevelONE $\HADreprs{6}\ \subset$ 
        \begin{description}
          \item  \SPACESlevelTWO $\{ \AUFfamily{F}{6}{2} \}\ \cup $
          \item  \SPACESlevelTWO $\left\{
\TRANSPOSE{\AUFfamily{F}{6}{2}}\right\}\ \cup $
          \item  \SPACESlevelTWO $\{ \AUFfamily{D}{6}{1} \}\ \cup  $
          \item  \SPACESlevelTWO $\{ \AUFfamily{C}{6}{0} \}\ \cup  $        
          \item  \SPACESlevelTWO $\{ \AUFfamily{S}{6}{0} \}\ \cup\ \ldots$
        \end{description}
  \item
  \item[$N=7$] 
        \SPACESlevelONE $\HADreprs{7}\ \subset$
        \begin{description}
          \item \SPACESlevelTWO $\{ \AUFfamily{F}{7}{0} \}\ \cup$
          \item \SPACESlevelTWO $\{ \AUFfamily{P}{7}{1} \}\ \cup$
          \item \SPACESlevelTWO $\{ \AUFfamily[A]{C}{7}{0} \}\ \cup$
          \item \SPACESlevelTWO $\{ \AUFfamily[B]{C}{7}{0} \}\ \cup$
          \item \SPACESlevelTWO $\{ \AUFfamily[C]{C}{7}{0} \}\ \cup$
          \item \SPACESlevelTWO $\{ \AUFfamily[D]{C}{7}{0} \}\ \cup\ \ldots$
        \end{description}
  \item
  \item[$N=8$] 
        \SPACESlevelONE $\HADreprs{8}\ \subset$
        \begin{description}
          \item \SPACESlevelTWO $\{ \AUFfamily{F}{8}{5} \}\ \cup\ \ldots$
        \end{description}
  \item
  \item[$N=9$]
        \SPACESlevelONE $\HADreprs{9}\ \subset$
        \begin{description}
          \item \SPACESlevelTWO $\{ \AUFfamily{F}{9}{4} \}\ \cup\ \ldots$
        \end{description}
  \item
  \item[$N=10$]
        \SPACESlevelONE $\HADreprs{10}\ \subset$
        \begin{description}
          \item \SPACESlevelTWO $\{ \AUFfamily{F}{10}{4} \}\ \cup$
          \item \SPACESlevelTWO $\left\{ \TRANSPOSE{\AUFfamily{F}{10}{4}}
\right\}\ \cup\ \ldots$
        \end{description}
  \item
  \item[$N=11$]
        \SPACESlevelONE $\HADreprs{11}\ \subset$
        \begin{description}
          \item \SPACESlevelTWO $\{ \AUFfamily{F}{11}{0} \}\ \cup$
          \item \SPACESlevelTWO $\{ \AUFfamily[A]{C}{11}{0} \}\ \cup$
          \item \SPACESlevelTWO $\{ \AUFfamily[B]{C}{11}{0} \}\ \cup$
          \item \SPACESlevelTWO $\{ \AUFfamily{N}{11}{0} \}\ \cup\ \ldots$
        \end{description}
  \item
  \item[$N=12$]
        \SPACESlevelONE $\HADreprs{12}\ \subset$
        \begin{description}
          \item \SPACESlevelTWO $\{ \AUFfamily[A]{F}{12}{9} \}\ \cup$
          \item \SPACESlevelTWO $\{ \AUFfamily[B]{F}{12}{9} \}\ \cup$
          \item \SPACESlevelTWO $\{ \AUFfamily[C]{F}{12}{9} \}\ \cup$
          \item \SPACESlevelTWO $\{ \AUFfamily[D]{F}{12}{9} \}\ \cup$
          \item \SPACESlevelTWO $\left\{ \TRANSPOSE{\AUFfamily[B]{F}{12}{9}}
\right\}\ \cup$
          \item \SPACESlevelTWO $\left\{ \TRANSPOSE{\AUFfamily[C]{F}{12}{9}}
\right\}\ \cup$
          \item \SPACESlevelTWO $\left\{ \TRANSPOSE{\AUFfamily[D]{F}{12}{9}}
\right\}\ \cup$
          \item \SPACESlevelTWO $\{ \AUFfamily{FD}{12}{8}            \}\ \cup$
          \item \SPACESlevelTWO $\{ \AUFfamily{FC}{12}{7}            \}\ \cup$
          \item \SPACESlevelTWO $\{ \AUFfamily{FS}{12}{7}            \}\ \cup$
          \item \SPACESlevelTWO $\{ \AUFfamily{DD}{12}{7}            \}\ \cup$
          \item \SPACESlevelTWO $\{ \AUFfamily{DC}{12}{6}            \}\ \cup$
          \item \SPACESlevelTWO $\{ \AUFfamily{DS}{12}{6}            \}\ \cup$
          \item \SPACESlevelTWO $\{ \AUFfamily{CC}{12}{5}            \}\ \cup$
          \item \SPACESlevelTWO $\{ \AUFfamily{CS}{12}{5}            \}\ \cup$
          \item \SPACESlevelTWO $\{ \AUFfamily{SS}{12}{5}            \}\ \cup\
\ldots$
        \end{description}
  \item
  \item[$N=13$]
        \SPACESlevelONE $\HADreprs{13} \subset$
        \begin{description}
          \item \SPACESlevelTWO $\{ \AUFfamily{F}{13}{0} \}\ \cup$
          \item \SPACESlevelTWO $\{ \AUFfamily[A]{C}{13}{0} \}\ \cup$
          \item \SPACESlevelTWO $\{ \AUFfamily[B]{C}{13}{0} \}\ \cup$                
          \item \SPACESlevelTWO $\{ \AUFfamily{P}{13}{2} \}\ \cup \ldots$
        \end{description}
  \item
  \item[$N=14$]
        \SPACESlevelONE $\HADreprs{14}\ \subset$
        \begin{description}
          \item \SPACESlevelTWO $\{ \AUFfamily{F}{14}{6} \}\ \cup$
          \item \SPACESlevelTWO $ \left\{ \TRANSPOSE{\AUFfamily{F}{14}{6}}
\right\}\ \cup$
          \item \SPACESlevelTWO $\{ \AUFfamily{FP}{14}{7} \}\ \cup$
          \item \SPACESlevelTWO $\bigcup_{k \in \{A,B,C,D\}} 
                                   \{ \AUFfamily[k]{FC}{14}{6} \}\ \ \cup$
          \item \SPACESlevelTWO $\{ \AUFfamily{PP}{14}{8} \}\ \cup$
          \item \SPACESlevelTWO $\bigcup_{k \in \{A,B,C,D\}} 
                                   \{ \AUFfamily[k]{PC}{14}{7} \}\ \ \cup$
          \item \SPACESlevelTWO $\bigcup_{(l,m) \in
                  \{(A,A),(A,B),(A,C),(A,D),(B,B),(B,C),(B,D),(C,C),(C,D),(D,D)\}}
                                   \{ \AUFfamily[lm]{CC}{14}{6} \}\ \ \cup\
\ldots$
        \end{description}
  \item
  \item[$N=15$]
        \SPACESlevelONE $\HADreprs{15}\ \subset$
        \begin{description}
          \item \SPACESlevelTWO $\{ \AUFfamily{F}{15}{8} \}\ \cup$
          \item \SPACESlevelTWO $\left\{ \TRANSPOSE{\AUFfamily{F}{15}{8}}
\right\}\ \cup\ \ldots$
        \end{description}
  \item
  \item[$N=16$]
        \SPACESlevelONE $\HADreprs{16}\ \subset$
        \begin{description}
          \item \SPACESlevelTWO $\{ \AUFfamily{F}{16}{17} \}\ \cup\ \ldots$
        \end{description}
  \item
\end{description}

Note that the presented list of equivalence classes
is complete only for $N=2,3,4,5$, while
for $N\ge 6$ the full set of solutions remains  unknown.
The list of open questions could be rather long,  
but let us mention here some most relevant.

\smallskip 

\noindent \hangafter=1 \hangindent=0.8cm
{\bf i).\  } Check if there exist other inequivalent 
        complex Hadamard matrices of size $N=6$.

\noindent \hangafter=1 \hangindent=0.8cm
{\bf ii).  } Find the ranges of parameters of the existing $N=6$ families 
           such that all cases included are not equivalent. 

\noindent \hangafter=1 \hangindent=0.8cm
{\bf iii).} Check whether there exists a continuous family of
         complex Hadamard matrices for $N=11$.

\noindent \hangafter=1 \hangindent=0.8cm
{\bf iv).} Investigate if all inequivalent real Hadamard matrices
           of size $N=16,20$ belong to continuous families
           or if some of them are isolated.

\noindent \hangafter=1 \hangindent=0.8cm
{\bf v).} Find for which $N$ there exist continuous families
          of complex Hadamard matrices which are not affine, and which
          are not contained in \AUFs\ of a larger dimension.

\noindent \hangafter=1 \hangindent=0.8cm
{\bf vi).} Find  the dimensionalities of continuous orbits of 
           inequivalent Hadamard matrices stemming from $F_N$
           if $N$ is {\it not} a power of prime.

\smallskip 

Problems analogous to {\bf i)--iv)}
 are obviously open for higher dimensions.
Thus a lot of work is still required
to get a full understanding of the properties of the set of complex
Hadamard matrices, even for one--digit dimensions.
In spite of this fact we tend to believe that
the above collection of matrices will be useful
to tackle different physical problems,
in particular these motivated by the theory of quantum information
\cite{We01}. Interestingly,
the dimension $N=6$,
the smallest product of two different primes, is the 
first case for which not all complex Hadamard matrices are known,
as well as the simplest case 
for which the MUB problem remains open \cite{Gr04}.

\medskip

It is a pleasure to thank B.-G. Englert for
his encouraging remarks which inspired this project.
We enjoyed fruitful collaboration on Hadamard matrices
with I. Bengtsson, {\AA}. Ericsson, M. Ku{\'s} and W. S{\l}omczy{\'n}ski.
We would like to thank C.D. Godsil, M. Grassl, M. R{\"o}tteler,
and R. Werner  for fruitful discussions and
P.~Di{\c t}{\v a}, M.~Matolcsi, R. Nicoara,  W. Orrick,
 and T.~Tao for helpful correspondence.
K.{\.Z}. is grateful for hospitality
of the Perimeter Institute for Theoretical Physics in Waterloo,
where this work was initiated.
We acknowledge financial support by
Polish Ministry of Science and Information Technology
under the grant PBZ-MIN-008/P03/2003.
We are indebted to W. Bruzda for creating a web version of 
the catalogue of complex Hadamard matrices which may be updated in future --
see http://chaos.if.uj.edu.pl/${\sim}$karol/hadamard

\end{document}